\documentclass[11pt,a4paper]{article}
\pdfoutput=1
 \usepackage{jheppub}
\allowdisplaybreaks
\usepackage{graphicx,psfrag}
\usepackage{bm}
\usepackage{mathbbol,verbatim}
\usepackage{slashed}
\usepackage{graphics}
\usepackage{color,ulem}
\usepackage{multirow}
\usepackage{tikz}
\usetikzlibrary{trees}
\usetikzlibrary{decorations.pathmorphing}
\usetikzlibrary{decorations.markings}
\usetikzlibrary{decorations, decorations.markings, decorations.pathmorphing, arrows, graphs, shapes.geometric, snakes}
\usetikzlibrary{arrows}

\tikzset{
    photon/.style={decorate, decoration={snake}, draw=red},
    dark/.style={draw=gray, postaction={decorate},
        decoration={markings,mark=at position .55 with {\arrow[draw=gray]{>}}}},
antidark/.style={draw=gray, postaction={decorate},
        decoration={markings,mark=at position .55 with {\arrow[draw=gray]{<}}}},
electron/.style={draw=violet, postaction={decorate},
        decoration={markings,mark=at position .55 with {\arrow[draw=violet]{>}}}},
quark/.style={draw=blue, postaction={decorate},
        decoration={markings,mark=at position .55 with {\arrow[draw=blue]{>}}}},
antiquark/.style={draw=blue, postaction={decorate},
        decoration={markings,mark=at position .55 with {\arrow[draw=blue]{<}}}},
        gluon/.style={decorate, draw=or,
        decoration={coil,amplitude=2pt, segment length=3pt}},
  ZZ/.style={decorate, decoration={snake,amplitude=1.5pt, segment length=5pt}, draw=greeen},
left,
  }

\definecolor{greeen}{rgb}{0.03,0.84,0.13}
\definecolor{test}{rgb}{0.03,0.74,0.33}
\definecolor{viol}{rgb}{0.44,0,0.94}
\definecolor{or}{rgb}{0.95,0.65,0}

\begin{document}

\preprint{UMD-PP-016-002, ULB-TH/16-03}

\title{Probing the Higgs Sector of the Minimal Left-Right Symmetric Model at Future Hadron Colliders}
\author[a]{P. S. Bhupal Dev,}
\author[b]{Rabindra N. Mohapatra,}
\author[c]{and Yongchao Zhang}

\affiliation[a]{ Max-Planck-Institut f\"{u}r Kernphysik, Saupfercheckweg 1, D-69117 Heidelberg, Germany}
\affiliation[b]{Maryland Center for Fundamental Physics, Department of Physics, University of Maryland, College Park, MD 20742, USA}
\affiliation[c]{Service de Physique Th\'{e}orique, Universit\'{e} Libre de Bruxelles, Boulevard du Triomphe, CP225, 1050 Brussels, Belgium}

\emailAdd{bhupal.dev@mpi-hd.mpg.de}
\emailAdd{rmohapat@umd.edu}
\emailAdd{yongchao.zhang@ulb.ac.be}

\date{\today}

\abstract{
If neutrino masses arise from a TeV-scale minimal Left-Right seesaw model, the ensuing extended Higgs sector with neutral, singly and doubly-charged scalars has a plethora of implications for new Higgs boson searches beyond the Standard Model at future hadron colliders, such as the $\sqrt s=14$ TeV High-Luminosity Large Hadron Collider (HL-LHC) and the proposed $\sqrt s=100$ TeV collider (FCC-hh or SPPC). In this article, we provide a glimpse of this new physics in the Higgs sector. Our discussion focuses on the minimal non-supersymmetric version of the Left-Right model with high-scale parity breaking but TeV-scale $SU(2)_R$-breaking, a property desirable to suppress the type-II seesaw contribution to neutrino masses. We analyze the masses and couplings of the physical Higgs bosons in this model, and discuss their dominant production and decay modes at hadron colliders. We identify the best discovery channels for each of the non-SM Higgs bosons and estimate the expected SM backgrounds in these channels to derive the  sensitivity reaches for the new Higgs sector at future hadron colliders under discussion. Following a rather conservative approach, we estimate that the heavy Higgs sector can be effectively probed up to 15 TeV at the $\sqrt s=100$ TeV machine. We also discuss how the LR Higgs sector can be distinguished from other extended Higgs sectors.
}

\maketitle

\section{Introduction}
The neutrino oscillation data have unambiguously established that neutrinos have tiny but non-zero masses as well as mixing between different flavors. 
Understanding these observed features necessarily requires some new physics beyond the Standard Model (SM). 
Since the origin of masses for all the SM charged fermions has now been clarified by the discovery of the Higgs boson~\cite{Aad:2012tfa, Chatrchyan:2012xdj}, an important question is where the neutrino masses come from. If we simply add three right-handed (RH) neutrinos $N$ to the SM, one can write Yukawa couplings of the form ${\cal L}_{\nu,Y}=Y_{\nu N} \bar{L} H N$ ($L$ and $H$ being the SM lepton and scalar $SU(2)_L$-doublets) which, via the same SM Higgs vacuum expectation value (VEV) at the electroweak (EW) scale $\langle H^0 \rangle\equiv v_{\rm EW}$, give Dirac masses to neutrinos of magnitude $m_D = Y_{\nu N} v_{\rm EW}$. To get sub-eV neutrino masses, however, we need to have $Y_{\nu N} \lesssim 10^{-12}$, which is an ``unnaturally" small number, unless there is any symmetry reason behind it. So the strong suspicion among theorists is that there is some new physics beyond the SM Higgs that is responsible for the small neutrino masses. Such an approach is likely to involve new Higgs bosons. The goal of this article is to make a case that there exists a natural class of TeV-scale models for neutrino masses whose Higgs sector can be probed at the proposed $\sqrt s=100$ TeV Future Circular Collider (FCC-hh) at CERN~\cite{fcc-hh} or the Super Proton-Proton Collider (SPPC) in China~\cite{Tang:2015qga}, as well as at the $\sqrt s=14$ TeV Large Hadron Collider (LHC) to some extent. For a review on the physics potential of the 100 TeV collider, see Ref.~\cite{Arkani-Hamed:2015vfh} and for the prospects of Higgs physics at such energies, see Ref.~\cite{Baglio:2015wcg}.

The model we will consider uses the type-I seesaw paradigm for neutrino masses~\cite{type1a, type1b, type1c, type1d, type1e} where the RH neutrinos alluded to above have Majorana masses in the multi-TeV range in addition to Yukawa couplings like all charged fermions in the SM. Neutrinos being electrically neutral allows for this possibility, making them different from the charged fermions.  This new mass generation paradigm could also be at the root of such different mass and mixing patterns for leptons compared to quarks.  The starting point of this physics is the seesaw matrix with the generic form in the $ (\nu, N)$ flavor space:
\begin{eqnarray}
{\cal M}_\nu \ = \ \left(\begin{array}{cc}0 & m_D\\ m^{\sf T}_D & M_N\end{array}\right), \end{eqnarray}
where $m_D$ mixes the $\nu$ and $N$ states and is generated by the SM Higgs VEV and $M_N$ is the Majorana mass matrix for the heavy RH neutrinos $N$ which embodies the new neutrino mass physics. The masses of the light neutrinos are then given by the seesaw formula:
\begin{eqnarray}
m_\nu \ \simeq \ -m_DM_N^{-1}m_D^{\sf T} \, .
\label{eq:seesaw}
\end{eqnarray}
The ``unnaturalness" of the Yukawa couplings alluded to above is now considerably ameliorated due to two features of the seesaw formula~\eqref{eq:seesaw}: first, it now depends on the square of the Yukawa couplings $Y_{\nu N}$, unlike in the Dirac neutrino case where $m_\nu \propto Y_{\nu N}$, and secondly, it is suppressed by the heavy Majorana masses $M_N$. If $M_N\sim 10^{14}$ GeV as in grand-unified  theories (GUTs), $Y_{\nu N}$ can indeed be of order one, whereas if $M_N \sim 1-10$ TeV, $Y_{\nu N} \sim 10^{-11/2}$ is enough to explain the neutrino oscillation data.\footnote{There also exist a natural class of TeV-scale left-right models, where the Dirac Yukawa couplings could be larger if the neutrino mass matrices have some specific textures; see e.g. Ref.~\cite{Dev:2013oxa}.} The former possibility, i.e. near GUT scale $M_N$, though quite attractive theoretically, is hard to test experimentally. We therefore consider the multi-TeV scale possibility which can be probed not only at the 14 TeV LHC but also at future 100 TeV machines under discussion, as well as in low-energy lepton number violation (LNV) searches e.g. neutrinoless double beta decay ($0\nu\beta\beta$) and lepton flavor violation (LFV) searches at the intensity frontier. For a phenomenological review of TeV-scale seesaw models, see e.g. Refs.~\cite{Drewes:2013gca, Deppisch:2015qwa}.


A natural class of models that provides a possible ultraviolet (UV)-completion of the TeV-scale seesaw models is the Left-Right Symmetric Model (``LR model'' for short throughout this paper) of weak interactions~\cite{LR1, LR2, LR3}, originally introduced to understand parity violation observed in weak decays starting from a short-distance theory that conserves parity. The LR model is based on the gauge group $SU(2)_L\times SU(2)_R\times U(1)_{B-L}$, where the RH fermions $(u_R, d_R)$ and $(e_R, N_R)$ are assigned in a parity-symmetric way to the RH doublets of $SU(2)_R$. The RH neutrinos (and three generations of them) are therefore a necessary part of the model and do not have to be added adhocly just to implement the seesaw mechanism. An important point is that the RH neutrinos acquire a Majorana mass as soon as the $SU(2)_R$ symmetry is broken at the scale $v_R$. This is quite analogous to the way the charged fermions get mass as soon as the SM gauge symmetry $SU(2)_L$ is broken by $\langle H^0\rangle$. The $SU(2)_R$ and electroweak symmetry breaking in the LR model necessarily require that there must exist new Higgs bosons in addition to the 125 GeV Higgs boson discovered at the LHC. The physics of these extra Higgs fields, namely the bi-doublet scalars (denoted here by $H_1^0, A_1^0, H_1^\pm$) and the triplet scalars ($H_{2,3}^0$, $A_2^0$, $H_2^\pm$ and $H_{1,2}^{\pm\pm}$), are determined to a large extent by the fact that they must explain neutrino masses, and therefore, probing their properties in colliders may provide some crucial insight into the nature of neutrino masses. Some specific aspects of the non-supersymmetric LR Higgs sector relevant to our collider analysis have been studied in Refs.~\cite{Gunion:1986im, Gunion:1989in, Deshpande:1990ip, Polak:1991vf, Barenboim:2001vu, Azuelos:2004mwa, Jung:2008pz, Bambhaniya:2013wza, Dutta:2014dba, Bambhaniya:2014cia, Maiezza:2015lza, Bambhaniya:2015wna}. 
Here we provide an extensive study of the Higgs properties in the minimal LR model and their high energy tests, i.e. couplings, production and decays, to determine their signals and mass reach at future hadron colliders.

Before proceeding further, we note that since LR seesaw models lead to new effects and add new contributions to already known low energy weak processes, it is necessary to know whether TeV scale $SU(2)_R$ breaking is compatible with low energy observations. It turns out that the $\Delta F=2$ hadronic flavor changing neutral current (FCNC) effects such as $K_L-K_S$, $\epsilon_{K}$, $B_S-\bar{B}_S$ mixings, as well as $b \to s \gamma$, receive significant contributions from RH charged current effects~\cite{Beall:1981ze, Ecker:1983uh, Branco:1982wp, Zhang:2007da, Maiezza:2010ic, Blanke:2011ry, Bertolini:2014sua, Bernard:2015boz,Ball:1999mb,Bigi:1983bpa,Babu:1993hx}, and, therefore, provide the most stringent constraints on the $SU(2)_R$ breaking scale $v_R$, as well as on the LR Higgs boson masses. In particular, they restrict the mass of the RH charged $W_R$ boson  to be $M_{W_R} \gtrsim 3$ TeV (assuming $g_L=g_R$ for the $SU(2)_{L,R}$ gauge couplings) and the masses of the heavy bi-doublet Higgs bosons $\gtrsim$ 10 TeV~\cite{Zhang:2007da}.\footnote{The FCNC constraints on the bi-doublet fields can be relaxed by introducing an extra $SU(2)_R$ quark doublet to generate the quark mixings in the SM~\cite{Mohapatra:2013cia}.} Since the maximum available center-of-mass energy at the LHC goes up to 14 TeV, the LR  model as a theory of neutrino masses can be probed at the LHC using the smoking-gun signal of same-sign dilepton plus dijet with no missing transverse energy~\cite{Keung:1983uu, Ferrari:2000sp, Schmaltz:2010xr, Nemevsek:2011hz,Chen:2011hc, Chakrabortty:2012pp, Das:2012ii, AguilarSaavedra:2012gf, Han:2012vk, Chen:2013fna,  Rizzo:2014xma, Gluza:2015goa, Ng:2015hba, Dev:2015kca} (for a review, see e.g. Ref.~\cite{Deppisch:2015qwa}) as long as $M_{W_R}$ is below 4-5 TeV~\cite{Ferrari:2000sp}, whereas part of the LR Higgs sector not constrained by the FCNC constraints (e.g. the neutral and doubly-charged Higgs bosons mostly from the RH triplet) can also be probed at the LHC. The current direct limits on the RH gauge bosons~\cite{Khachatryan:2014dka, Aad:2015xaa} are of the same order as the indirect FCNC constraints, i.e. around 3 TeV, whereas the lower limits on the non-standard Higgs boson masses are much weaker, roughly varying from 100 to 500 GeV, depending on the search mode for the neutral~\cite{Aad:2014vgg, Aad:2015kna, Aad:2015agg, neutral1, Khachatryan:2014wca, Khachatryan:2015tha}, singly-charged~\cite{Aad:2014kga, Aad:2015nfa, Aad:2015typ, CMS:2014cdp, Khachatryan:2015qxa} and doubly-charged~\cite{ATLAS:2014kca, CMS:2016cpz} Higgs sectors. On the other hand, a 100 TeV $pp$ collider provides an unprecedented opportunity to probe the RH gauge boson masses up to $\sim 30-35$ TeV~\cite{Rizzo:2014xma, Dev:2015kca, Arkani-Hamed:2015vfh}, as well as the entire Higgs sector of the LR model, as demonstrated in this paper. There are also low energy tests of the model in the domain of leptonic physics, such as the LFV processes of $\mu\to e\gamma$, $\mu\to 3e$ and $\mu-e$ conversion in nuclei~\cite{Riazuddin:1981hz, Pal:1983bf, Mohapatra:1992uu, Cirigliano:2004mv, Cirigliano:2004tc, Bajc:2009ft, Tello:2010am, Das:2012ii, Barry:2013xxa, Dev:2013oxa, Vasquez:2015una, Awasthi:2015ota, Dev:2015kca, Bambhaniya:2015ipg}, electric dipole moment of neutron~\cite{Ecker:1983dj,Frere:1991jt, Maiezza:2014ala} and electron~\cite{Nieves:1986uk, Nemevsek:2012iq}, as well as the LNV process of $0\nu\beta\beta$~\cite{Mohapatra:1980yp, Mohapatra:1981pm, Picciotto:1982qe, Hirsch:1996qw, Tello:2010am, Arnold:2010tu, Chakrabortty:2012mh, Barry:2013xxa, Dev:2013vxa, Dev:2013oxa, Huang:2013kma, Dev:2014xea, Mahajan:2014nca, Ge:2015yqa, Borah:2015ufa, Awasthi:2015ota, Bambhaniya:2015ipg} which are the focus of the intensity frontier. Thus, the TeV-scale LR models straddle both the energy as well as the intensity frontier, although our main focus here will be on the energy frontier aspects of the LR Higgs sector.

The main new results of this paper are summarized below:
\begin{enumerate}
\item [(i)] We point out that there exists a theoretical lower bound on the ratio of the $SU(2)_R$ and $SU(2)_L$ gauge couplings, $g_R/g_L$, regardless of the way LR symmetry is broken [see Section~\ref{sec:2.3}].

\item [(ii)] We derive all relevant leading-order, tree-level couplings involving the heavy scalars in the minimal LR model [see Appendix~\ref{app:A}] and identify the most important ones for their collider phenomenology [see Tables~\ref{tab:H10} to \ref{tab:H2pp}].

\item [(iii)] For the bi-doublet scalars $H_1^0$, $A_1^0$ and $H_1^\pm$, both their production and decay at the 100 TeV collider are mainly dictated by their Yukawa couplings to the third generation SM fermions. For the scenario considered here, their key discovery channels are $bb \to  H_1^0 / A_1^0 \to bb$ and $bg \to H_1^\pm t \to ttb$ [see Figures \ref{fig:feynman1} to \ref{fig:production2}, and Tables \ref{tab:decay} and \ref{tab:bkg1}]. Using these modes, we can have a $3\sigma$-level sensitivity for the neutral bi-doublet scalars up to a mass of 15 TeV and for the singly-charged scalars up to a mass of 7 TeV at the 100 TeV collider [see Figure~\ref{fig:sens-bi}], independent of the other model parameters. Their prospects are not so promising at the LHC, mainly because of the FCNC constraints.

\item [(iv)] The hadrophobic scalars $H_3^0$ and $H_2^{\pm\pm}$ from the RH triplet can be dominantly produced either through the Higgs portal (for $H_3^0$), the Drell-Yan channel (for $H_2^{\pm\pm}$), or the vector boson fusion (VBF) process mediated by RH gauge bosons [see Figures \ref{fig:feynman3} to \ref{fig:prod-lhc}]. After being produced, they decay predominantly into two SM Higgs bosons ($H_3^0 \to hh$) and pairs of same-sign leptons ($H_2^{\pm\pm} \to \ell^\pm \ell^\pm$), respectively, as long as other decay modes involving heavy scalars and RH gauge bosons are not kinematically allowed [see Tables \ref{tab:decay} and \ref{tab:bkg2}]. Their sensitivities depend on the RH scale, and the scalar and gauge couplings [see Figures \ref{fig:discovery1} and \ref{fig:discovery2}]. According to our conservative estimate,  they can be probed up to a few TeV scale at the 100 TeV machine, and below TeV scale at the LHC.

\item [(v)] We discuss some possible distinctions of the Higgs signals in the minimal LR model from those arising in other popular multi-Higgs scenarios, such as the two Higgs doublet model (2HDM), which includes the Minimal Supersymmetric Standard Model (MSSM) Higgs sector [see Section~\ref{sec:7}].
\end{enumerate}

The rest of the article is organized as follows: in Section~\ref{sec:2}, we give a brief overview of the minimal TeV-scale LR models for type-I seesaw mechanism. In Section~\ref{sec:3}, we analyze the masses and couplings of the various Higgs fields in the minimal LR model.  In Sections~\ref{sec:4} and \ref{sec:5}, we discuss the dominant production and decay modes of the new Higgs fields, respectively. In Section~\ref{sec:6}, we identify the key discovery channels for each of the new Higgs bosons, and estimate the dominant SM background to calculate the sensitivity reach at future hadron colliders. In Section~\ref{sec:7}, we point out some possible ways one can distinguish the LR Higgs sector from the MSSM Higgs sector at colliders. Section~\ref{sec:8} summarizes our main results. In Appendix~\ref{app:A}, we give all the couplings of the SM and heavy Higgs bosons to the fermions, vector bosons and among themselves. In Appendix~\ref{app:B}, we give the exact formulas at leading order (LO) for various partial decay widths of the heavy Higgs bosons in the minimal LR model.


\section{Minimal TeV-scale Left-Right model} \label{sec:2}
The LR model~\cite{LR1, LR2, LR3} extends the SM gauge group ${\cal G}_{\rm SM}\equiv SU(3)_c\times SU(2)_L\times U(1)_Y$ to ${\cal G}_{\rm LR}\equiv SU(3)_c\times SU(2)_L\times SU(2)_R\times U(1)_{B-L}$. The quarks and leptons are assigned to the following irreducible representations of ${\cal G}_{\rm LR}$:
\begin{align}
& Q_{L,i} \ = \ \left(\begin{array}{c}u_L\\d_L \end{array}\right)_i : \: \left({ \bf 3}, {\bf 2}, {\bf 1}, \frac{1}{3}\right), \qquad \qquad
Q_{R,i} \ = \ \left(\begin{array}{c}u_R\\d_R \end{array}\right)_i : \: \left({ \bf 3}, {\bf 1}, {\bf 2}, \frac{1}{3}\right), \\
& \psi_{L,i} \ = \  \left(\begin{array}{c}\nu_L \\ e_L \end{array}\right)_i : \: \left({ \bf 1}, {\bf 2}, {\bf 1}, -1 \right), \qquad \qquad
\psi_{R,i} \ = \ \left(\begin{array}{c} N_R \\ e_R \end{array}\right)_i : \: \left({ \bf 1}, {\bf 1}, {\bf 2}, -1 \right),
\label{lrSM}
\end{align}
where $i=1,2,3$ represents the family index, and the subscripts $L,R$ denote  the left- and right-handed chiral projection operators $P_{L,R} = (1\mp \gamma_5)/2$, respectively. The $B$ and $L$ charges are fixed using the electric charge formula~\cite{Marshak:1979fm, Davidson:1978pm}
\begin{align}
Q \ = \ I_{3L}+I_{3R}+\frac{B-L}{2} \, .
\label{eq:charge}
\end{align}
In the minimal version of LR model the Higgs sector consists of the following multiplets:
\begin{eqnarray}
\Phi \ = \ \left(\begin{array}{cc}\phi^0_1 & \phi^+_2\\\phi^-_1 & \phi^0_2\end{array}\right) : ({\bf 1}, {\bf 2}, {\bf 2}, 0), \qquad
\Delta_R \ = \ \left(\begin{array}{cc}\Delta^+_R/\sqrt{2} & \Delta^{++}_R\\\Delta^0_R & -\Delta^+_R/\sqrt{2}\end{array}\right) : ({\bf 1}, {\bf 1}, {\bf 3}, 2).
\label{eq:scalar}
\end{eqnarray}
The gauge symmetry $SU(2)_R\times U(1)_{B-L}$ is broken by the VEV of the  neutral component of the $SU(2)_R$ triplet, $\langle \Delta^0_R\rangle \equiv v_R$, to the SM group $U(1)_Y$. There is also a left-handed counterpart $\Delta_L$ to $\Delta_R$, but we do not include this field here for two reasons: (i) There are versions of the model where parity and $SU(2)_R$ gauge symmetry scales are decoupled so that $\Delta_L$ fields are absent from the low-energy theory~\cite{CMP}, and the Higgs sector is simpler, as just noted in Eq.~\eqref{eq:scalar}; (ii) The presence of the $\Delta_L$ field in TeV-scale LR models generates a type-II seesaw~\cite{Mohapatra:1980yp, type2a, type2b, type2c, type2d} contribution to the neutrino masses which is large, requiring heavy fine-tuning of couplings in the scalar potential to understand neutrino masses. Thus, the decoupling of the $\Delta_L$ fields avoids this problem and provides a natural way to realize the type-I seesaw ~\cite{type1a, type1b, type1c, type1d, type1e} for neutrino masses in the minimal LR model. 
The SM electroweak gauge group is broken to $U(1)_{\rm EM}$ by the VEVs of the bi-doublet Higgs field $\Phi$:
\begin{align}
\langle \phi \rangle \ = \ \left(\begin{array}{cc}
\kappa & 0 \\
0 & \kappa' e^{i\alpha}
\end{array}\right) \, , \label{eq:phi}
\end{align}
where $\alpha$ is a CP phase and $\kappa^2+\kappa'^2 \equiv v_{\rm EW}^2 \simeq (174 \, {\rm GeV})^2$. In our discussion below, we will assume that $\kappa'\ll \kappa$, which largely simplifies our analytic expressions for the LR Higgs masses and couplings discussed later.

\subsection{Fermion Masses} \label{sec:2.1}
To see how the fermions get their masses and how seesaw mechanism arises in this model, we write down the Yukawa Lagrangian:
\begin{eqnarray}
\label{eqn:Lyukawa}
{\cal L}_Y \ & = & \  h^a_{q,ij}\bar{Q}_{L, i}\Phi_a Q_{R, j} +\tilde{h}^a_{q,ij}\bar{Q}_{L, i}\tilde{\Phi}_a Q_{R,j}+ h^a_{\ell, ij}\bar{\psi}_{L,i}\Phi_a \psi_{R, j} + \tilde{h}^a_{\ell, ij}\bar{\psi}_{L, i}\tilde{\Phi}_a \psi_{R, j} \nonumber \\
&& 
+ f_{ij} \psi_{R,i}^{\sf T} C i\tau_2 \Delta_R \psi_{R,j}  ~+~ {\rm H.c.}
\end{eqnarray}
where $a$ is for labeling the Higgs bi-doublets, $\tilde{\Phi}=\sigma_2\Phi^*\sigma_2$ ($\sigma_2$ being  the  second Pauli matrix) and $C=i\gamma_2\gamma_0$ is the charge conjugation operator ($\gamma_{\mu}$ being the Dirac matrices). After symmetry breaking, the quark and charged lepton masses are given by the generic formulas $M_u = h_u\kappa + \tilde{h}_u e^{-i\alpha} \kappa'$ for up-type quarks, $M_{d} = h_de^{i\alpha} \kappa' + \tilde{h}_d \kappa$ for down-type quarks, and similarly for the charged leptons.  The above Yukawa Lagrangian \eqref{eqn:Lyukawa} leads to the Dirac mass matrix  for neutrinos $m_D = h_{\ell}\kappa + \tilde{h}_{\ell} e^{-i\alpha} \kappa'$ and the Majorana mass matrix $M_N=fv_R$ for the RH neutrinos, which go into Eq.~(\ref{eq:seesaw}) for calculating the light neutrino masses and mixing.

\subsection{Gauge Boson Masses} \label{sec:2.2}
Given the above symmetry breaking pattern, the gauge boson masses can be easily obtained from the canonical kinetic term of the Lagrangian:
\begin{eqnarray}
\mathcal{L}_{\rm kinetic} \  = \ {\rm Tr}\left[(D_\mu \Phi)^{\dagger}(D^\mu \Phi)\right] + 
{\rm Tr}\left[(D_\mu \Delta_R)^{\dagger}(D^\mu \Delta_R)\right]  \,,
\end{eqnarray}
with the covariant derivatives
\begin{eqnarray}
D_\mu \Phi & \ = \ & \partial_\mu \Phi
- \frac{i}{2} g_L  \,    \overrightarrow{W}_{L\mu}\cdot \vec{\sigma} \, \Phi
+ \frac{i}{2} g_R  \, \Phi \, \overrightarrow{W}_{R\mu}\cdot \vec{\sigma} \,,  \\
D_\mu\Delta_{R} & \ = \ & \partial_\mu\Delta_{R}
- \frac{i}{2} g_{R} \left[ \overrightarrow{W}_{R\mu} \cdot \vec{\sigma} ,\, \Delta_{R} \right]
- i g_{BL} B_\mu \Delta_{R} \,,
\end{eqnarray}
where $\vec{\sigma}\equiv (\sigma_1, \sigma_2,\sigma_3)$ denotes the three Pauli matrices.

In the approximation $\kappa, \kappa'\ll v_R$, we get the following mass eigenvalues for the massive charged and neutral gauge bosons:
\begin{eqnarray}
M^2_{W} & \ = \ & \frac{g^2_L}{2}(\kappa^2+\kappa'^2)\, , \qquad \qquad M^2_{W_R}  \ = \  g^2_R  v^2_R \, , \label{eqn:gbosonmass} \\
M^2_{Z} & \ = \ & \frac{g^2_L}{2\cos^2\theta_w}(\kappa^2+\kappa'^2) \, , \qquad
M^2_{Z_R}  \ = \  2(g^2_R+g^2_{BL})v^2_R \, , \label{eqn:gbosonmassZ}
\end{eqnarray}
where $\theta_w$ is the weak mixing angle, defined as in the SM as $e=g_L\sin\theta_w$ ($e$ being the electromagnetic coupling), and we have used the relation~\cite{Georgi:1977wk}
\begin{align}
\frac{1}{g_Y^2} \ = \ \frac{1}{g_R^2}+\frac{1}{g_{BL}^2}\, .
\end{align}

The mixing between the SM $W$ boson and the heavy $W_R$ is given by
\begin{equation}
\tan\zeta_W \ = \ - \frac{g_L}{g_R} \frac{ \kappa \kappa' }{v^2_R}
\ \simeq \ - \xi_W
\left(\frac{M_{W}}{M_{W_R}}\right)^2 \,,
\end{equation}
with the parameter
\begin{eqnarray}
\xi_W \ = \ \frac{g_R}{g_L} \frac{ 2 \tan\beta }{ 1 + \tan^2\beta } \,,
\end{eqnarray}
where $\tan\beta = \kappa' / \kappa$ the ratio between the two EW-scale VEVs.
In the neutral gauge sector, the $3\times 3$ mass matrix in the basis $(W_L^3,W_R^3,B)$ is diagonalized by
\begin{eqnarray}
\left(\begin{array}{ccc}
W_H^3 \\ W_Z^3 \\ A
\end{array} \right) \ = \
\left(\begin{array}{ccc}
0 & \cos\phi & -\sin\phi \\
\cos\theta_w & - \sin\theta_w \sin\phi & - \sin\theta_w \cos\phi \\
\sin\theta_w & \cos\theta_w \sin\phi &  \cos\theta_w \cos\phi \\
\end{array} \right)
\left(\begin{array}{ccc}
W_L^3 \\ W_R^3 \\ B
\end{array} \right) \,,
\end{eqnarray}
where $\cos^2\phi \equiv \frac{g_{R}^2}{ g_{R}^2 + g_{BL}^2 }$.
The mixing between the SM $Z$ boson and the heavy $Z_R$ is also suppressed by the VEV ratio $\kappa^2 /v_R^2$:
\begin{eqnarray}
\zeta_Z \ \simeq \
 - \frac{\kappa^2}{4 v_R^2} \frac{\sin\phi\cos^3\phi}{\sin\theta_w}
\ = \ - \xi_Z \left( \frac{M_Z}{M_{Z_R}} \right)^2
\end{eqnarray}
with the parameter
\begin{eqnarray}
\xi_Z \ = \ \left[ \frac{g_R^2}{g_L^2} - \left(1+\frac{g_R^2}{g_L^2}\right) \sin^2\theta_w \right]^{1/2} \,.
\end{eqnarray}

\subsection{Lower limit on $g_R/g_L$} \label{sec:2.3}
Using the definitions of the mixing angles $\theta_w$ and $\phi$ given in Section~\ref{sec:2.2}, we find that to the leading order in $\kappa/v_R$, the masses of heavy gauge bosons are related via
\begin{eqnarray}
\label{eq:heavyWZratio}
\frac{M_{Z_R}}{M_{W_R}} \ \simeq \ \frac{\sqrt2}{\cos\phi} \ = \ \sqrt2 \frac{g_R}{g_L}\left(\frac{g_R^2}{g_L^2} - \tan^2\theta_w\right)^{-1/2} \,.
\end{eqnarray}
From Eq.~(\ref{eq:heavyWZratio}), we see that
to keep the masses of heavy gauge bosons from becoming imaginary~\cite{Brehmer:2015cia, Deppisch:2015cua}, it is theoretically required that in the LR model
\begin{eqnarray}
\frac{g_R}{g_L} \ \geq \ \tan\theta_w \ \simeq \ 0.55 \,.
\label{grl}
\end{eqnarray}
When the ratio $g_R / g_L$ goes to this theoretical limit, $Z_R$ becomes infinitely heavy and decouples from the EW scale and RH breaking physics. It might appear from this derivation that the lower limit on $g_R/g_L$ depends on the symmetry breaking pattern; however this is not so and we have found a completely general derivation of this bound, as given below.

Our starting point is the electric charge formula~\eqref{eq:charge} which implies that in the final theory after symmetry breaking, regardless of how symmetry breaking is implemented, there is a generic relation between the electric charge and the gauge couplings~\cite{Georgi:1977wk}:
\begin{eqnarray}
\frac{1}{e^2} \ = \ \frac{1}{g_L^2}+\frac{1}{g_R^2}+\frac{1}{g_{BL}^2} \, .
\label{eq:match}
\end{eqnarray}
Now using the definition of the mixing angle $\phi$ given in Section~\ref{sec:2.2}, we can write $g^2_R = {g^2_{BL}}\cot^2\phi$, and therefore, from Eq.~\eqref{eq:match}, we get
 \begin{eqnarray}
 \frac{g^2_R}{e^2} \ = \ \frac{g^2_R}{g^2_L}+ \frac{1}{\sin^2\phi} \, .
\label{eq:match2}
 \end{eqnarray}
 Now using the fact that $e= g_L\sin\theta_w$, we get from Eq.~\eqref{eq:match2}
 \begin{eqnarray}
\frac{g^2_R}{g^2_L} \ = \ \frac{\tan^2 \theta_w}{\sin^2\phi} \, ,
\end{eqnarray}
which {\it always} implies that $g_R^2/g_L^2 \geq \tan^2\theta_w$, as in Eq.~\eqref{grl}. This has important implications for the phenomenology of the LR models~\cite{Nemevsek:2011hz, Patra:2015bga, Aydemir:2015nfa}. Especially in the context of the recent CMS $eejj$~\cite{Khachatryan:2014dka} and ATLAS diboson~\cite{Aad:2015owa} excesses, an LR model interpretation necessarily requires $g_R<g_L$, which has interesting consequences for both LNV and LFV processes~\cite{Gluza:2015goa, Awasthi:2015ota, Bambhaniya:2015ipg, Brehmer:2015cia, Deppisch:2015cua, Deppisch:2014zta, Dobrescu:2015qna, Gao:2015irw, Dev:2015pga, Coloma:2015una, Dobrescu:2015jvn, Sajjad:2015urz, Das:2015ysz, Das:2016akd, Shu:2016exh, Aydemir:2015oob}. In addition, the lower limit on $W_R$ from leptogenesis constraints~\cite{Frere:2008ct, Dev:2014iva, Dev:2015vra, Dhuria:2015cfa} can be relaxed in LR models with $g_R<g_L$, thus opening up more parameter space compatible with the observed matter-antimatter asymmetry in our Universe.

\section{Heavy Higgs bosons in the minimal LR model}
\label{sec:3}
In order to discuss the Higgs sector of the minimal LR model, we need to write down the Higgs potential of the parity-symmetric theory involving the bi-doublet and triplet Higgs fields. The most general renormalizable scalar potential for the
$\Phi$ and $\Delta_{R}$ fields, which is invariant under the gauge group ${\cal G}_{\rm LR}$, is given by
\begin{eqnarray}
\label{eqn:potential}
\mathcal{V} & \ = \ & - \mu_1^2 \: {\rm Tr} (\Phi^{\dag} \Phi) - \mu_2^2
\left[ {\rm Tr} (\tilde{\Phi} \Phi^{\dag}) + {\rm Tr} (\tilde{\Phi}^{\dag} \Phi) \right]
- \mu_3^2 \:  {\rm Tr} (\Delta_R
\Delta_R^{\dag}) \nonumber
\\
&&+ \lambda_1 \left[ {\rm Tr} (\Phi^{\dag} \Phi) \right]^2 + \lambda_2 \left\{ \left[
{\rm Tr} (\tilde{\Phi} \Phi^{\dag}) \right]^2 + \left[ {\rm Tr}
(\tilde{\Phi}^{\dag} \Phi) \right]^2 \right\} \nonumber \\
&&+ \lambda_3 \: {\rm Tr} (\tilde{\Phi} \Phi^{\dag}) {\rm Tr} (\tilde{\Phi}^{\dag} \Phi) +
\lambda_4 \: {\rm Tr} (\Phi^{\dag} \Phi) \left[ {\rm Tr} (\tilde{\Phi} \Phi^{\dag}) + {\rm Tr}
(\tilde{\Phi}^{\dag} \Phi) \right]  \\
&& + \rho_1  \left[ {\rm
Tr} (\Delta_R \Delta_R^{\dag}) \right]^2 
+ \rho_2 \: {\rm Tr} (\Delta_R
\Delta_R) {\rm Tr} (\Delta_R^{\dag} \Delta_R^{\dag}) \nonumber
\\
&&+ \alpha_1 \: {\rm Tr} (\Phi^{\dag} \Phi) {\rm Tr} (\Delta_R \Delta_R^{\dag})
+ \left[  \alpha_2 e^{i \delta_2}  {\rm Tr} (\tilde{\Phi}^{\dag} \Phi) {\rm Tr} (\Delta_R
\Delta_R^{\dag}) + {\rm H.c.} \right]
+ \alpha_3 \: {\rm
Tr}(\Phi^{\dag} \Phi \Delta_R \Delta_R^{\dag}) \,.  \nonumber
\end{eqnarray}
Due to the LR symmetry, all the 12 parameters $\mu^2_{1,2,3}$, $\lambda_{1,2,3,4}$, $\rho_{1,2}$, $\alpha_{1,2,3}$ are real, and the only CP-violating phase is $\delta_2$ associated with the coupling $\alpha_2$, as explicitly shown in Eq.~\eqref{eqn:potential}. Minimizing the potential with respect to the three VEVs $\kappa$, $\kappa'$, $v_{R}$ and the phase $\alpha$ associated with the VEV $\kappa'$ [cf. Eq.~\eqref{eq:phi}]
leads to four relations among them and the coefficients in the potential:
\begin{eqnarray}
\label{f9}
\frac{\mu_1^2}{v_R^2} & \ = \ & \alpha_1 - \frac{\alpha_3 \xi^2}{1-\xi^2}
 + 2\left[ \lambda_1 (1+\xi^2)
+ 2 \lambda_4 \xi \cos \alpha \right] \epsilon^2 \,,  \\
\frac{\mu_2^2}{v_R^2} & \ = \ & \frac{\alpha_2 \cos(
\alpha + \delta_2 ) }{\cos \alpha} + \frac{\alpha_3 \xi}{2
(1-\xi^2)\cos
\alpha}
 +\left[ 4 \lambda_2 \xi \frac{\cos 2 \alpha}{\cos\alpha} + \frac{2 \lambda_3 \xi}{\cos\alpha} +
\lambda_4 (1+ \xi^2) \right]
\epsilon^2 \,, \\
\frac{\mu_3^2}{v_R^2} & \ = \ & 2\rho_1  +
\left[ \alpha_1 (1 + \xi^2) + \alpha_3 \xi^2 \right] \epsilon^2
+ 4 \alpha_2 \xi \epsilon^2 \cos( \alpha + \delta_2 )  \,, \\
&& 2 \alpha_2 (1- \xi^2)  \sin \delta_2 \ = \
\xi \sin \alpha \left[ \alpha_3 + (4 \lambda_3 - 8 \lambda_2) (1- \xi^2)
\epsilon^2 \right] \,,
\label{relation6}
\end{eqnarray}
where $\xi \equiv \kappa' / \kappa = \tan\beta$, $\epsilon \equiv \kappa/v_R$. These conditions can be used to remove the three mass parameters $\mu^2_{1,\,2,\,3}$ from the potential.
In the limit of $\xi,\, \epsilon \ll 1$, Eq.~\eqref{relation6} reduces to
\begin{eqnarray}
2 \alpha_2 \sin \delta_2 \ \simeq \ \xi \alpha_3 \sin \alpha \,,
\end{eqnarray}
which implies that in the CP-conserving limit of the potential with $\delta_2 \rightarrow 0$, the phase $\alpha \simeq 0$ in the VEV of $\Phi$.

The parity-symmetric theory has important implications for neutrino masses~\cite{Mohapatra:1980yp}. Note that in presence of the $\Delta_L$ field in the low-energy effective scalar potential, the additional minimization condition of the above Higgs potential with respect to $\langle \Delta_L^0\rangle \equiv v_L$ would require that $v_L \sim \kappa^2/v_R$. Thus, for $v_R\sim {\cal O}$(TeV), we have $v_L\sim {\cal O}$(MeV), which gives an unacceptably large type-II seesaw contribution $\sim fv_L$ to the Majorana mass of the left-handed neutrinos for $f\sim {\cal O}(1)$. One solution is to invoke huge cancellations between the type-I and type-II contributions to keep the left-handed neutrino masses at the sub-eV level. A more natural way is to eliminate the type-II seesaw contribution altogether, e.g. in a theory with D-parity breaking~\cite{CMP} where by introducing a parity-odd singlet with high-scale VEV, one can give a large mass to $\Delta_L$ so that it decouples from the low-energy theory. In that case, we can simply drop the $\Delta_L$ field from our analysis, as done throughout this paper.

As far as the Higgs bosons are concerned, two new kinds of physical Higgs bosons arise in the minimal LR theory, as given by Eq.~\eqref{eq:scalar}. The first class arises from the extension of the SM Higgs doublet to the LR model, i.e.~the bi-doublet field $\Phi$ and the second class from the RH triplet field $\Delta_R$ that breaks the $SU(2)_R$ symmetry. We will call the latter {\it hadrophobic} Higgs bosons, since they do not couple to quarks prior to symmetry breaking [cf.~Eq.~\eqref{eqn:Lyukawa}]. They are also responsible for the type-I seesaw scale and maintain their hadrophobic nature even after symmetry breaking, i.e. coupling only  to the lepton sector in the limit of $\kappa,\kappa'\ll v_R$. There emerge couplings to quarks only through their mixing to the bi-doublet Higgs sector, which are proportional to $\kappa/v_R$ or $\kappa'/v_R$.
Since our ultimate goal in this paper is the exploration of the scalars in LR model at the 14 TeV LHC and future 100 TeV collider, we will assume that the RH neutrinos and the RH symmetry breaking are both in the multi-TeV range (or going up to the few times 10 TeV range for the 100 TeV collider). This generally means that the new Higgs fields are also in the TeV range (or going up to few times 10 TeV range). The above assumption of $\kappa/v_R \ll1$ is therefore a very good one and we can safely neglect the bidoublet-triplet Higgs mixing in our subsequent analysis.

Considering only the bidoublet $\Phi$ and triplet $\Delta_R$ Higgs fields in the minimal model, there are a total number of 14 degrees of freedom in the scalar sector, of which two neutral components and two pairs of singly-charged states are eaten by the massive gauge bosons ($W^\pm, \, Z, \, W^\pm_R, \, Z_R$), thus leaving the remaining eight as the physical scalars. Taking the second derivative of the potential with respect to the dynamical fields in the linear decomposition around their VEVs, i.e.
\begin{eqnarray}
\phi_1^0 & \ = \ & \kappa + \frac{1}{\sqrt2} \phi_1^{\rm 0 \, Re} + \frac{i}{\sqrt2} \phi_1^{\rm 0 \, Im} \,, \nonumber \\
\phi_2^0 & \ = \ & \kappa' e^{i\alpha} + \frac{1}{\sqrt2} \phi_2^{\rm 0 \, Re} + \frac{i}{\sqrt2} \phi_2^{\rm 0 \, Im} \,, \nonumber \\
\Delta_R^0 & \ = \ & v_R + \frac{1}{\sqrt2} \Delta_R^{\rm 0 \, Re} + \frac{i}{\sqrt2} \Delta_R^{\rm 0 \, Im} \, ,
\end{eqnarray}
and the charged fields $\phi_{1,2}^\pm,\, \Delta_R^\pm,\, \Delta_R^{\pm\pm}$, we can arrive at the mass matrices for the neutral, singly and doubly-charged scalars, in the basis of the components below, respectively,
\begin{eqnarray}
\label{eqn:states}
&& \{ \phi_1^{\rm 0 \, Re},\, \phi_2^{\rm 0 \, Re},\,\, \Delta_R^{\rm 0 \, Re},\,
\phi_1^{\rm 0 \, Im},\, \phi_2^{\rm 0 \, Im},\, \, \Delta_R^{\rm 0 \, Im} \} \,, \quad
\{ \phi_1^\pm,\, \phi_2^\pm,\, \Delta_R^\pm \} \,, \quad
 \{  \Delta_R^{\pm\pm} \} \,.
\end{eqnarray}
Due to the large numbers of parameters in the potential~\eqref{eqn:potential}, the mass matrices are quite complicated.
From the phenomenological point of view, it is however helpful to make some reasonable and appropriate approximations to capture the main features of the theory in the interesting regions of the parameter space. To this end, we take into account the observation that $\epsilon = \kappa / v_R \ll 1$ for a multi-TeV scale seesaw. Also, in light of the third generation fermion mass dominance in the SM, it is a reasonable assumption that $\xi = \kappa' / \kappa \ll 1$. Furthermore, CP observables require that the phase $\alpha \ll 1$. For the ease of perturbative expansions,  we assume the small parameters $\epsilon$, $\xi$ and $\alpha$ are about of the same order.

In this simplified scenario, we first consider the neutral scalars. To obtain the mass of 125 GeV for the SM Higgs, we expand the whole $6\times6$ mass matrix in the basis of the neutral states in Eq.~(\ref{eqn:states}) to the second order of $\epsilon^2 v_R^2 \simeq \kappa^2 = v_{\rm EW}^2$. In doing this, it becomes clear that the state $\Delta_R^{\rm 0 \, Im}$ is just the longitudinal component for the heavy $Z_R$ boson. The remaining five states correspond to the SM Higgs boson $h$, the Goldstone boson $G_Z$ for the SM $Z$ boson, and two heavy CP-even states $H_{1,\,3}^0$ and one heavy CP-odd scalar $A_1^0$.\footnote{We have denoted the real part of $\Delta^0_R$ as $H_3^0$ and real part of the heavy decoupled field $\Delta^0_L$ as $H_2^0$; because of decoupling of $\Delta_L$ fields, $H^0_2$ does not appear in the low-energy spectrum. Similarly for the doubly-charged scalars, the $H_1^{\pm\pm}$ from $\Delta_L$ decouples and we are left only with $H_2^{\pm\pm}$ from $\Delta_R$.} To the LO in $v_R^2$, the $5\times5$ matrix reads
\begin{eqnarray}
\mathcal{M}^0_{(0)} \ = \ v_R^2 \left(
\begin{array}{ccccc}
 0 & 0 & 0 &  0 &0\\
 0 & \alpha _3 & 0&0  & 0 \\
 0 & 0 & 4 \rho _1  & 0&0 \\
 0 & 0 & 0 & 0 & 0 \\
 0 & 0 & 0  &0 & \alpha _3
\end{array}
\right) \,,
\end{eqnarray}
which implies that the new scalars $H_{1,\,3}^0$ and $A_1^0$ are all at the $v_R$ scale, if the relevant quartic couplings $\alpha_3$ and $\rho_1$ are of order one. The mass matrices at the linear and quadratic orders of the small parameters $\xi \sim \epsilon \sim \alpha$ are, respectively,
\begin{align}
\label{eqn:scalarmatrix2}
\mathcal{M}^0_{(1)} & \ = \ v_R^2 \left(
\begin{array}{ccccc}
 0 & -\alpha _3\xi   & 2 \alpha _1\epsilon  &0 & 0 \\
 -\alpha _3\xi   & 0 & 4 \alpha _2\epsilon  &0 & 0 \\
 2 \alpha _1\epsilon  & 4 \alpha _2\epsilon  &0 & 0 & 0 \\
 0 & 0 & 0  & 0 & \alpha _3 \xi   \\
 0 & 0 & 0 & \alpha _3 \xi   & 0
\end{array}
\right) \,, \\
\mathcal{M}^0_{(2)} & \ = \ v_R^2 \left(
\begin{array}{ccccc}
 4 \lambda _1 \epsilon ^2+ \alpha _3 \xi ^2& 4 \lambda _4 \epsilon ^2 & 4\alpha _2 \epsilon  \xi  & 0 &
   -\alpha _3\alpha  \xi  \\
 4 \lambda _4\epsilon ^2 & 4 \left(2 \lambda _2+\lambda _3\right) \epsilon ^2+\alpha _3\xi ^2 & 2
   \left(\alpha _1+\alpha _3\right)\epsilon  \xi & -\alpha _3\alpha  \xi  & 0 \\
 4 \alpha _2\epsilon  \xi  & 2 \left(\alpha _1+\alpha _3\right)\epsilon  \xi & 0 & 0 & 0 \\
 0 & -\alpha _3\alpha  \xi  & 0 & \alpha _3\xi ^2 & 0 \\
 -\alpha _3\alpha  \xi  & 0 & 0 & 0 & \left(4 \lambda _3-8 \lambda _2\right) \epsilon ^2+\alpha _3\xi ^2
\end{array}
\right). \nonumber \\
\end{align}
Then the full mass matrix up to the order of $\mathcal{O} (\epsilon^2)$,
\begin{eqnarray}
\mathcal{M}^0_{} \ = \
\mathcal{M}^0_{(0)} + \mathcal{M}^0_{(1)} + \mathcal{M}^0_{(2)}
\end{eqnarray}
can be diagonalized by a rotation matrix up to $\mathcal{O} (\epsilon^2)$, which renders
\begin{eqnarray}
\small
\label{eqn:higgsmixing}
\left( \begin{array}{c}
h \\ H_1^0 \\ H_3^0 \\ G_Z \\ A_1^0 \\
\end{array} \right) =
\left(
\begin{array}{ccccc}
 1 -\frac12 \xi^2 -\frac{  \alpha _1^2\epsilon^2}{8 \rho _1^2} & \xi  & -\frac{  \alpha _1\epsilon}{2 \rho _1} & 0 & \alpha  \xi  \\
 -\xi  & 1 -\frac12 \xi^2 -\frac{8 \alpha _2^2\epsilon^2}{(4 \rho _1-\alpha _3)^2} & -\frac{4 \alpha _2\epsilon}{4 \rho _1-\alpha _3} & -\alpha  \xi  & 0 \\
 \frac{  \alpha _1\epsilon}{2 \rho _1} & \frac{4 \alpha _2\epsilon}{4 \rho _1-\alpha _3} & 1 -\frac{  \alpha _1^2\epsilon^2}{8 \rho _1^2} -\frac{8 \alpha _2^2\epsilon^2}{(4 \rho _1-\alpha _3)^2} & 0 & 0 \\
 0 & \alpha  \xi  & 0 & 1 -\frac12 \xi^2 & -\xi  \\
 -\alpha  \xi  & 0 & 0 & \xi  & 1 -\frac12 \xi^2
\end{array} \right)
\left( \begin{array}{c}
\phi_1^{\rm 0 \, Re} \\ \phi_2^{\rm 0 \, Re} \\ \Delta_R^{\rm 0 \, Re} \\ \phi_1^{\rm 0 \, Im} \\ \phi_2^{\rm 0 \, Im} \\
\end{array} \right) \,. \nonumber \\ \small
\end{eqnarray}
After the diagonalization, we arrive at the SM Higgs $h$, the two CP-even scalars $H^0_{1,\,3}$ and the CP-odd $A_{1}^0$, as well as the two Goldstone bosons $G_Z$ and $G_{Z_R}$, with the physical masses given by
\begin{align}
M_h^2 & \ = \  \left(  4 \lambda _1-\frac{\alpha _1^2}{\rho _1} \right) \kappa^2 \,, \label{eqn:hmass} \\
M_{H_1^0}^2 & \ = \  \alpha _3 ( 1 + 2 \xi ^2 ) v_R^2 + 4 \left( 2 \lambda
   _2+\lambda _3 + \frac{4 \alpha _2^2 }{\alpha _3-4 \rho _1} \right) \kappa^2 \,, \label{eqn:H10mass} \\
M_{H_3^0}^2 & \ = \ 4 \rho_1 v_R^2 + \left( \frac{\alpha_1^2}{\rho_1} - \frac{16 \alpha_2^2}{\alpha_3 -4\rho_1} \right) \kappa^2 \,, \label{eqn:H30mass} \\
M_{A_1^0}^2 & \ = \ \alpha _3 ( 1 + 2 \xi ^2 ) v_R^2 +4 \left(\lambda _3-2 \lambda _2\right) \kappa^2 \,. \label{eqn:A10mass}
\end{align}
It should be noted here that in the minimal version of LR model, none of these heavy neutral Higgs bosons can act as a viable candidate for the recently observed diphoton excess at 750 GeV~\cite{ATLAS:2015, CMS:2015dxe}, as the bi-doublet components $H_1^0$ and $A_1^0$ are stringently constrained by the FCNC data, and the neutral triplet scalar $H_3^0$ can not be produced abundantly enough at the LHC to explain the diphoton events~\cite{Dasgupta:2015pbr,Deppisch:2016scs,Berlin:2016hqw}.\footnote{However, in an alternative minimal version of the LR models with only two doublets to break the $SU(2)_L \times SU(2)_R \times U(1)_{B-L}$ gauge group down to the electromagnetic gauge group $U(1)_{\rm EM}$~\cite{Babu:1988mw,Babu:1989rb}, heavy vector-like fermions and a singlet scalar can be introduced to generate the SM fermion masses via the generalized seesaw mechanism, and the diphoton events can be explained in a natural manner~\cite{Dev:2015vjd}. The phenomenology of the heavy Higgs bosons in the alternative version of minimal LR model can be found in, e.g. Ref.~\cite{Mohapatra:2014qva}. }

For the singly-charged scalars, the mass matrix is given by, up to the quadratic order in $\epsilon$, and in the basis of the singly-charged states in Eq.~(\ref{eqn:states}),
\begin{eqnarray}
{\cal M}^+_{(2)} \ = \ v_R^2 \left(
\begin{array}{cccc}
 \xi ^2 \alpha _3 &  \xi  \alpha _3 (1-i \alpha )  & \frac{1}{\sqrt{2}} \epsilon  \xi  \alpha _3 \\
 \xi  \alpha _3 (1+ i \alpha ) & \alpha _3 \left(1+ \xi ^2\right)  & \frac{1}{\sqrt{2}} \epsilon  \alpha _3
   \\
 \frac{1}{\sqrt{2}} \epsilon  \xi  \alpha _3 & \frac{1}{\sqrt{2}} \epsilon  \alpha _3 & \frac{\epsilon ^2
   \alpha _3}{2}
\end{array}
\right) \,.
\end{eqnarray}
The corresponding rotation is
\begin{eqnarray}
\label{eqn:higgsmixing2}
\left( \begin{array}{c}
G_L^+ \\ H_1^+ \\ G_R^+ \\
\end{array} \right) \ = \
\left(
\begin{array}{ccc}
 1 -\frac12 \xi^2 & - \xi (1-i\alpha)    & 0 \\
 \xi (1+ i \alpha )  & 1  -\frac12 \xi^2 -\frac14 \epsilon^2 & \frac{1 }{\sqrt{2}} \epsilon \\
 0 & -\frac{1 }{\sqrt{2}} \epsilon  & 1 -\frac14 \epsilon^2
\end{array}
\right)
\left( \begin{array}{c}
\phi_1^+ \\ \phi_2^+ \\  \Delta_R^+ \\
\end{array} \right) \,,
\end{eqnarray}
where $G^+_{L,\,R}$ are the Goldstone bosons eaten by the SM $W^+$ and heavy $W_R^+$ gauge bosons, and $H_{1}^+$ is the singly-charged Higgs mass eigenstate with mass naturally at the $v_R$ scale:
\begin{eqnarray}
\label{eqn:scalarmass2}
M^2_{H_1^\pm} \ = \ \alpha _3 \left[(1 +2 \xi ^2) v_R^2 + \frac12 \kappa^2\right] \,.
\end{eqnarray}
It is obvious that the scalar eigenstates ($H_1^0$, $A_1^0$, $H_1^\pm$) from the same doublet in $\Phi$ have nearly degenerate masses $\sqrt{\alpha_3} v_R$ at the LO. It is trivial to get the mass for the doubly charged scalar:
\begin{eqnarray}
\label{eqn:scalarmass3}
M^2_{H_2^{\pm\pm}} \ = \ 4 \rho_2 v_R^2 + \alpha_3 \kappa^2 \,.
\end{eqnarray}

It is straightforward to obtain all the couplings of the SM and heavy Higgs bosons to the fermions, vector bosons and among themselves, and the full lists of couplings are collected in Tables~\ref{tab:triple} to \ref{tab:gauge2} in Appendix~\ref{app:A}. Due to the large number of heavy Higgs bosons and the quartic couplings in the scalar potential (\ref{eqn:potential}), most of the couplings look rather complicated; to simplify them, we have expanded in terms of the small VEV ratios $\epsilon$ and $\xi$, and the small CP-violating phase $\alpha$, as done for the scalar masses above. The most relevant couplings for the production and decays of the heavy Higgs bosons in the minimal LR model as discussed below are collected in Tables \ref{tab:H10} to \ref{tab:H2pp}.
\begin{table}[t!]
  \centering
  \caption[]{The couplings relevant to $H_1^0$ production and decay at hadron colliders. 
}
  \label{tab:H10}
  \begin{tabular}{lll}
  \hline\hline
  coupling & value \\ \hline
  $H^0_1 hh$ & $ \sqrt2 \left[ 3 \lambda _4  +\alpha _1 \alpha _2 \left(\frac{2}{\alpha _3-4 \rho _1}-\frac{1}{\rho _1}\right)\right] \kappa$ \\
  $H^0_1 h H^0_3$ & $ 2 \sqrt{2} \alpha _2 v_R $  \\
  $h H^0_1 H^0_1$ & $ \sqrt2 \left[
   \lambda _1 +4 \lambda _2 +2 \lambda _3 +\frac{8 \alpha _2^2}{\alpha _3-4 \rho _1}-\frac{\alpha _1 \left(\alpha _1+\alpha _3\right)}{4\rho _1} \right] \kappa $ \\
$H_1^0 H^0_1 H^0_1$ & $ \sqrt{2}  \left[ \lambda _4 +\frac{2 \alpha _2 \left(\alpha _1+\alpha _3\right)}{\alpha _3-4 \rho _1} \right] \kappa$ \\
$H_3^0 H^0_1 H^0_1$ & $ \frac{1}{\sqrt{2}} \left(\alpha _1+\alpha _3 \right) v_R $  \\
\hline
  $H_1^0 h hh$ & $\lambda_4$  \\ \hline
  $H_1^0 \bar{t}t$ & $- \frac{m_b}{\sqrt2\kappa}$  \\
  $H_1^0 \bar{b}b$ & $\frac{m_t}{\sqrt2\kappa}$  \\
  $H_1^0 N N$ & $\frac{M_N}{\sqrt2 v_R} \left( -\frac{4 \alpha _2\epsilon}{4 \rho _1-\alpha _3} \right)$  \\ \hline
  $H_1^0  W_{R }^+ W_R^{-}$ & $\frac{g_R^2 v_R }{\sqrt2} \left(\frac{8 \alpha _2 \epsilon }{\alpha _3-4 \rho _1} \right)$ \\
  $H_1^0  Z_{R } Z_R^{}$ & $\frac{\sqrt2 g_R^2 v_R }{\cos^2\phi} \left(\frac{4 \alpha _2 \epsilon }{\alpha _3-4 \rho _1} \right)$ \\
  $H_1^0  W_{}^+ W_R^{-}$ & $-\frac{ g_L g_R \kappa }{\sqrt2} $ \\
  \hline\hline
  \end{tabular}
\end{table}

\begin{table}[t!]
  \centering
  \caption[]{The couplings relevant to $A_1^0$ production and decay at hadron colliders.
For $HHV$ couplings, $k$ is the corresponding momentum of the scalar field pointing into the vertex. }
  \label{tab:A10}
  \begin{tabular}{lll}
  \hline\hline
  couplings & values  \\ \hline
  $A^0_1 A^0_1 h$ & $\sqrt2 \left[ \left(\lambda _1-4 \lambda _2+2 \lambda _3\right)-\frac{\alpha _1 \left(\alpha _1+\alpha _3\right)}{4\rho_1}  \right]   \kappa$ &  \\
$H_1^0 A^0_1 A^0_1$ & $\sqrt2 \left[ \lambda _4  +\frac{2 \alpha _2 \left(\alpha _1+\alpha _3\right)}{\alpha _3-4 \rho _1} \right] \kappa $ \\
$H_3^0 A^0_1 A^0_1$ & $\frac{1}{\sqrt{2}} \left(\alpha _1+\alpha _3 \right) v_R$  \\
 \hline
  $A_1^0 \bar{t}t$ & $\frac{i m_b}{\sqrt2 \kappa}$ \\
  $A_1^0 \bar{b}b$ & $\frac{i m_t}{\sqrt2 \kappa}$ \\  \hline
  $A_1^0  W_{}^+ W_R^{-}$ & $-\frac{ ig_L g_R \kappa}{\sqrt2} $ \\ \hline
  $A_1^0 H_1^0 Z$ & $  -\frac{ig_L}{2\cos\theta_w} \times \left[ k(A_1^0) - k(H_1^0)\right]$  \\
  $A_1^0 H_1^+ W_{}^- $ & $ - \frac{i}{2} g_L \times \left[ k(A_1^0) - k(H_1^+)\right]$ \\
  $A_1^0 H_1^0 Z_R$ & $ \frac{ig_L}{\cos\theta_w} \left(
  \frac{1}{2} \sin\theta_w \cot\phi \right) \times \left[ k(A_1^0) - k(H_1^0)\right]$ \\
  \hline\hline
  \end{tabular}
\end{table}

\begin{table}[t!]
  \centering
  \caption[]{The couplings relevant to $H_1^\pm$ production and decay at hadron colliders. For $HHV$ couplings, $k$ is the corresponding momentum of the scalar field pointing into the vertex. }
  \label{tab:H1p}
  \begin{tabular}{lll}
  \hline\hline
  couplings & values \\ \hline
$h H^+_1 H^-_1$ & $\frac{1}{\sqrt2} \left[ \alpha _3+4 \lambda _1 -\frac{\alpha _1 \left(\alpha _1+\alpha _3\right)}{\rho _1}  \right] \kappa$ \\
$H_1^0 H_1^+ H_1^-$ & $2\sqrt2 \left[ \lambda _4 + \frac{2 \alpha _2 \left(\alpha _1+\alpha _3\right)}{\alpha _3-4 \rho _1} \right] \kappa$ \\
$H_3^0 H_1^+ H_1^-$ & $\sqrt{2} \left( \alpha _1+\alpha _3\right) v_R$  \\ \hline
  $H_1^+ \bar{t}_L b_R$ & $\frac{m_t}{\sqrt2\kappa}$  \\
  $H_1^+ \bar{t}_R b_L$ & $\frac{m_b}{\sqrt2\kappa}$  \\
  $H_1^+ \bar{N} \tau_L$ & $- \frac{m_\tau}{\sqrt2\kappa}$  \\
  $H_1^+ N e_R$ & $- \frac{\epsilon M_N}{\sqrt2 v_R}$  \\
\hline
  $H_1^+  W_{R }^{-} Z_R^{}$ & $\frac{g_L^2 \kappa}{\sqrt2} \left[ - \frac{\tan\theta_w (1+\sin^2\phi)}{\sin^2\phi\cos\phi} \right]$ \\
  $H_1^+  W_{R }^- Z^{}$ & $-\frac{ g_L g_R \kappa }{\sqrt2 \cos\theta_w} $ \\ \hline
  $H_1^{+} H_1^{-} \gamma$ & $e \times \left[ k(H_1^{+}) - k(H_1^{-})\right]$ \\
  $H_1^{+} H_1^{-} Z$ & $\frac{g_L \cos2\theta_w }{2\cos\theta_w} \times \left[ k(H_1^{+}) - k(H_1^{-})\right]$ \\
  $H_1^{+} H_1^{-} Z_R$ & $ \frac12 g_L \tan\theta_w\cot\phi  \times \left[ k(H_1^{+}) - k(H_1^{-})\right]$ \\
  $H_1^+ h  W_{R}^- $ & $- \frac12 g_R \times \left[ k(H_1^+) - k(h)\right]$ \\
  \hline\hline
  \end{tabular}
\end{table}

\begin{table}[t!]
  \centering
  \caption[]{The couplings relevant to $H_3^0$ production and decay at hadron colliders.
}
  \label{tab:H30}
  \begin{tabular}{lll}
  \hline\hline
  couplings & values \\ \hline
  $H^0_3 hh$ & $ \frac{1}{\sqrt{2}} \alpha _1  v_R $ \\
  $H^0_3 h H^0_1$ & $ 2 \sqrt{2} \alpha _2 v_R $ \\
  $h H^0_3 H^0_3$ & $ -\sqrt2 \alpha _1 \kappa \left[ 1 - \frac{\alpha _1}{2\rho _1} +\frac{8 \alpha _2^2}{\alpha_1(\alpha _3-4 \rho _1)}  \right] $ \\
  $H_1^0 H^0_3 H^0_3$ & $ -2\sqrt2 \alpha_2 \kappa \left[ 1 - \frac{\alpha _1}{2\rho _1}+\frac{4 \alpha _1+\alpha _3}{2(\alpha _3-4 \rho _1)} \right] $ \\
  $H_3^0 H^0_1 H^0_1$ & $ \frac{1}{\sqrt{2}} \left(\alpha _1+\alpha _3 \right) v_R $  \\
  $H_3^0 A^0_1 A^0_1$ & $\frac{1}{\sqrt{2}} \left(\alpha _1+\alpha _3 \right) v_R$  \\
$H_3^0 H^0_3 H^0_3$ & $ \sqrt{2} \rho_1 v_R $ \\
$H_3^0 H_1^+ H_1^-$ & $\sqrt{2} \left( \alpha _1+\alpha _3\right) v_R$  \\
  $H_3^0 H_2^{++} H_2^{--}$ & $2 \sqrt{2} \left(\rho _1+2 \rho _2\right)v_R $ \\
\hline
  $H_3^0 \bar{t} t$ & $ \frac{m_t}{\sqrt2 \kappa}\frac{\alpha _1 \epsilon }{2 \rho _1}$ \\
  $H_3^0 \bar{b}b$  & $- \frac{m_t}{\sqrt2 \kappa} \frac{4 \alpha _2 \epsilon }{\alpha_3-4 \rho _1}$ \\
  $H_3^0 NN$  & $\frac{M_N}{\sqrt2 v_R} $ \\ \hline
  $H_3^0  W_{R }^+ W_R^{-}$ & $\sqrt2 g_R^2 v_R $ \\
  $H_3^0  Z_{R } Z_R^{}$ & $\frac{\sqrt2 g_R^2 v_R }{\cos^2\phi} $ \\
  \hline\hline
  \end{tabular}
  \vspace{1.5cm}
\end{table}

\begin{table}[t!]
  \centering
  \caption[]{The couplings relevant to $H_2^{\pm\pm}$ production and decay at hadron colliders. For $HHV$ couplings, $k$ is the corresponding momentum of the scalar field pointing into the vertex. }
  \label{tab:H2pp}
  \begin{tabular}{lll}
  \hline\hline
  couplings & values \\ \hline
  $h H^{++}_2 H^{--}_2$ & $\sqrt2 \left[ \alpha _3-\frac{2 \alpha _1 \rho _2}{\rho _1} \right] \kappa $ \\
  $H_1^0 H_2^{++} H_2^{--}$ & $2\sqrt2 \alpha_2 \kappa \, \frac{\alpha _3+8 \rho _2}{\alpha _3-4 \rho _1} $ \\
  $H_3^0 H_2^{++} H_2^{--} $ & $2 \sqrt{2} \left(\rho _1+2 \rho _2\right)v_R $ & \\ \hline
  $H_2^{++} e_R e_R$ & $- \frac{M_N}{\sqrt2 v_R}$  \\ \hline
  $H_2^{++} W^{-}_{R } W^{-}_R$ & $-2 g_R^2 v_R$ \\
\hline
  $H_2^{++} H_2^{--} \gamma$ & $2e \times \left[ k(H_2^{++}) - k(H_2^{--})\right]$ \\
  $H_2^{++} H_2^{--} Z$ & $   \frac{g_L} {\cos\theta_w} \left( -2 \sin^2\theta_w \right)\times \left[ k(H_2^{++}) - k(H_2^{--})\right]$ \\
  $H_2^{++} H_2^{--} Z_R$ & $   \frac{g_L} {\cos\theta_w} \left[  \sin\theta_w (\cot\phi-\tan\phi) \right]\times \left[ k(H_2^{++}) - k(H_2^{--})\right]$ \\
  \hline\hline
  \end{tabular}
\end{table}

\section{Production of the heavy Higgs bosons} \label{sec:4}

In this section we give the parton-level production cross sections for the heavy scalar fields in the minimal LR model at the 14 TeV LHC and future $100$ TeV FCC-hh/SPPC. The new Higgs fields $(H^0_1, A^0_1, H^\pm_1)$ from the bi-fundamental representation, being in the same SM doublet, are quasi-degenerate in mass and in terms of the fields in Eq. (\ref{eqn:states}), they are given by
\begin{eqnarray}
H_1^0 \ \equiv \phi^{\rm 0 \, Re}_2 \,, \quad
A_1^0 \ \equiv \ \phi^{\rm 0 \, Im}_2 \,, \quad
H_1^\pm \ \equiv \ \phi^\pm_2 \, ,
\end{eqnarray}
in the limit of $\kappa \ll v_R$.
It turns out that the masses of the $H_1^0, A_1^0$ fields are constrained to be $M_{H_1^0}\geq 8-10$ TeV by low energy FCNC effects~\cite{Zhang:2007da}, which is also applicable to the mass of $H_1^\pm$, as argued above. This indirect limit on their masses is much stronger than the direct search limits from the LHC data~\cite{Aad:2014vgg, Aad:2015kna, Aad:2015agg, neutral1, Khachatryan:2014wca, Khachatryan:2015tha, Aad:2014kga, Aad:2015nfa, Aad:2015typ, CMS:2014cdp, Khachatryan:2015qxa}. These bi-doublet fields are therefore not accessible at the LHC, but ripe for searches at the 100 TeV collider.

As for the hadrophobic Higgs fields from the components of $\Delta_R$, they will mix with bi-doublet Higgs components by a small amount and in the limit of $\kappa/v_R\ll 1$, we can identify these fields as
\begin{eqnarray}
H_3^0 \ \equiv \ \Delta^{\rm 0 \, Re}_R \,, \quad
H_2^{\pm\pm} \ \equiv \ \Delta^{\pm\pm}_R \,.
\end{eqnarray}
The current direct search limits for the doubly-charged scalars are in the range of 500-600 GeV~\cite{ATLAS:2014kca, CMS:2016cpz}. One should also keep in mind the lower limit $v_R\gtrsim 5$ TeV is derived from the constraints on $M_{W_R}$~\cite{Khachatryan:2014dka, Aad:2015xaa} and $M_{Z_R}$~\cite{Patra:2015bga}. In addition, there are various other constraints on RH neutrinos at sub-TeV scale, which can be extended to TeV-scale $M_N$ at future colliders, depending on the light-heavy neutrino mixing~\cite{Deppisch:2015qwa, Atre:2009rg, Dev:2012zg, Cely:2012bz, Dev:2013wba, Das:2014jxa, Alva:2014gxa, Antusch:2015mia, Banerjee:2015gca, Izaguirre:2015pga, Gago:2015vma, Asaka:2015oia, Das:2015toa, deGouvea:2015euy, Antusch:2015gjw, Basso:2015aee, Dev:2016vif}. The hadrophobic Higgs sector of the LR model provides a complementary probe of the seesaw scale at future colliders, independent of the neutrino mixing.

Using the couplings collected in Tables \ref{tab:H10} to \ref{tab:H2pp}, we calculate the main collider signals of the heavy Higgs sector in the minimal LR model. We require $M_{H^0_1}, M_{H^\pm_1}, M_{A^0_1}\geq 10$ TeV to satisfy the FCNC constraints, whereas  $M_{H_2^{\pm\pm}}, M_{H^0_3}$ can be as light as a few hundred GeV, since there are no such stringent flavor constraints on them.
Some representative Feynman diagrams for the dominant production channels of the heavy bi-doublet and hadrophobic scalars are presented in Figures~\ref{fig:feynman1}, \ref{fig:feynman2}, \ref{fig:feynman3} and \ref{fig:feynman4}.
We use {\tt CalcHEP3.6.25}~\cite{Belyaev:2012qa} to do the LO parton-level simulations with all the relevant couplings implemented into the model files, which is linked to {\tt LHAPDF6}~\cite{Buckley:2014ana} to use the {\tt CT14}~\cite{Dulat:2015mca} parton distribution functions (PDFs). For those channels where the next-to-leading order (NLO) or next-to-next-to-leading order (NNLO) corrections in quantum chromodynamics (QCD) are important, we estimate the relevant $K$-factors and multiply them with the LO cross sections to obtain the appropriate NLO or NNLO cross sections. We do not include the electroweak radiative corrections which are expected to be smaller than the QCD corrections. Also we do not study the scale dependence of the higher-order QCD corrections, but simply set both factorization and renormalization scales equal to the invariant mass of the heavy Higgs boson, e.g. $\mu_F=\mu_R=M_{H_1^0}$ for $H_1^0$ production. The final results for the dominant production channels of all the heavy scalars at the 100 TeV collider are depicted in Figures~\ref{fig:production1}, \ref{fig:production2}, \ref{fig:production3} and \ref{fig:production4}. The corresponding cross sections for the production of hadrophobic scalars at the 14 TeV LHC are shown in Figure~\ref{fig:prod-lhc}.\footnote{The 13 TeV LHC cross sections are not shown here, since there is not much difference between these two sets of numbers.} 
Some details of the production channels are given below.


\subsection{Bi-doublet Higgs Production}
\label{sec:4.1}
Here we discuss the production of the neutral CP-even $H^0_1$ and CP-odd $A^0_1$ fields, as well as the singly-charged $H_1^\pm$ fields.
\subsubsection{$H_1^0/A_1^0$}
Unlike the case of the SM Higgs boson $h$, where the gluon fusion process $gg\to h$ through the top-quark loop gives the dominant contribution due to the large Yukawa coupling, the $H^0_1$ and $A^0_1$ couplings to the top-quark in the LR model are suppressed by $m_b/m_t$ in the limit of vanishing $\kappa'/\kappa$, whereas the couplings to the bottom-quark are enhanced [cf. Tables~\ref{tab:H10} and \ref{tab:A10}]. Therefore, the loop-induced gluon fusion contribution to the production of $H_1^0$ and $A_1^0$ will be mainly through the bottom-quark loop, and is therefore suppressed by the absolute square of the loop factor (see e.g. Ref.~\cite{Djouadi:2005gi})
\begin{align}
A_{1/2}(\tau_b) \ = \ \frac{2}{\tau_b^2} \,\big[\tau_b\,+\,(\tau_b-1)f(\tau_b)\big] \;,
\label{loop1}
\end{align}
where $\tau_b = 4 m_b^2 / M_{H_1^0}^2$ and $f(\tau_b)={\rm \arcsin}^2\sqrt{\tau_b}$. 
%
%
For instance, for a heavy $H_1^0$ with mass of 10 TeV and the bottom-quark mass of $m_b = 4.2$ GeV, the bottom-quark induced loop factor $|A_{1/2}(\tau_b)| = 8.7 \times 10^{-5}$, while for the top-quark with $m_t = 173.2$ GeV, the corresponding effective loop factor $|A_{1/2}(\tau_t)| m_b^2/ m_t^2 = 2.6 \times 10^{-5}$, where $\tau_t = 4 m_t^2 / M^2_{H_1^0}$. Therefore, the gluon fusion processes induced by both bottom and top-quark loops can be safely neglected here. Since the couplings of $H_1^0,~A_1^0$ to light quarks are Yukawa-suppressed, the dominant production channel will involve bottom-quark induced tree-level processes, as shown in Figure~\ref{fig:feynman1}. Note that the bottom-quark content of the proton is not negligible at higher center-of-mass energies,\footnote{In this context, it is crucial to use one of the modern PDF sets with an  accurate bottom-quark PDF. We have cross-checked some of our results for two different recently released PDF sets, namely, {\tt CT14}~\cite{Dulat:2015mca} and {\tt NNPDF3.0}~\cite{Ball:2014uwa}, and found good agreement.} and this is the main reason for the sizable contribution from the $b\bar{b}$ initial states in Figure~\ref{fig:feynman1}.
\begin{figure}[t!]
  \def\topdiff{0.25}
  \def\toppos{-1.5}
  \def\vertexstart{-4}
  \def\vertex{\vertexstart+1.5}
  \def\topoffset{0.75}

  \centering

\begin{tabular}{ccc}
&   \begin{tikzpicture}[]
  \draw[dashed](\vertexstart,0)--(\vertex,0)node[right]{{\footnotesize$H_1^0 / A_1^0$}}; 
  \draw[quark] (\vertexstart-1.5,1)node[left]{{\footnotesize$b$}} -- (\vertexstart,0);
  \draw[antiquark] (\vertexstart-1.5,-1)node[left]{{\footnotesize$\bar{b}$}} -- (\vertexstart,0);
  \end{tikzpicture} & \\
& (a) & \\
  \begin{tikzpicture}[]
  \draw[dashed](\vertexstart,-1)--(\vertex,-1)node[right]{{\footnotesize$H_1^0 / A_1^0$}};
  \draw[gluon](\vertexstart,1)--(\vertex,1)node[right]{{\footnotesize$g$}};
  \draw[quark] (\vertexstart-1.5,1)node[left]{{\footnotesize$b$}} -- (\vertexstart,1);
  \draw[antiquark] (\vertexstart-1.5,-1)node[left]{{\footnotesize$\bar b$}} -- (\vertexstart,-1);
  \draw[quark] (\vertexstart,1) -- (\vertexstart,0)node[right]{{\footnotesize$b$}} -- (\vertexstart,-1);
  \end{tikzpicture} &
  \begin{tikzpicture}[]
  \draw[dashed](\vertexstart,-1)--(\vertex,-1)node[right]{{\footnotesize$H_1^0 / A_1^0$}};
  \draw[quark](\vertexstart,1)--(\vertex,1)node[right]{{\footnotesize$b$}};
  \draw[gluon] (\vertexstart-1.5,1)node[left]{{\footnotesize$g$}} -- (\vertexstart,1);
  \draw[quark] (\vertexstart-1.5,-1)node[left]{{\footnotesize$b$}} -- (\vertexstart,-1);
  \draw[quark] (\vertexstart,-1) -- (\vertexstart,0)node[right]{{\footnotesize$b$}} -- (\vertexstart,1);
  \end{tikzpicture} &
  \begin{tikzpicture}[]
  \draw[antiquark](\vertexstart,-1)--(\vertex,-1)node[right]{{\footnotesize$\bar b$}};
  \draw[dashed](\vertexstart,0)node[left]{{\footnotesize$b$}}--(\vertex,0)node[right]{{\footnotesize$H_1^0 / A_1^0$}};
  \draw[quark](\vertexstart,1)--(\vertex,1)node[right]{{\footnotesize$b$}};
  \draw[gluon](\vertexstart-1.5,1)node[left]{{\footnotesize$g$}} -- (\vertexstart,1);
  \draw[gluon] (\vertexstart-1.5,-1)node[left]{{\footnotesize$g$}} -- (\vertexstart,-1);
  \draw[quark] (\vertexstart,-1) -- (\vertexstart,0);
  \draw[quark] (\vertexstart, 0) -- (\vertexstart,1);
  \end{tikzpicture} \\
(b) & (c) & (d)
\end{tabular}
  \caption{Representative Feynman diagrams for the dominant production processes of $H_1^0$ and $A_1^0$  from bottom-quark annihilation. Formally, (b) and (c) are part of the NLO corrections to (a), and (d) is part of the NNLO correction to (a) in the {\it inclusive} production cross section for $pp\to H_1^0/A_1^0+X$, ignoring the final state jets in $X$.
}
  \label{fig:feynman1}
\end{figure}
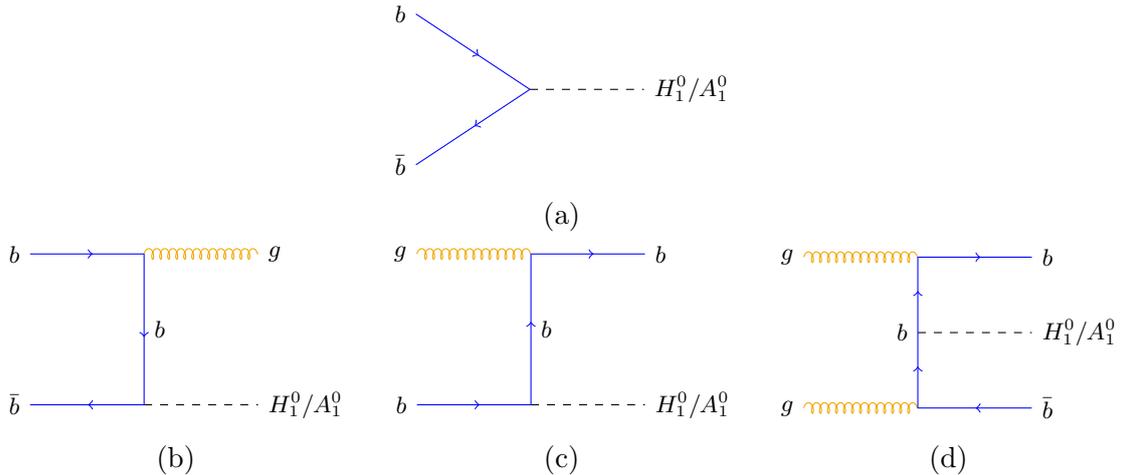
\begin{figure}[!t]
  \centering
  \includegraphics[width=0.5\textwidth]{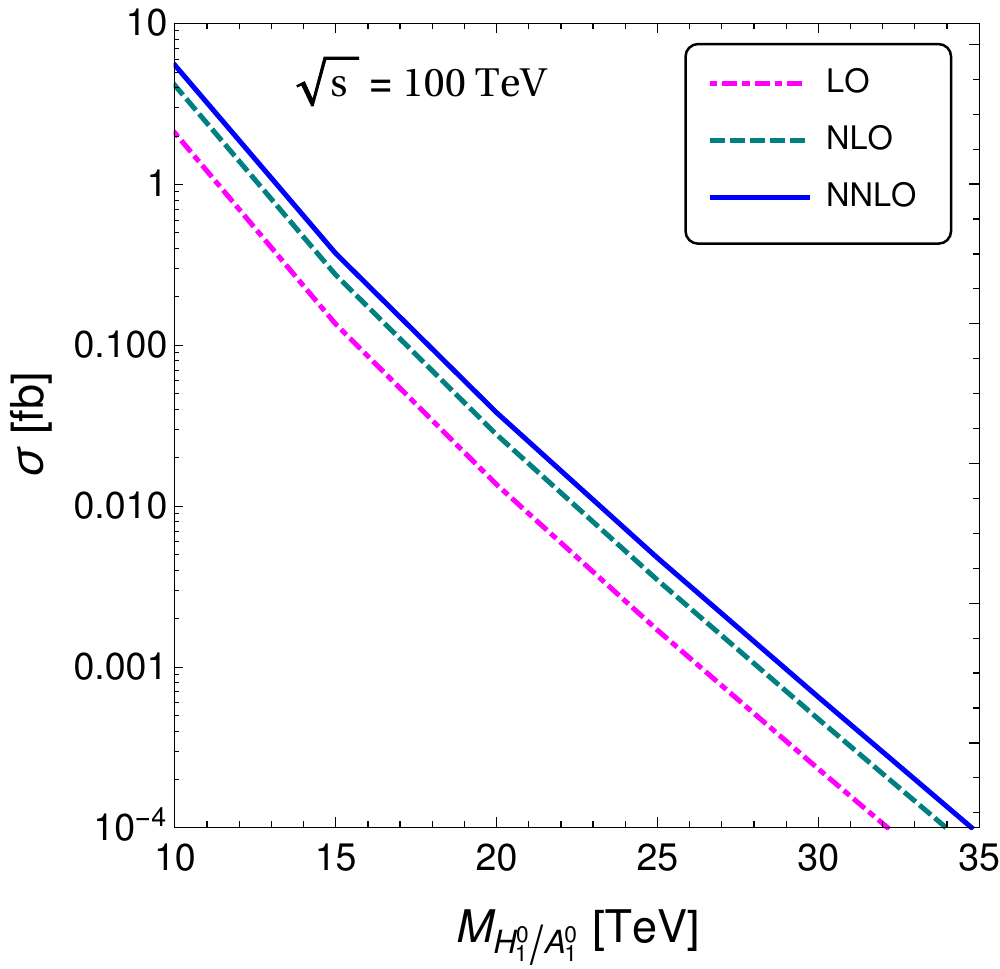}
  \caption{Dominant production cross sections for the heavy neutral bi-doublet Higgs bosons $H_1^0$ and $A_1^0$ in the minimal LR model at $\sqrt s= 100$ TeV $pp$ collider. 
  }
  \label{fig:production1}
\end{figure}

The parton-level cross sections for $pp\to H_1^0/A_1^0$ at $\sqrt s=100$ TeV are calculated at LO using {\tt CalcHEP3.6.25}~\cite{Belyaev:2012qa}. For such heavy scalars at 100 TeV center-of-mass energy, the average momentum fraction $x$ carried by the partons in the colliding protons can be as large as $\sim  0.3$, although it suffers from large experimental uncertainties, possibly of order 50\% or even larger~\cite{lhchiggs}. Another subtle point is that for a more accurate estimate of the production cross sections, the large QCD logarithmic terms $\alpha_s \log (M_{H_1^0} / m_b)$ which are of order one,  have to be resummed properly. These issues should be addressed, if one wants to make a more precise calculation of the cross section. As an initial step in this direction, we estimate the parton-level production cross sections at NLO and NNLO, using an appropriately modified version of the public code {\tt SusHi}~\cite{Harlander:2012pb}, which takes into account the virtual corrections with gluon exchange in the $b\bar{b}$ vertex and bottom-quark self-energy  corrections, as well as the emission of additional gluons from any of the bottom-quark or gluon legs, in addition to the higher-order tree-level processes shown in Figure~\ref{fig:feynman1}. In our parton-level simulations, we have applied the basic jet transverse momentum cut of $p_T(j) > 50$ GeV and jet separation $\Delta R (jj) > 0.4$ for the final states with one or more jets (including $b$-jets) for 100 TeV center-of-mass energy collisions. Our final results are shown in Figure~\ref{fig:production1}. It is worth noting that the NLO and NNLO $K$-factors turn out to be quite large for the inclusive $H_1^0 / A_1^0$ production $pp\to H_1^0/A_1^0 X$, mainly due to the sizable contributions of the tree-level processes listed in Figures~\ref{fig:feynman1} (b)-(d).


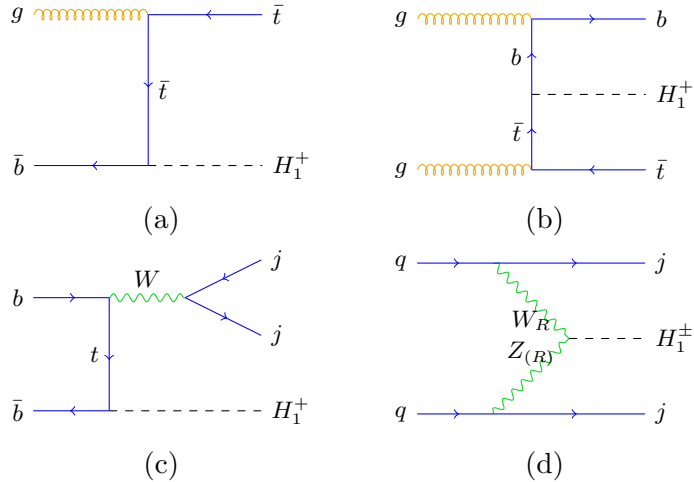
\begin{figure}[t!]
  \def\topdiff{0.25}
  \def\toppos{-1.5}
  \def\vertexstart{-4}
  \def\vertex{\vertexstart+1.5}
  \def\topoffset{0.75}

  \centering

  \noindent
\begin{tabular}{ccc}
  \begin{tikzpicture}[]
  \draw[dashed](\vertexstart,-1)--(\vertex,-1)node[right]{{\footnotesize$H_1^+$}};
  \draw[antiquark](\vertexstart,1)--(\vertex,1)node[right]{{\footnotesize$\bar t$}};
  \draw[gluon] (\vertexstart-1.5,1)node[left]{{\footnotesize$g$}} -- (\vertexstart,1);
  \draw[antiquark] (\vertexstart-1.5,-1)node[left]{{\footnotesize$\bar b$}} -- (\vertexstart,-1);
  \draw[antiquark] (\vertexstart,-1) -- (\vertexstart,0)node[right]{{\footnotesize$\bar t$}} -- (\vertexstart,1);
  \end{tikzpicture}
& &
  \begin{tikzpicture}[]
  \draw[antiquark](\vertexstart,-1)--(\vertex,-1)node[right]{{\footnotesize$\bar{t}$}};
  \draw[dashed](\vertexstart,0)--(\vertex,0)node[right]{{\footnotesize$H_1^+$}};
  \draw[quark](\vertexstart,1)--(\vertex,1)node[right]{{\footnotesize$b$}};
  \draw[gluon](\vertexstart-1.5,1)node[left]{{\footnotesize$g$}} -- (\vertexstart,1);
  \draw[gluon] (\vertexstart-1.5,-1)node[left]{{\footnotesize$g$}} -- (\vertexstart,-1);
  \draw[quark] (\vertexstart,-1) -- (\vertexstart,-0.5)node[left]{{\footnotesize$\bar t$}} -- (\vertexstart,0);
  \draw[quark] (\vertexstart, 0) -- (\vertexstart,0.5)node[left]{{\footnotesize$b$}} -- (\vertexstart,1);
  \end{tikzpicture} \\
(a) & &  (b)  \\
  \begin{tikzpicture}[]
  \draw[dashed](\vertexstart-0.5,-1)--(\vertex,-1)node[right]{{\footnotesize$H_1^+$}};
  \draw[quark](\vertexstart+0.5,0.5)--(\vertex,0)node[right]{{\footnotesize$j$}};
  \draw[antiquark](\vertexstart+0.5,0.5)--(\vertex,1)node[right]{{\footnotesize$j$}};
  \draw[ZZ](\vertexstart-0.5,0.5) -- (\vertexstart,0.5)node[above]{{\footnotesize$W$}}  --(\vertexstart+0.5,0.5);
  \draw[quark] (\vertexstart-1.5,0.5)node[left]{{\footnotesize$b$}} -- (\vertexstart-0.5,0.5);
  \draw[antiquark] (\vertexstart-1.5,-1)node[left]{{\footnotesize$\bar b$}} -- (\vertexstart-0.5,-1);
  \draw[quark] (\vertexstart-0.5,0.5) -- (\vertexstart-0.5,-0.25)node[left]{{\footnotesize$t$}} -- (\vertexstart-0.5,-1);
  \end{tikzpicture}
& &
\begin{tikzpicture}[]
  \draw[ZZ](\vertexstart-0.5,1)--(\vertexstart,0.5)node[below]{{\footnotesize$W_R$}} -- (\vertexstart+0.5,0);
  \draw[ZZ](\vertexstart-0.5,-1)--(\vertexstart,-0.5)node[above]{{\footnotesize$Z_{(R)}$}} -- (\vertexstart+0.5,0);
  \draw[dashed](\vertexstart+0.5,0) -- (\vertex,0)node[right]{{\footnotesize$H_1^\pm$}};
  \draw[quark] (\vertexstart-0.5,1) -- (\vertex,1)node[right]{{\footnotesize$j$}};
  \draw[quark] (\vertexstart-0.5,-1) -- (\vertex,-1)node[right]{{\footnotesize$j$}};
  \draw[quark] (\vertexstart-1.5,1)node[left]{{\footnotesize$q$}} -- (\vertexstart-0.5,1);
  \draw[quark] (\vertexstart-1.5,-1)node[left]{{\footnotesize$q$}} -- (\vertexstart-0.5,-1);
  \end{tikzpicture} \\
(c) & & (d)
\end{tabular}
  \caption{Representative Feynman diagrams for the dominant production processes of $H_1^\pm$: (a) the associated  production with a top quark, $g\bar{b} \to H_1^\pm t$; (b) the associated production with top and bottom quark jets, $gg \to H_1^\pm t b$, which is formally an NLO correction to (a); (c) the production with two light quark jets, $b\bar{b} \to H_1^\pm jj$ with $j=u,d,c,s$; and (d) VBF process, $qq \to H_1^\pm jj$, where $j$ can be any of the six quarks.}
  \label{fig:feynman2}
\end{figure}
\begin{figure}[!t]
  \centering
   \includegraphics[width=0.5\textwidth]{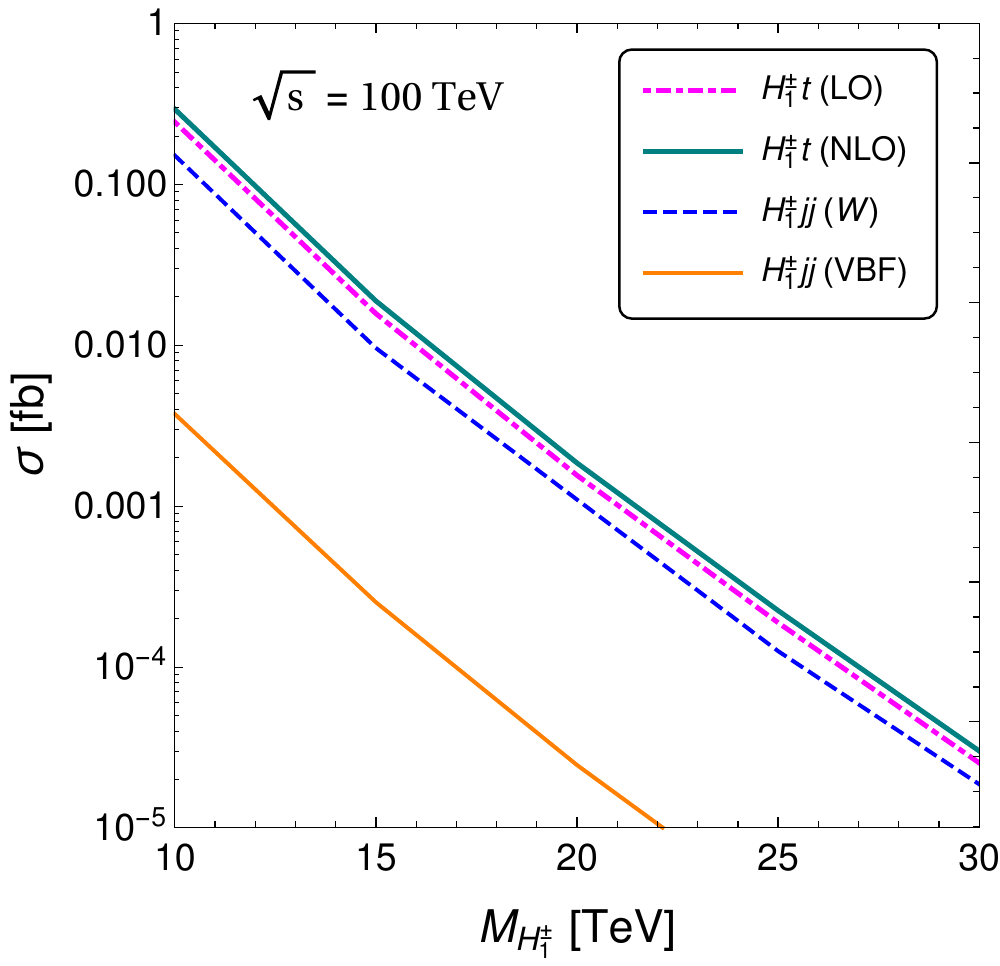}
  \caption{Dominant production cross sections for the singly-charged  Higgs bosons $H_1^\pm$ in the minimal LR model at $\sqrt s= 100$ TeV $pp$ collider. Here we have chosen $\alpha_3=3$ and $g_L=g_R$ for the VBF process.  }
  \label{fig:production2}
\end{figure}

\subsubsection{$H_1^\pm$}
Turning now to the singly-charged Higgs field $H_1^\pm$, the dominant production process is via associated production with a highly boosted top quark jet, e.g. $\bar{b}g \rightarrow H_1^+ \bar{t}$, as shown in Figure~\ref{fig:feynman2} (a). This is mainly due to the large (sizable) gluon (bottom-quark) content of the colliding protons and the large Yukawa coupling of $H_1^{\pm}$ to third-generation fermions [cf.~Table~\ref{tab:H1p}].
The large Yukawa coupling $H_1^\pm t b$, as well as the strong coupling, renders the NLO correction shown in Figure~\ref{fig:feynman2} (b) also important. We find that the NLO $K$-factor for the process $pp\to H_1^\pm t$ at 100 TeV collider is 1.6, as shown in Figure~\ref{fig:production2}.

Another important channel for $H_1^\pm$ production at $\sqrt s=100$ TeV is in association with two light-quark  jets (up, down, charm, strange and their anti-particles). There are two contributing processes: one via the associated production with the SM $W$ boson from bottom-quark annihilation with the $W$ boson decaying into two light quark jets, i.e. $\bar{b}b \to H_1^\pm W^\mp \to H_1^\pm jj$, as shown in Figure~\ref{fig:feynman2} (c); the second is from the VBF process $qq \to W_R Z_{(R)}jj \to H_1^\pm jj$, as shown in Figure~\ref{fig:feynman2} (d), which are suppressed by the heavy gauge boson masses. We find that $H_1^\pm jj$ production is dominated by the $W$-boson mediated process, with cross section about 3/5 that of the dominant $H_1^\pm t$ channel at LO, and over 10 times larger than the VBF channel, as shown in Figure~\ref{fig:production2}. Here we have chosen $\alpha_3=3$ and $g_L=g_R$, and varied $v_R$ as a function of $M_{H_1^\pm}$ [cf. Eq.~\eqref{eqn:scalarmass2}].\footnote{For a fixed value of $M_{H_1^\pm}$, with a larger $\alpha_3$, the RH VEV $v_R$ can be comparatively smaller and thus $W_R$ is lighter, which could enhance to some extent the VBF production cross section.}


\subsection{Hadrophobic Higgs Production}
\label{sec:4.2}
In this section, we discuss the dominant production channels for the hadrophobic Higgs sector comprising of the neutral CP-even scalar $H_3^0$ and the doubly-charged scalar $H_2^{\pm\pm}$.
\subsubsection{$H_3^0$}
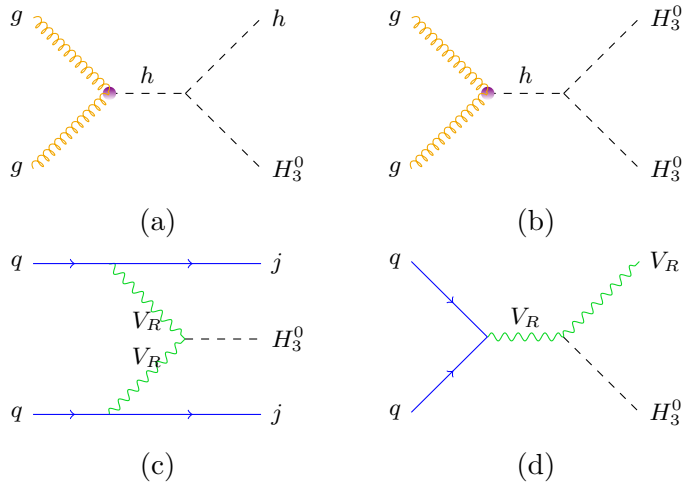
\begin{figure}[t!]
  \def\topdiff{0.25}
  \def\toppos{-1.5}
  \def\vertexstart{-4}
  \def\vertex{\vertexstart+1.5}
  \def\topoffset{0.75}

  \centering
\begin{tabular}{ccc}
  \noindent
  \begin{tikzpicture}[]
  \draw[dashed](\vertexstart+0.5,0)--(\vertex,-1)node[right]{{\footnotesize$H_3^0$}};
  \draw[dashed](\vertexstart+0.5,0)--(\vertex,1)node[right]{{\footnotesize$h$}};
  \draw[dashed](\vertexstart-0.5,0)--(\vertexstart,0)node[above]{{\footnotesize$h$}} -- (\vertexstart+0.5,0);
  \shade[top color=violet, bottom color=white] (\vertexstart-0.5,0)circle(2.5pt);
  \draw[gluon] (\vertexstart-1.5,1)node[left]{{\footnotesize$g$}} -- (\vertexstart-0.5,0);
  \draw[gluon] (\vertexstart-1.5,-1)node[left]{{\footnotesize$g$}} -- (\vertexstart-0.5,0);
  \end{tikzpicture} & &
  \begin{tikzpicture}[]
  \draw[dashed](\vertexstart+0.5,0)--(\vertex,-1)node[right]{{\footnotesize$H_3^0$}};
  \draw[dashed](\vertexstart+0.5,0)--(\vertex,1)node[right]{{\footnotesize$H_3^0$}};
  \draw[dashed](\vertexstart-0.5,0)--(\vertexstart,0)node[above]{{\footnotesize$h$}} -- (\vertexstart+0.5,0);
  \shade[top color=violet, bottom color=white] (\vertexstart-0.5,0)circle(2.5pt);
  \draw[gluon] (\vertexstart-1.5,1)node[left]{{\footnotesize$g$}} -- (\vertexstart-0.5,0);
  \draw[gluon] (\vertexstart-1.5,-1)node[left]{{\footnotesize$g$}} -- (\vertexstart-0.5,0);
  \end{tikzpicture} \\
(a) & & (b) \\
  \begin{tikzpicture}[]
  \draw[ZZ](\vertexstart-0.5,1)--(\vertexstart,0.5)node[below]{{\footnotesize$V_R$}} -- (\vertexstart+0.5,0);
  \draw[ZZ](\vertexstart-0.5,-1)--(\vertexstart,-0.5)node[above]{{\footnotesize$V_R$}} -- (\vertexstart+0.5,0);
  \draw[dashed](\vertexstart+0.5,0) -- (\vertex,0)node[right]{{\footnotesize$H_3^0$}};
  \draw[quark] (\vertexstart-0.5,1) -- (\vertex,1)node[right]{{\footnotesize$j$}};
  \draw[quark] (\vertexstart-0.5,-1) -- (\vertex,-1)node[right]{{\footnotesize$j$}};
  \draw[quark] (\vertexstart-1.5,1)node[left]{{\footnotesize$q$}} -- (\vertexstart-0.5,1);
  \draw[quark] (\vertexstart-1.5,-1)node[left]{{\footnotesize$q$}} -- (\vertexstart-0.5,-1);
  \end{tikzpicture} & &
\begin{tikzpicture}[]
  \draw[dashed](\vertexstart+0.5,0)--(\vertex,-1)node[right]{{\footnotesize$H_3^0$}};
  \draw[ZZ](\vertexstart+0.5,0)--(\vertex,1)node[right]{{\footnotesize$V_R$}};
  \draw[ZZ](\vertexstart-0.5,0)--(\vertexstart,0)node[above]{{\footnotesize$V_R$}} -- (\vertexstart+0.5,0);
  \draw[quark] (\vertexstart-1.5,1)node[left]{{\footnotesize$q$}} -- (\vertexstart-0.5,0);
  \draw[quark] (\vertexstart-1.5,-1)node[left]{{\footnotesize$q$}} -- (\vertexstart-0.5,0);
  \end{tikzpicture}
\\
(c) & & (d)
\end{tabular}
  \caption{Representative Feynman diagrams for the dominant production processes of $H_3^0$: (a) the associated production with the SM Higgs, $pp \to h^{\ast} / H_1^{0 \, (\ast)} \to H_3^0 h$; (b) pair production, $pp \to h^{\ast} / H_1^{0 \, (\ast)} \to H_3^0 H_3^0$; (c) heavy VBF, $qq \to H_3^0 jj$ mediated by a pair of $V_R$ ($=W_R,Z_R$) in the $t$-channel; and (d) Higgsstrahlung process, $qq \to V^\ast_R \to H_3^0 V_R$. In (a) and (b),  the LO effective $hgg$ vertex is predominantly from the top-quark loop induced SM coupling.}
  \label{fig:feynman3}
\end{figure}

The dominant production mode for $H_3^0$ is either via the VBF process involving RH gauge bosons $W_R$ or $Z_R$ in the $t$-channel, or via associated production with the SM Higgs, or via pair production, 
depending on the mass spectrum. The corresponding Feynman diagrams are shown in Figure~\ref{fig:feynman3}. The associated and pair-production channels are mediated by an $s$-channel SM Higgs $h$ or the heavy $H_1^0$ through their effective coupling to gluon induced by the third-generation quark loop. The $H_1^0$ portal is generally suppressed by the heavy bi-doublet mass [cf.~Eq.~\eqref{eqn:H10mass}], as well as by the bottom-quark loop factor [cf. Eq.~\eqref{loop1}], with significant contribution only in the resonance region $M_{H_1^0} \simeq 2 M_{H_3^0}$.

\begin{figure}[!t]
  \centering
   \includegraphics[width=0.5\textwidth]{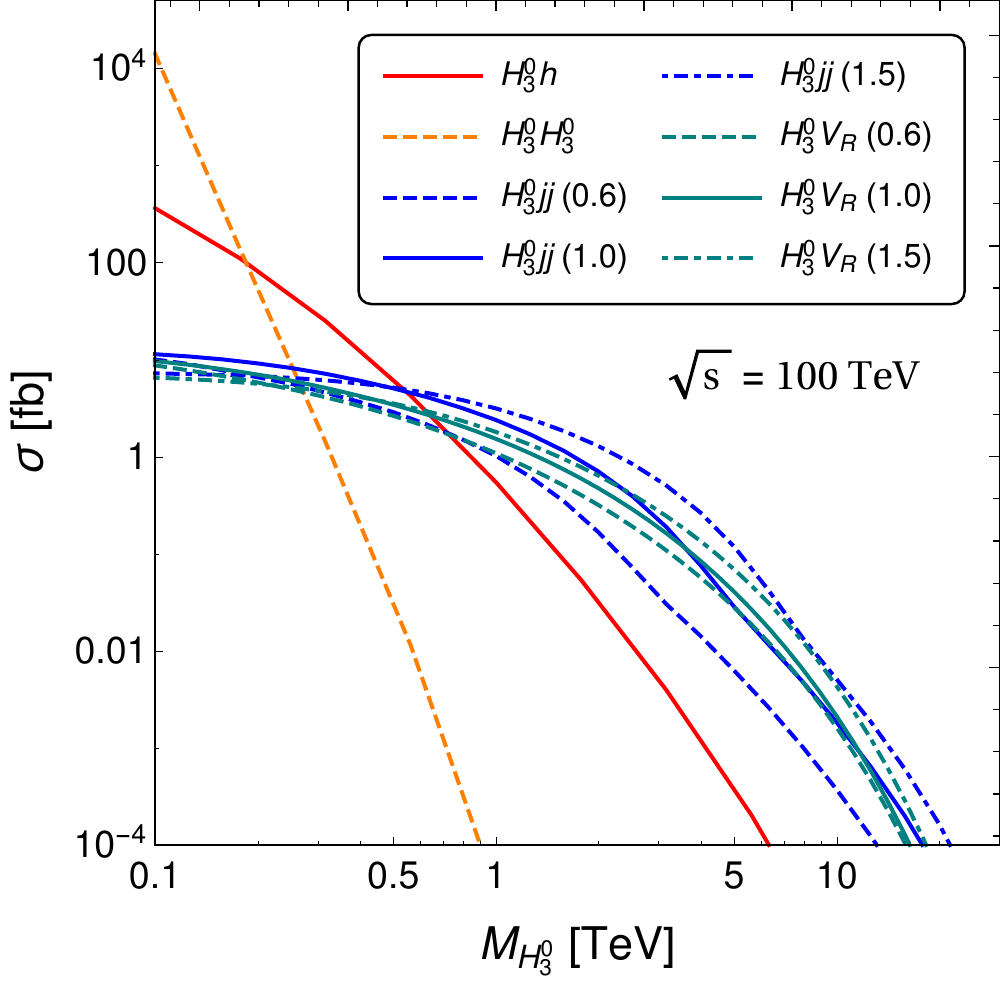}
   \caption{Dominant production cross sections for the neutral hadrophobic  Higgs boson $H_3^0$ in the minimal LR model at $\sqrt s= 100$ TeV $pp$ collider. Here we have chosen $v_R=10$ TeV, $\alpha_1=0.01$ and $\alpha_2=0$. The VBF ($H_3^0jj$) and Higgsstrahlung ($H_3^0V_R$) cross sections are shown for three different values of the gauge coupling ratio $g_R/g_L$ (in parenthesis).}
  \label{fig:production3}
\end{figure}

For the SM Higgs portal, the couplings $hhH_3^0$ and $hH_3^0 H_3^0$ [cf.~Table~\ref{tab:H30}] are related to the masses of the SM Higgs and $H_3^0$ via Eqs.~(\ref{eqn:hmass}) and (\ref{eqn:H30mass}), and also to the trilinear SM Higgs coupling $\lambda_{hhh}$ at the LO [cf.~Table~\ref{tab:triple}]:
\begin{eqnarray}
\label{eqn:triple_SM}
\lambda_{hhh} \ = \ \frac{1}{2\sqrt2} \left( 4\lambda_1 - \frac{\alpha_1^2}{\rho_1} \right)\kappa \ \simeq \frac{1}{2\sqrt2} \frac{M^2_h}{v_{\rm EW}} \,.
\end{eqnarray}
Note that the $\lambda_{hhh}$ coupling in Eq.~\eqref{eqn:triple_SM} is the {\it same} as in the SM. As for the quartic coupling $\lambda_1$, we have
\begin{eqnarray}
\lambda_1 \ = \ \frac{M_h^2}{4 \kappa^2} + \frac{\alpha_1^2 v_R^2}{M_{H_3^0}^2} \,,
\end{eqnarray}
which approaches the SM quartic coupling $\lambda_{4h} = M_h^2/4v^2_{\rm EW}$ in the limit of $\alpha_1 \to 0$, i.e. when the ``correction'' term $\alpha_1^2/\rho_1$ from interactions with the RH triplet $\Delta_R$ vanishes. For a light $H_3^0$ with $M_{H_3^0} \ll v_R$, a large $\alpha_1$ would potentially push $\lambda_1$ to be deep in the non-perturbative region.
As an illustrative example, we work in the simple benchmark scenario with the heavy $H_1^0$ portal switched off (i.e. $\alpha_2 = 0$) and only the SM Higgs portal turned on with a small coupling parameter $\alpha_1 = 0.01$. Note that the NNLO $K$-factor for the $gg$ fusion processes in Figures~\ref{fig:feynman3} (a,b) induced by the top-quark loop is known to be large, of order 2~\cite{lhchiggs}, and therefore must be included in the calculation. Our parton-level results for the associated and pair-production cross sections at NNLO are presented in Figure~\ref{fig:production3} (red and orange curves).  For a value of $\alpha_1$ different from 0.01 (and with $\alpha_2 = 0$), one can estimate the cross sections in these two channels by simply rescaling the corresponding cross sections given in Figure~\ref{fig:production3} by numerical factors of
\begin{align}
\left(\frac{\alpha_1}{0.01}\right)^2 \quad {\rm and} \quad \left(\frac{\alpha_1}{0.01}\right)^2 \left[\frac{1 -  2 \alpha_1 v_R^2/M^2_{H_3^0}}{1 -  2 \times 0.01 \times v_R^2/M^2_{H_3^0}}  \right]^{2} \, , \nonumber
\end{align}
for the $hH_3^0$ and $H_3^0H_3^0$ production, respectively.


The VBF process $pp\to H_3^0 jj$ shown in Figure~\ref{fig:feynman3} (c) is dominated by the $W_R$ fusion channel, and dictated by the gauge coupling $g_R$, the $H_3^0W_RW_R$ coupling [cf.~Table~\ref{tab:H30}] and the heavy charged gauge boson mass [cf.~Eq.~\eqref{eqn:gbosonmass}]. The $Z_R$ fusion process is sub-dominant, since the $M_{Z_R}>M_{W_R}$ in the minimal version of the LR model [cf.~Eq.~\eqref{eq:heavyWZratio}]. To illustrate the effect of the RH gauge coupling on the VBF production of $H_3^0$, we present in Figure~\ref{fig:production3} the LO cross sections with three benchmark values for the RH coupling $g_R$ with $g_R/g_L=0.6 ,\, 1.0 ,\, 1.5$,\footnote{Note that there is a theoretical lower limit on $g_R/g_L \gtrsim 0.55$ [cf.~Eq.~\eqref{grl}]. Although we are not aware of any LR model with $g_R>g_L$ to be compatible with GUTs, we have considered one benchmark value of $g_R/g_L = 1.5$ just for the comparison sake.} and with $v_R = 10$ TeV.  Here we have just imposed the basic trigger cuts on the jets: $p_T (j) > 50$ GeV and $\Delta R_{jj} > 0.4$, and no specialized VBF selection cuts like large rapidity gap and high invariant mass for the dijet system. The NLO corrections to the heavy VBF process turn out to be much smaller than the corresponding SM Higgs case, as estimated using {\tt VBFNLO2.7.1}~\cite{Baglio:2014uba}.

Similarly, the Higgsstrahlung process $pp\to V_R^* \to H_3^0 V_R$ shown in Figure~\ref{fig:feynman3} (d) is also dominated by the $W_R$ channel, and dictated by the gauge coupling $g_R$, the $H_3^0W_RW_R$ coupling [cf.~Table~\ref{tab:H30}] and the heavy charged gauge boson mass [cf.~Eq.~\eqref{eqn:gbosonmass}]. The effect of the RH gauge coupling $g_R$ on this production process is also illustrated in Figure~\ref{fig:production3}.

It is clear that in our benchmark scenario for a light $H_3^0$ with $M_{H_3^0} \lesssim 500$ GeV, the production at 100 TeV collider is dominated by the SM Higgs portal. When $H_3^0$ is heavier, either the heavy VBF or the Higgsstrahlung process takes over as the dominant channel. The latter cross sections also increase with $g_R$ for a given $v_R$ (and $M_{H_3^0} \gtrsim 1$ TeV), because although $W_R$ becomes heavier, the stronger scattering amplitude dependence on the gauge coupling $g_R^4$ (two powers from the couplings to SM fermions and two from couplings to $H_3^0$) can overcome easily the phase space suppression due to the larger mediator mass. For instance, for a 5 TeV $H_3^0$ the gauge coupling $g_R = 1.5 g_L$ can enhance the VBF cross section by a factor of 2.7 with respect to the $g_R = g_L$ case and by a factor of 16.7 with respect to the $g_R = 0.6 g_L$ case, as shown in Figure~\ref{fig:production3} (blue curves).

\subsubsection{$H_2^{\pm\pm}$}
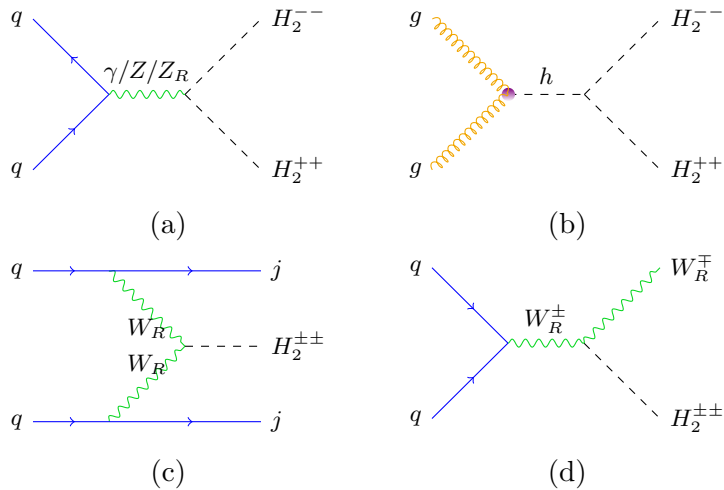
\begin{figure}[t!]
  \def\topdiff{0.25}
  \def\toppos{-1.5}
  \def\vertexstart{-4}
  \def\vertex{\vertexstart+1.5}
  \def\topoffset{0.75}

  \centering
\begin{tabular}{ccc}
  \noindent
  \begin{tikzpicture}[]
  \draw[dashed](\vertexstart+0.5,0)--(\vertex,-1)node[right]{{\footnotesize$H_2^{++}$}};
  \draw[dashed](\vertexstart+0.5,0)--(\vertex,1)node[right]{{\footnotesize$H_2^{--}$}};
  \draw[ZZ](\vertexstart-0.5,0)--(\vertexstart,0)node[above]{{\footnotesize$\gamma/Z/Z_R$}} -- (\vertexstart+0.5,0);
  \draw[antiquark] (\vertexstart-1.5,1)node[left]{{\footnotesize$q$}} -- (\vertexstart-0.5,0);
  \draw[quark] (\vertexstart-1.5,-1)node[left]{{\footnotesize$q$}} -- (\vertexstart-0.5,0);
  \end{tikzpicture} & &
  \begin{tikzpicture}[]
  \draw[dashed](\vertexstart+0.5,0)--(\vertex,-1)node[right]{{\footnotesize$H_2^{++}$}};
  \draw[dashed](\vertexstart+0.5,0)--(\vertex,1)node[right]{{\footnotesize$H_2^{--}$}};
  \draw[dashed](\vertexstart-0.5,0)--(\vertexstart,0)node[above]{{\footnotesize$h$}} -- (\vertexstart+0.5,0);
  \shade[top color=violet, bottom color=white] (\vertexstart-0.5,0)circle(2.5pt);
  \draw[gluon] (\vertexstart-1.5,1)node[left]{{\footnotesize$g$}} -- (\vertexstart-0.5,0);
  \draw[gluon] (\vertexstart-1.5,-1)node[left]{{\footnotesize$g$}} -- (\vertexstart-0.5,0);
  \end{tikzpicture} \\
(a) && (b) \\
  \begin{tikzpicture}[]
  \draw[ZZ](\vertexstart-0.5,1)--(\vertexstart,0.5)node[below]{{\footnotesize$W_R$}} -- (\vertexstart+0.5,0);
  \draw[ZZ](\vertexstart-0.5,-1)--(\vertexstart,-0.5)node[above]{{\footnotesize$W_R$}} -- (\vertexstart+0.5,0);
  \draw[dashed](\vertexstart+0.5,0) -- (\vertex,0)node[right]{{\footnotesize$H_2^{\pm\pm}$}};
  \draw[quark] (\vertexstart-0.5,1) -- (\vertex,1)node[right]{{\footnotesize$j$}};
  \draw[quark] (\vertexstart-0.5,-1) -- (\vertex,-1)node[right]{{\footnotesize$j$}};
  \draw[quark] (\vertexstart-1.5,1)node[left]{{\footnotesize$q$}} -- (\vertexstart-0.5,1);
  \draw[quark] (\vertexstart-1.5,-1)node[left]{{\footnotesize$q$}} -- (\vertexstart-0.5,-1);
  \end{tikzpicture} & &
\begin{tikzpicture}[]
  \draw[dashed](\vertexstart+0.5,0)--(\vertex,-1)node[right]{{\footnotesize$H_2^{\pm\pm}$}};
  \draw[ZZ](\vertexstart+0.5,0)--(\vertex,1)node[right]{{\footnotesize$W_R^\mp$}};
  \draw[ZZ](\vertexstart-0.5,0)--(\vertexstart,0)node[above]{{\footnotesize$W_R^\pm$}} -- (\vertexstart+0.5,0);
  \draw[quark] (\vertexstart-1.5,1)node[left]{{\footnotesize$q$}} -- (\vertexstart-0.5,0);
  \draw[quark] (\vertexstart-1.5,-1)node[left]{{\footnotesize$q$}} -- (\vertexstart-0.5,0);
  \end{tikzpicture}
\\
(c) & & (d)
\end{tabular}
  \caption{Representative Feynman diagrams for the dominant production of $H_2^{\pm\pm}$: (a) Drell-Yan pair production; (b) Higgs-portal pair production; (c) heavy VBF; and (d) Higgsstrahlung.  In (b),  the LO effective $hgg$ vertex is predominantly from the top-quark loop induced SM coupling.}
  \label{fig:feynman4}
\end{figure}

From Table~\ref{tab:H2pp}, we find that for the doubly-charged Higgs $H_2^{\pm\pm}$ production, one dominant channel is pair production via the Drell-Yan (DY) process, as shown in Figure~\ref{fig:feynman4} (a), with an $s$-channel photon or $Z$ boson, and potentially resonance enhancement from the heavy $H_1^0$, $H_3^0$ or $Z_R$ bosons. A sub-leading contribution comes from the SM Higgs portal, as shown in Figure~\ref{fig:feynman4} (b). The VBF process mediated by RH gauge bosons $W_R^\pm$ in the $t$-channel, as shown in Figure~\ref{fig:feynman4} (c), and the Higgsstrahlung process mediated by $W_R^\pm$ in the $s$-channel, as shown in Figure~\ref{fig:feynman4} (d), are also important.

To calculate the cross sections for all these channels, we adopt the same set parameter as for the $H_3^0$ case, i.e. $\alpha_1 = 0.01 \, , \alpha_2 = 0 \, , v_R = 10 \, {\rm TeV}$ and $g_R/g_L=0.6 ,\, 1.0 ,\, 1.5$ for the VBF and Higgsstrahlung processes, while for the DY processes, we keep $g_R=g_L$. In the DY mode, the $H_1^0$ and $H_3^0$ portals are again turned off, since in most of the parameter space of interest this is dominated by the $\gamma/Z$-mediated process. For the sub-leading SM Higgs portal DY channel, we choose  $M_{H_3^0} = 5$ TeV to fix the coupling $\rho_1$ and multiply the LO cross section by the NNLO $K$-factor of 2 (for the gluon fusion via top-quark loop). The $K$-factors for the DY and VBF processes are not so large, so their LO cross sections are sufficient for our purpose. We apply the basic trigger cuts  $p_T (j) > 50$ GeV and $\Delta R (jj) > 0.4$ for the VBF process.

\begin{figure}[!t]
  \centering
    \includegraphics[width=0.5\textwidth]{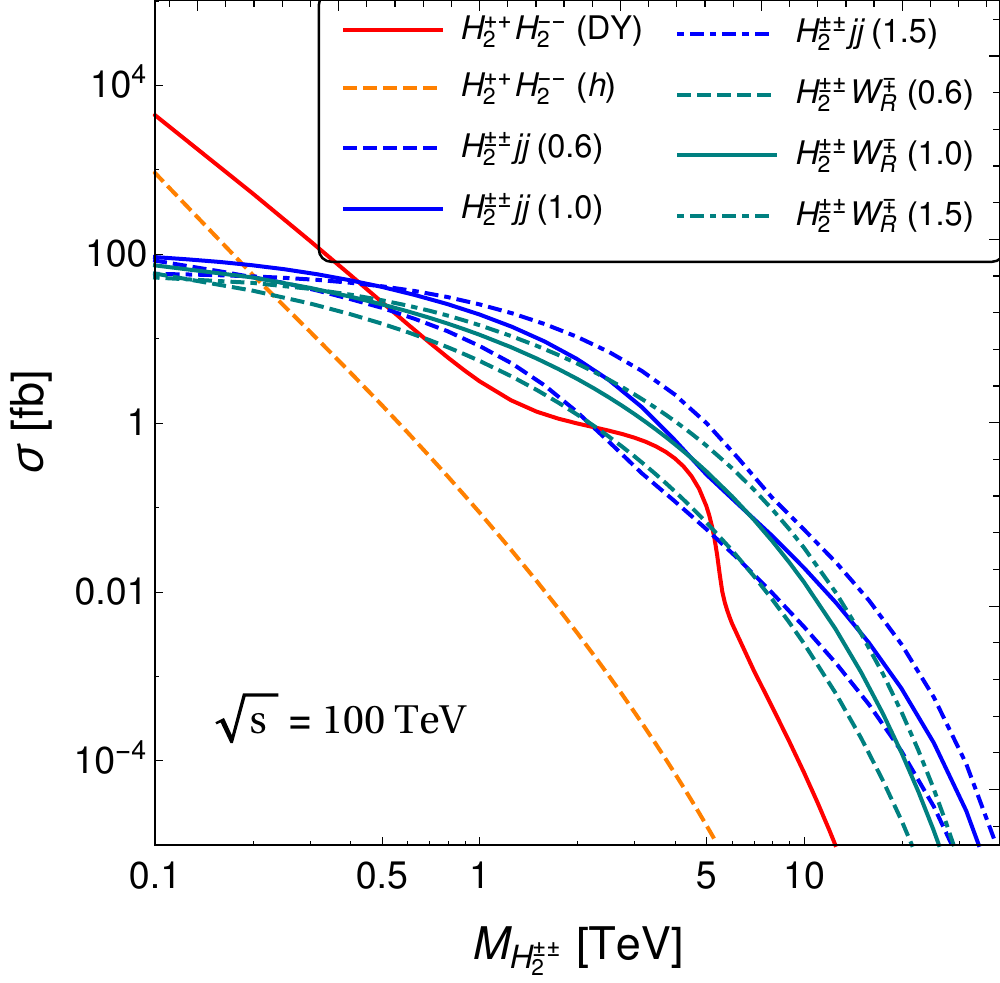}
  \caption{Dominant production cross sections for the doubly-charged Higgs bosons $H_2^{\pm\pm}$ in the minimal LR model at $\sqrt s= 100$ TeV $pp$ collider. Here we have chosen $\alpha_1 = 0.01 \, , \alpha_2 = 0 \, , v_R = 10 \, {\rm TeV}$ and $M_{H_3^0} = 5$ TeV. The VBF ($H_2^{\pm\pm}jj$) and Higgsstrahlung ($H_2^{\pm\pm}W_R^\mp$) cross sections are shown for three different values of the gauge coupling ratio $g_R/g_L$ (in parenthesis).}
  \label{fig:production4}
\end{figure}

Our results for the $H_2^{\pm\pm}$ production are shown in Figure~\ref{fig:production4}.  The bump around 5 TeV in the DY case is due to the resonance-like production at $M_{Z_R} \simeq 2 M_{H_2^{\pm\pm}}$. The effect of $g_R$ on the DY production of $H_2^{\pm\pm}$ is significant only in this resonance region and is not shown in the plot. As for the neutral hadrophobic scalar $H_3^0$, for smaller $M_{H_2^{\pm\pm}} \lesssim 500$ GeV, the DY process is dominant, whereas for relatively larger $M_{H_2^{\pm\pm}}$, this is kinematically suppressed compared to the VBF/Higgsstrahlung process. Also, the RH gauge coupling can  largely enhance the latter channels; for instance, the cross section for $g_R = 1.5 g_L$ is 2.7 times larger than the $g_R = g_L$ case and 16.6 than the $g_R = 0.6 g_L$ case. The Higgs portal is always found to be sub-dominant compared to the DY process.

\subsection{Hadrophobic Higgs at LHC Run II}
\label{sec:4.3}

\begin{figure}[!t]
  \centering
  \includegraphics[width=0.49\textwidth]{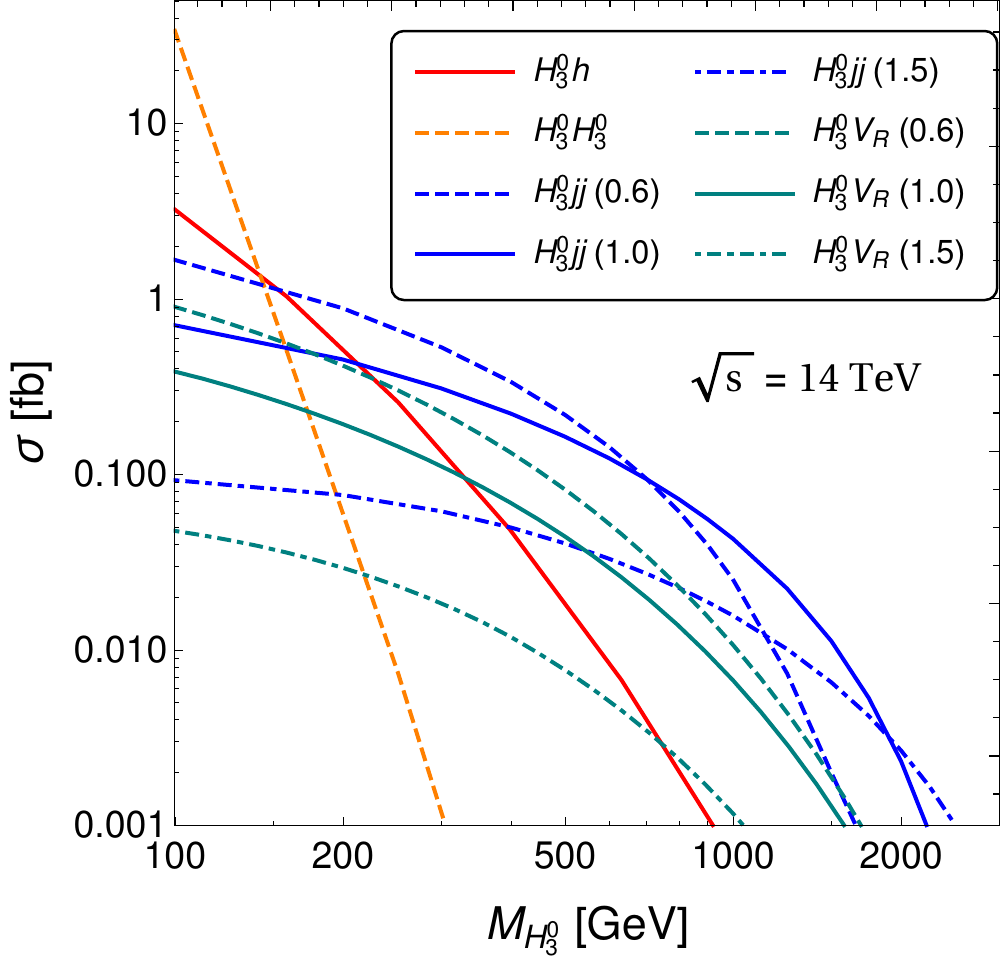}
  \includegraphics[width=0.49\textwidth]{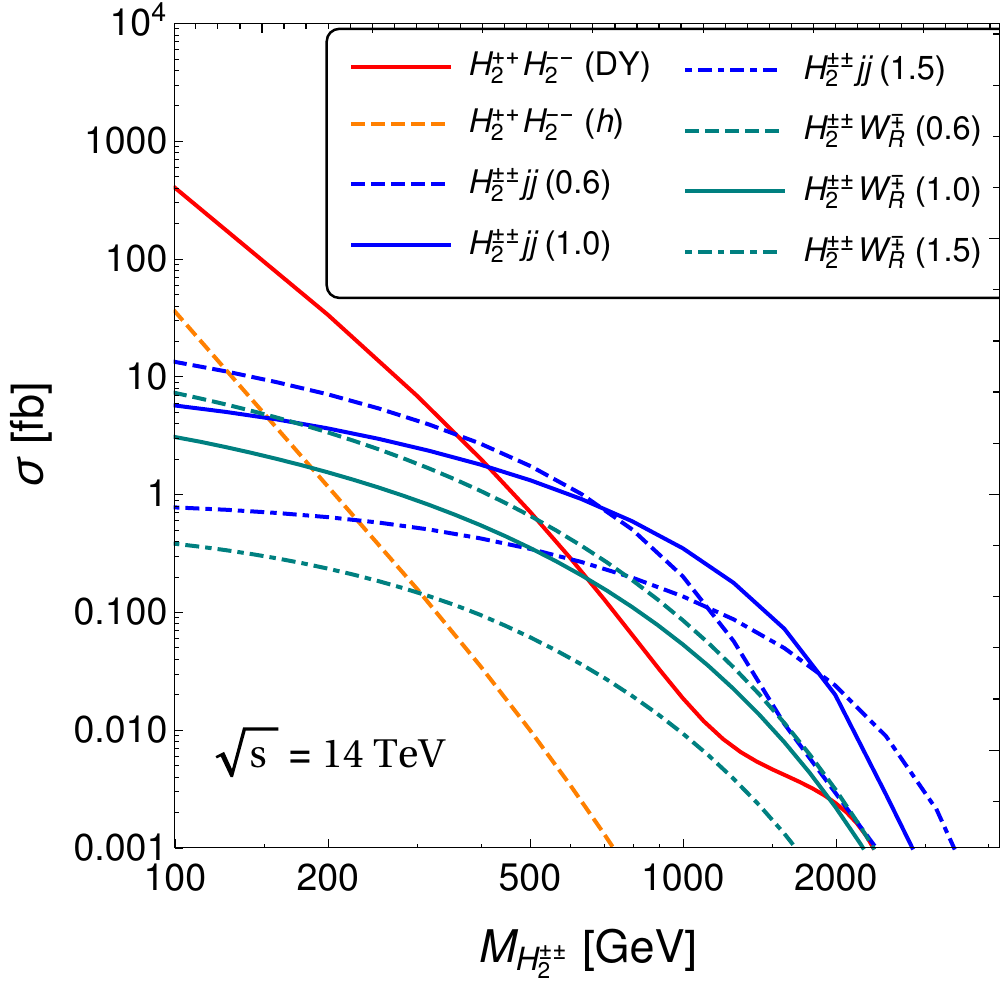}
  \caption{Dominant production cross sections for the hadrophobic Higgs bosons $H_3^0$ (left panel) and $H_2^{\pm\pm}$ (right panel) in the minimal LR model at the 14 TeV LHC. Here we have chosen $\alpha_1 = 0.01 \, , \alpha_2 = 0 \, ,v_R = 5$ TeV. The values in parenthesis are for different ratios of $g_R/g_L$.}
  \label{fig:prod-lhc}
\end{figure}

As the hadrophobic scalars $H_3^0$ and $H_2^{\pm\pm}$ are not constrained by the FCNC effects and can be as light as sub-TeV scale, they could be accessible at the LHC Run II.\footnote{There were arguments that, due to the interactions of $H_3^0$ to the heavy gauge bosons and $H_1^0$ at one-loop level, the mass of $H_3^0$ is indirectly constrained by experimental limits on the masses of $W_R$, $Z_R$ and $H_1^0$ such that $M_{H_3^0} \gtrsim 4$ TeV~\cite{Basecq:1988cv}. However, as an effective phenomenological scenario at the TeV scale, the minimal LR model is always embedded into some GUTs at super-high energy scale and we neglect such constraints on the $H_3^0$ mass throughout this paper.} The dominant production channels are the same as at the 100 TeV collider and are shown in Figures~\ref{fig:feynman3} and \ref{fig:feynman4}. The corresponding production cross sections at the LHC with 14 TeV center-of-mass energy are presented in Figure~\ref{fig:prod-lhc}, for which we adopt the same set of couplings as in the 100 TeV case, i.e. $\alpha_1 = 0.01 \, , \alpha_2 = 0$ and $g_R/g_L=0.6, \,  1\, , 1.5$. We use milder trigger cuts on jets with $p_T (j) > 25$ GeV and $\Delta R (jj) > 0.4$. Due to the severe kinematic suppression, the Higgs portal channels for $H_3^0$ (DY channel for $H_2^{\pm\pm}$) are the dominant ones only for a hadrophobic scalar below 150 GeV (400 GeV).

In Figure~\ref{fig:prod-lhc}, we have chosen a lower RH scale of $v_R = 5$ TeV, so that we can have a heavy $W_R$ boson still accessible at the LHC~\cite{Ferrari:2000sp},\footnote{Given $v_R = 5$ TeV and the RH gauge coupling $g_R=0.6g_L$, the $W_R$ mass is close to 2 TeV, which can explain the recent CMS $eejj$~\cite{Khachatryan:2014dka} and ATLAS diboson~\cite{Aad:2015xaa} excesses~\cite{Deppisch:2015cua, Brehmer:2015cia, Dobrescu:2015qna, Gao:2015irw, Dev:2015pga}.} and the hadrophobic scalars can be produced much more abundantly at LHC Run II even if both the SM Higgs portal and $H_1^0$ portal are off for $H_3^0$.\footnote{For such a lower scale of $v_R$, the quartic coupling $\alpha_3 = M^2_{H_1^0}/v_R^2$ is pushed to be large by the FCNC constraints. }  With regard to the case with a 10 TeV $v_R$, the lower RH scale renders a much larger cross section in the VBF/Higgsstrahlung channels for both $H_3^0$ and $H_2^{\pm\pm}$. In addition, when the hadrophobic scalars are light, e.g. few hundred GeV, these channels can benefit from a smaller $g_R$ due to the smaller $W_R$ mass, while when these scalars are heavy, e.g. beyond 2 TeV scale, the scenarios with a larger $g_R$ eventually overcome due to the larger gauge couplings of $W_R$ to the SM fermions and the hadrophobic scalars.



\section{Decays of the heavy Higgs bosons} \label{sec:5}


From the couplings in Tables~\ref{tab:H10} to \ref{tab:H2pp}, it is easy to identify the dominant decay channels of the heavy bi-doublet and hadrophobic scalars in the minimal LR model, all of which are collected in Table \ref{tab:decay} (in the limit of massless decay products). The corresponding formulas for these decay widths up to the LO are listed in Appendix~\ref{app:B}. It is remarkable that since all the masses of heavy scalars and heavy vector bosons are proportional to the RH scale $v_R$, the mass dependence of phase space and squared amplitudes can be largely canceled out, and as a result, all the dominant decay widths are proportional to $v_R$, as it is the only relevant energy scale in the high-energy limit of the theory. Thus up to some phase space factors which approach unity in the massless limit of the decay products, the decay branching ratios are simply the ratios among the combinations of Yukawa couplings, quartic scalar couplings and some numerical factors.

\begin{table}[t]
  \centering
  \caption[]{Dominant decay channels (with some of the important further decay channels of the decay products) of the heavy Higgs bosons in the minimal LR model and their corresponding branching fractions (without including the secondary decays) in the limit when the decay products are much lighter than the parent particle. For $H_3^0$ we have defined the ``rescaled'' total width $\Gamma_{H_3^0} \equiv {\alpha_1^2 + 8\alpha_2^2 + 4(\alpha_1+\alpha_3)^2 + 8(\rho_1+2\rho_2)^2 + 12 \rho_1^2 + 12 \rho_1 f^2}$, and the factor $\delta_H = 1$ for $H_1^0 H_1^0$ and $A_1^0 A_1^0$ and $2$ for the charged pairs $H_1^+ H_1^-$. If some of the heavy product channels are not kinematically allowed such as $H_3^0 \to H_1^0 H_1^0$ and $H_2^{\pm\pm} \to W_R^\pm W_R^\pm$, they have to be eliminated from the dominate decay channels and the approximate total widths.   See text and Appendix~\ref{app:B} for more details and comments.}
  \label{tab:decay}
  \begin{tabular}{lll}
  \hline\hline
  Scalar & Decay Channels & Branching Ratio \\ \hline
  $H_1^0$  & $b\bar{b}$ & $\frac{6\alpha_3 y_t^2}{6\alpha_3 y_t^2 + 16 \alpha_2^2 + \alpha_3^2}$  \\
  & $hH_3^0 ~(\rightarrow hhh \rightarrow 6b / 4b2\gamma$) & $\frac{16\alpha_2^2}{6\alpha_3 y_t^2 + 16 \alpha_2^2 + \alpha_3^2}$ \\
  & $WW_R ~(\rightarrow 4j / \ell^\pm \ell^\pm 4j) $ & $\frac{\alpha_3^2}{6\alpha_3 y_t^2 + 16 \alpha_2^2 + \alpha_3^2}$  \\ \hline
  $A_1^0$ & $b\bar{b}$ & $\frac{6 y_t^2}{6 y_t^2 + \alpha_3}$  \\
  & $WW_R~( \rightarrow 4j / \ell^\pm \ell^\pm 4j) $ & $\frac{\alpha_3}{6y_t^2 + \alpha_3}$  \\ \hline
  $H_1^+$ & $t\bar{b}~( \rightarrow bbjj / bb\ell\nu)$ & $\frac{3 y_t^2}{3 y_t^2 + \alpha_3}$ \\
  & $ZW_R ~(\rightarrow 4j / \ell^\pm \ell^\mp\ell^\pm \ell^\pm jj)$ &$\frac{ \alpha_3}{6 y_t^2 + 2\alpha_3}$ \\
  & $hW_R~( \rightarrow bbjj/ \ell^\pm \ell^\pm bb jj)$ & $\frac{\alpha_3}{6 y_t^2 + 2\alpha_3}$ \\ \hline
  $H_3^0$ & $hh~( \rightarrow 4b / 2b2\gamma$) &
  $\frac{\alpha_1^2}{ \Gamma_{H_3^0}}$ \\
  & $H_1^0 h~(\to 4b)$ & $\frac{8\alpha_2^2}{\Gamma_{H_3^0}}$ \\
  & $H_1^0 H_1^0 / A_1^0 A_1^0 / H_1^+ H_1^- ~(\to 4b/ttbb)$ & $\frac{(\alpha_1+\alpha_3)^2 \delta_H}{ \Gamma_{H_3^0}}$ \\
  & $H_2^{++}H_2^{--}~( \rightarrow \ell^+ \ell^+ \ell^- \ell^-)$ & $\frac{8(\rho_1+2\rho_2)^2}{ \Gamma_{H_3^0}}$ \\
  & $W_R^\pm W_R^\mp ~(\rightarrow 4j / \ell^\pm \ell^\pm 4j)$ & $\frac{8\rho_1^2}{ \Gamma_{H_3^0}}$ \\
  & $Z_RZ_R~( \rightarrow 4j / \ell^\pm \ell^\mp jj)$ & $\frac{4\rho_1^2}{\Gamma_{H_3^0}}$ \\
  & $NN~( \rightarrow \ell^\pm\ell^\pm 4j)$ &  $\frac{12\rho_1 f^2}{ \Gamma_{H_3^0}}$ \\
   \hline
  $H_2^{++}$ & $\ell^+ \ell^+$ & $\frac{3}{3 + 8 \rho_2}$ \\
  & $W_R^+ W_R^+~( \rightarrow 4j / \ell^\pm \ell^\pm 4j)$ & $\frac{8 \rho_2}{3 + 8 \rho_2}$ \\
  \hline\hline
  \end{tabular}
\end{table}

For $H_1^0$, the dominant decay channels are $b\bar{b}$, $hH_3^0$ and $WW_R$ which almost saturate the total decay width, as long as the two latter channels are open. 
In most of the parameter region of interest, these three channels are comparable, depending on the relative values of the SM top-Yukawa coupling $y_t$ and the quartic couplings $\alpha_2$ and $\alpha_3$ in the potential (\ref{eqn:potential}), with $\alpha_3$ related to the mass of $H_1^0$. Specifically, we get
\begin{eqnarray}
\Gamma( H_1^0 \to b\bar{b} ) : \Gamma( H_1^0 \to H_3^0 h ) : \Gamma( H_1^0  \to W^\pm W_R^{\mp} ) \ \simeq \
\frac{3 \alpha_3^{1/2} y_t^2}{16\pi} :
\frac{\alpha_2^2 \alpha_3^{-1/2}}{2\pi} :
\frac{\alpha_3^{3/2}}{32\pi} \,.
\end{eqnarray}
Although the coupling $H_1^0 W W_R$ depends on the RH gauge coupling $g_R$ [cf.~Table~\ref{tab:H10}], this dependence is canceled out by the $W_R$ boson mass $M_{W_R} = g_R v_R$. Thus the dependence of the decay branching fractions of $H_1^0$ on the gauge coupling $g_R$ enters only through the $W_R$ mass in the velocity $\beta_2$ [cf.~Eq.~\eqref{eqn:beta2}] and the function $f_2$ [cf.~Eq.~\eqref{eqn:f2}] defined in Appendix \ref{app:B}.
The other decay channels are suppressed either by the small couplings, for instance the $hh$ channel by $\epsilon = \kappa / v_R$ and $t\bar{t}$ channel by $m_b / m_t$, or by the phase space such as the $W_RW_R$ and $Z_RZ_R$ channels. The three-body decay into the SM Higgs states $H_1^0 \to hhh$ can be used to measure directly the quartic coupling $\lambda_4$, but the branching ratio is at most $6\times10^{-3}$, suppressed by the phase space. Given the three dominant channels with large couplings of order one, the total decay width of $H_1^0$ is generally very large, up to 2 or 3 TeV or even larger for a $\sim 10$ TeV mass. Even only the $b\bar{b}$ channel can contribute a width of few hundred GeV, if the $H_3^0h$ and $WW_R$ channels are not open kinematically.  This implies that the CP-even scalar $H_1^0$ (and also $A_1^0$, $H_1^\pm$ and $H_2^{\pm\pm}$, as we will see below) in the minimal LR model will appear as a wide resonance if it is accessible to the future colliders.

The decay of $A_1^0$ is somewhat similar to $H_1^0$, and is dominated by the $b\bar{b}$ and $WW_R$ channels, with the partial decay widths for these two channels the same as that for $H_1^0$ at the LO, as they share the same Yukawa and gauge couplings up to a complex phase. All other channels are highly suppressed. The singly-charged Higgs $H_1^\pm$ comes from the same doublet as $H_1^0$ and $A_1^0$ and its decay is closely related to the two neutral scalars. From the couplings in Table~\ref{tab:H1p}, it is easily found that $H_1^\pm$ decays dominantly to $t\bar{b} \, (\bar{t}b)$ and $ZW_R$, with the partial width relations governed by the gauge and Yukawa interactions before the spontaneous symmetry breaking at the RH scale:
\begin{eqnarray}
\label{eqn:width}
&& \Gamma(H_1^0 \rightarrow b\bar{b}) \ \simeq \
\Gamma(A_1^0 \rightarrow b\bar{b}) \ \simeq \
2 \Gamma(H_1^+ \rightarrow t\bar{b}) \,, \\
&& \Gamma(H_1^0 \rightarrow WW_R) \ \simeq \
\Gamma(A_1^0 \rightarrow WW_R) \  \simeq \
2 \Gamma(H_1^+ \rightarrow ZW_R^+) \ \simeq \ 2\Gamma( H_1^+  \to h W_R^{+} ) \, .
\end{eqnarray}
These simple relations among the partial decay widths of the heavy neutral and charged scalars $H_1^0$, $A_1^0$ and $H_1^\pm$ are characteristic signals of the minimal LR model, and can be used as a way to distinguish the LR Higgs sector from other non-SM Higgs sectors, such as the MSSM.

For the neutral hadrophobic scalar $H_3^0$, if it is not heavy enough to produce the heavy pairs $NN$, $W_RW_R$, $Z_RZ_R$ or $H_2^{++}H_2^{--}$, it can decay only into a pair of SM Higgs states $hh$, since the $t\bar{t}$ and $b\bar{b}$ channels are suppressed by the small mixing parameter $\epsilon$.\footnote{Even for $M_{H_3^0}<2M_h$, the dominant tree-level decay mode of $H_3^0$ is still into (off-shell) SM Higgs bosons. If the $H_3^0hh$ coupling is really small, the loop-induced $H_3^0\to \gamma\gamma$ decay will take over. For the collider sensitivity study in the next section, we will simply assume that the $H_3^0hh$ coupling is large enough to ensure the decay of $H_3^0\to 4b$ inside the detector.} In this case, its total width is rather small (of order 10 GeV), depending on the quartic parameter $\rho_1$ which is related to the mass of $H_3^0$ via $M_{H_3^0} = 2 \sqrt{\rho_1} v_R$, and also on $\alpha_1$ which is directly related to the SM Higgs mass and trilinear coupling $\lambda_{hhh}$ [cf.~Eqs.~(\ref{eqn:hmass}) and (\ref{eqn:triple_SM})]. If the decays to heavy particles are open, the width would be largely enhanced, as none of those couplings are suppressed.
One interesting case is the decay of $H_3^0$ into a pair of doubly-charged Higgs, which decays further into four leptons: $H_3^0  \to H_2^{++} H_2^{--} \to \ell^+_i \ell^+_i \ell^-_j \ell^-_j$, where $i,\,j$ are the flavor indices. In this case we can study the two hadrophobic scalars simultaneously in one chain of production and decay processes. Note that in this channel, the trilinear coupling $(\rho_1 + 2\rho_2)$ for the vertex $H_3^0 H_2^{++} H_2^{--}$ is directly related to the masses of the two heavy hadrophobic scalars, cf.~Eqs.~\eqref{eqn:H30mass} and \eqref{eqn:scalarmass3}.

For the doubly-charged scalar $H_2^{\pm\pm}$, the dominant decay channel is to a pair of same-sign leptons. If its mass is larger than twice the $W_R$ mass, the $W_RW_R$ channel is also open and contributes sizably to the total width. Here again the dependence of width on the gauge coupling $g_R$ is only through the $W_R$ mass.

Finally, there are also LNV Higgs decays, such as $H_3^0\to NN\to \ell^\pm \ell^\pm+4j$~\cite{Maiezza:2015lza}  that could provide additional distinct signals of the LR model.


\section{Discovery potential}
\label{sec:6}


Given the dominant production and decay modes of the heavy Higgs states in the minimal LR model, we list here the key discovery channels for this new Higgs sector at future hadron colliders. As a brief guideline for future in-depth and sophisticated studies, we only calculate the collider signals and the relevant dominant SM backgrounds at parton-level for a conservative estimation of the sensitivity reach for the heavy Higgs masses at the FCC-hh/SPPC.
For concreteness, we focus mainly on the
decay modes to SM particles, unless  otherwise  specified. This choice is motivated by the fact that compared to the pure SM final states, the channels with non-SM heavy particles, e.g. to heavy gauge bosons $W_R/Z_R$ and heavy neutrinos $N$ [cf. Table~\ref{tab:decay}], are somewhat obscure due to the hitherto unknown model parameters, such as the $W_R/Z_R/N$ mass and the RH gauge coupling $g_R$. Moreover, the signals of these heavy gauge bosons will be easier to see in other channels involving their direct production, such as in dijet, dilepton or dilepton+dijet final states~\cite{Keung:1983uu}, {\it before} they can be detected in cascade decays from heavy Higgs production. Therefore, the Higgs decay modes involving non-SM particles might be more relevant to other exotic studies at future hadron colliders~\cite{Arkani-Hamed:2015vfh} and we do not discuss them here for simplicity.

\subsection{Bi-doublet Higgs Sector}

For the heavy bi-doublet sector, namely, $H_1^0,\, A_1^0$ and $H_1^\pm$, the dominant production channels are determined by the Yukawa couplings of these heavy scalars to the third generation quarks, independent of the quartic scalar couplings or the RH gauge coupling at the LO, as discussed in Section~\ref{sec:4.1}. Therefore, their sensitivity reach at future colliders can be determined more robustly without making any assumptions on the model parameters.

\subsubsection{$H_1^0/A_1^0$}
For the bidoublet neutral scalars $H_1^0/A_1^0$, the main discovery channel is $pp \to H_1^0/A_1^0 \to b\bar{b}$. Due to the high center-of-mass energy and large masses of $H_1^0/A_1^0$, as required by FCNC constraints, the $b$-jets are highly boosted, which could be helpful in distinguishing them to some extent from the otherwise huge SM $b\bar{b}$ background $\sim 10^6$ pb at $\sqrt s=100$ TeV, which is many orders of magnitude above the signal. Since we know the bi-doublet Higgs has to be beyond 10 TeV in the LR model, we simply apply an invariant mass cut on the hardest bottom quark jets, $M_{bb} > 10$ TeV, in addition to the basic $p_T$ and jet-separation cuts used in Section~\ref{sec:4.1}. The severe $M_{bb}$ cut significantly reduces the QCD background to about 1.5 pb at NNLO, without losing much of the signal.

For the CP-even $H_1^0$, there is an additional key channel, i.e. $pp\to H_1^0\to hH_3^0\to hhh$ [cf. Table~\ref{tab:decay}]. If $H_3^0$ is not very heavy, e.g. at the sub-TeV scale, the branching ratio of $H_1^0\to h H_3^0$ can be sizable (at the level of 10\% for 10 TeV $H_1^0$ and 1 TeV $H_3^0$) and this is a viable channel for both $H_1^0$ and $H_3^0$ discovery. The triple Higgs production can be searched for via distinct final states of $6b$ or $4b+2\gamma$~\cite{Papaefstathiou:2015paa,Chen:2015gva}.
The LO $gg\to hhh$ production cross section in the SM is 3.05 fb at $\sqrt s=100$ TeV, with a large NLO K-factor of $\sim 2$~\cite{Papaefstathiou:2015paa}. This has to be multiplied with the appropriate SM branching ratios of either $h\to b\bar{b}$ or $h\to \gamma\gamma$. For the $6b$ final state, another dominant SM background is from triple $Z$ production, which has a cross section of 260 fb at NLO~\cite{Torrielli:2014rqa, Zat100TeV}. For the $4b2\gamma$ final state, we should also take into account the backgrounds due to $hZZ$ and $Zhh$ production, which have NLO cross sections of 37 fb and 8.3 fb, respectively~\cite{Zat100TeV}. However, since ${\rm BR}(h\to \gamma\gamma)=2.27\times 10^{-3}$ is much smaller than ${\rm BR}(h\to bb)= 0.58$, the number of signal events for $4b2\gamma$ will be much smaller than $6b$ and it would be challenging to search for heavy scalars in the $4b2\gamma$ (or $2b4\gamma/6\gamma$) mode.

\subsubsection{$H_1^{\pm}$}
For the singly charged $H_1^{\pm}$, the key discovery channel is $pp \to H_1^\pm t \to ttb$. Again, due to the large mass of $H_1^\pm$, both $t$ and $b$-jets will be highly boosted, which will be a key feature to extract the signal from the irreducible QCD background. In particular, jet substructure analysis of the heavy quark jets and the kinematic observables could help to suppress the SM background and also to distinguish the LR model from other scenarios such as the MSSM. As it is more challenging to reconstruct a fat top-quark jet in our simulations, we apply only the simple selection cut of $M_{tb} > 5$ TeV on the hardest $tb$ pair to reduce the QCD background from $\sim 300$ pb level to about 20 pb at NLO, while retaining sizable number of signal events.

\begin{table}[t]
  \centering
  \caption[]{Key discovery channels and dominant SM backgrounds for the bi-doublet Higgs states in the minimal LR model at $\sqrt{s} = 100$ TeV $pp$ collider. The last column gives the cross sections for the SM backgrounds (calculated at an order same as or higher than the signal), after applying the selection cuts discussed in the text.}
  \label{tab:bkg1}
  \begin{tabular}{|l|l|l|r|}
  \hline\hline
  scalar & discovery channel & SM background & $\sigma_{\rm SM}$ [fb] \\ \hline
  \multirow{3}{*}{$H_1^0 / A_1^0$}  &
  \multirow{1}{*}{$H_1^0 / A_1^0 \to b\bar{b}$}
  & $b\bar{b}$ &  $1500$   \\
  \cline{2-4}
  & \multirow{2}{*}{$H_1^0 \to hH_3^0 \rightarrow hhh$}
  & $hhh \to 6b$ & 0.038  \\
  && $ZZZ \to 6b$  & 0.19  \\
  \hline
  $H_1^\pm$ & \multirow{1}{*}{$H^\pm t \to ttb$} & $ttb \to bbbjj\ell\nu$  & 984  \\
  \hline\hline
  \end{tabular}
\end{table}

\begin{figure}[!t]
  \centering
   \includegraphics[width=0.5\textwidth]{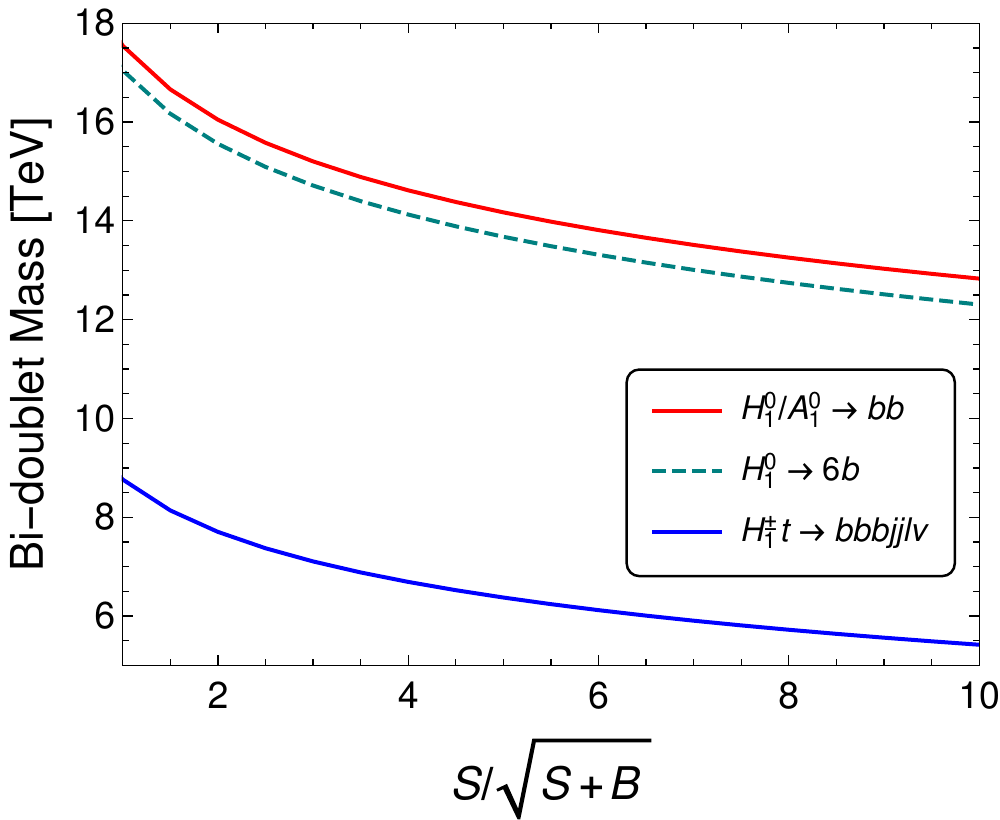}
  \caption{Sensitivity reach of the bi-doublet scalars in the minimal LR model at $\sqrt s= 100$ TeV $pp$ collider.}
  \label{fig:sens-bi}
\end{figure}

All the key discovery channels and dominant backgrounds for the bi-doublet heavy scalars are collected in Table~\ref{tab:bkg1}. To estimate the prospects of these heavy scalars at a future 100 TeV collider, we assume an optimistic integrated luminosity of 30 ab$^{-1}$ and calculate the expected number of corresponding signal ($S$) and background ($B$)  events. With this, we compute the expected signal sensitivity $S/\sqrt{S+B}$ and find that a $3\sigma$ sensitivity can be reached for $H_1^0 / A_1^0$ masses up to 15.2 TeV in the $b\bar{b}$ channel and up to 14.7 TeV for $H_1^0$ in the $hhh\to 6b$ channel, as shown in Figure~\ref{fig:sens-bi} (red and green curves, respectively). Although the number of $hhh$ signal events is expected to be smaller than the $b\bar{b}$ events, the corresponding SM background is also much smaller, as noted above, which makes the sensitivity reach in both channels comparable. Thus, as long as $H_3^0$ is not too heavy (below TeV scale) and the scalar coupling $\alpha_2$ for the vertex $h H_1^0 H_3^0$ is not small ($\gtrsim 0.1$), the $hhh$ final state is one of the primary channels to search for both the heavy CP-even scalars $H_1^0$ and $H_3^0$ simultaneously.

On the other hand, the prospects for the singly-charged Higgs bosons are not so promising, since the production cross section is smaller than the neutral bi-doublet case, and moreover, the $ttb$ background is much larger than the signal, even after imposing severe kinematic cuts. We find that a $3\sigma$ level sensitivity in the $ttb\to bbbjj\ell \nu$ (with $\ell=e,\mu$) channel can be reached up to $M_{H_1^\pm}=7.1$ TeV only, as shown in Figure~\ref{fig:sens-bi} (blue curve). This is a rather optimistic limit, since the multi-particle final state under consideration is rather difficult to analyze in practice.

\subsection{Hadrophobic Higgs Sector}

For the hadrophobic heavy scalars, it is more intricate to calculate the sensitivity reach, since it depends on other LR model parameters as well. As long as the Higgs portals are open, i.e. with $\alpha_1 \neq 0$ in the simple scenarios considered in Sections~\ref{sec:4.2} and \ref{sec:4.3},  the production cross sections for $H_3^0$ and $H_2^{\pm\pm}$ depend on the RH scale $v_R$ in all the dominant  channels. For the associated production channel $H_3^0 h$, the coupling is directly proportional to the scale $v_R$ [cf.~Table~\ref{tab:H30}]. On the other hand, for the SM Higgs-portal $H_3^0 H_3^0$ pair-production process, the dependence on $v_R$ comes through the mixing term $\alpha_1 / \rho_1$, which depends on $M_{H_3^0} = 2\sqrt{\rho_1} v_R$. For a fixed value of $\alpha_1$ and $M_{H_3^0}$, the trilinear scalar coupling $h H_3^0 H_3^0$ is a function of the RH scale $v_R$. When $v_R$ is larger, the quartic coupling $\rho_1$ is smaller and the mixing $\alpha_1 / \rho_1$ becomes larger, which can enhance the production of $H_3^0$ pairs. For the VBF production of $H_3^0$, for a fixed value of the RH gauge coupling $g_R$, when the RH scale $v_R$ goes higher, due to the huge suppression by the heavy gauge boson $W_R$ (and $Z_R$) masses, the production of $H_3^0$ drops rapidly beyond the pair-production of on-shell $W_R$ bosons.
For the VBF production of the doubly-charged scalar $H_2^{\pm\pm}$, the situation is rather similar to the $H_3^0$ case, with the only differences being the factor of 2 in the coupling to $W_R$ [cf.~Tables~\ref{tab:H30} and \ref{tab:H2pp}], a factor of 2 for identical particles in the vertex $H_2^{\pm\pm} W^\mp W^\mp$, and the absence of sub-leading contributions from $Z_RZ_R$ fusion.
The sensitivities for both $H_3$ and $H_2^{\pm\pm}$ in the Higgsstrahlung channel are also sensitive to the gauge coupling $g_R$ and the masses of RH gauge bosons; however, they are weaker than the VBF sensitivities for the same set of model parameters.

\subsubsection{$H_3^0$}
For relatively heavy hadrophobic scalar $H_3^0$, the key discovery channel is the VBF: $pp  \to  H_3^0 jj  \to  hh^{(*)}jj$, with a sub-dominant contribution from the Higgsstrahlung process $pp  \to  H_3^0 W_R  \to  hh^{(*)}jj$. This can be ideally searched for at $\sqrt s=100$ TeV in either $4b+jj$ or $bb\gamma\gamma jj$ channels. The dominant background is VBF production of SM Higgs pair, which is estimated to be 80 fb at 100 TeV~\cite{Baglio:2015wcg}. The $ZZW$ background is also sizable, about 1.4 pb~\cite{Torrielli:2014rqa}. The corresponding backgrounds at $\sqrt s=14$ TeV LHC are 12 fb for Higgs-pair production~\cite{Baglio:2015wcg} and 36 fb for $ZZW$~\cite{Yong-Bai:2015xna}.

For smaller $H_3^0$ masses, the triple Higgs channel $pp\to H_3^0h\to hh^{(*)}h$ becomes important. The $hhh$ channels are subject to the uncertainty in the quartic couplings in the scalar potential, but may benefit from the on-shell decay of $H_3^0$. Different from other heavy scalars in the minimal LR model, the decay width of $H_3^0$ could be comparatively rather small, say few times 10 GeV, as long as none of the heavy particle channels are not open. Thus by reconstructing the invariant mass of the right Higgs pair, one could expect a significant resonance-like peak above the SM backgrounds, which mostly come from triple Higgs~\cite{Papaefstathiou:2015paa, Baglio:2015wcg} or triple $Z$ decays~\cite{Torrielli:2014rqa, Binoth:2008kt,Lazopoulos:2007ix}. Moreover, the narrower triple Higgs invariant mass can be used to distinguish this channel from the same final states due to $H_1^0$ decay which is likely to be a broader resonance of order TeV.

Another interesting possibility is the pair-production of $H_3^0$ which leads to four-Higgs final states: $pp\to H_3^0H_3^0\to hh^{(*)}hh^{(*)}$. Here the dominant SM background comes from $4Z$ production with a NLO cross section of 0.8 fb at $\sqrt s=100$ TeV~\cite{Torrielli:2014rqa}. However, the $H_3^0 H_3^0$ channel is relevant for discovery of $H_3^0$ only at very small values of $M_{H_3^0}<200$ GeV [cf. Figures~\ref{fig:production3} and \ref{fig:prod-lhc} (left)], where one of the SM Higgs bosons in the $H_3^0\to hh$ decay must be off-shell, thus significantly decreasing the signal sensitivity.

\begin{table}[t]
  \centering
  \caption[]{Key discovery channels and dominant SM backgrounds for the hadrophobic Higgs states in the minimal LR model at hadron colliders. The last two columns give the cross sections for the SM background at $\sqrt s=100$ and 14 TeV respectively (calculated at an order same as or higher than the corresponding signal), after applying the selection cuts mentioned in the text.}
  \label{tab:bkg2}
  \begin{tabular}{|l|l|l|l|l|}
  \hline\hline
  scalar & discovery channel & SM background & $\begin{matrix}\sigma_{\rm SM} \, {\rm [fb]} \\ (100 \, {\rm TeV}) \end{matrix}$ & $\begin{matrix}\sigma_{\rm SM} \, {\rm [fb]} \\ (14 \, {\rm TeV}) \end{matrix}$ \\ \hline
  \multirow{5}{*}{$H_3^0$} & \multirow{2}{*}{$H_3^0 jj \to hhjj$} & $hhjj \to bbbbjj$ & 27 & 4.1  \\
  && $ZZW \to bbbbjj$ & 21 & 0.54 \\
  \cline{2-5}
  & \multirow{2}{*}{$H_3^0 h \to hhh$}
  & $hhh \to 6b$ & 1.2 & 0.016 \\
  && $ZZZ \to 6b$ & 0.91 & 0.054 \\
  \cline{2-5}
  & $H_3^0 H_3^0 \to hhhh$ & $ZZZZ \to 8b$ & $4.2 \times 10^{-4}$ & $1.5 \times 10^{-5}$ \\
  \hline
  \multirow{2}{*}{$H_2^{\pm\pm}$} & $H_2^{++} H_2^{--} \to \ell^+ \ell^+ \ell^- \ell^-$ &  $ZZ \to \ell^+ \ell^+ \ell^- \ell^-$ & 2.1 & 0.18  \\ \cline{2-5}
  & $H_2^{\pm\pm}jj\to \ell^\pm\ell^\pm jj$ & $WZ,ZZ,WW$ & 1000 & 71  \\
  \hline\hline
  \end{tabular}
\end{table}

\begin{figure}[t!]
  \centering
  \includegraphics[width=0.49\textwidth]{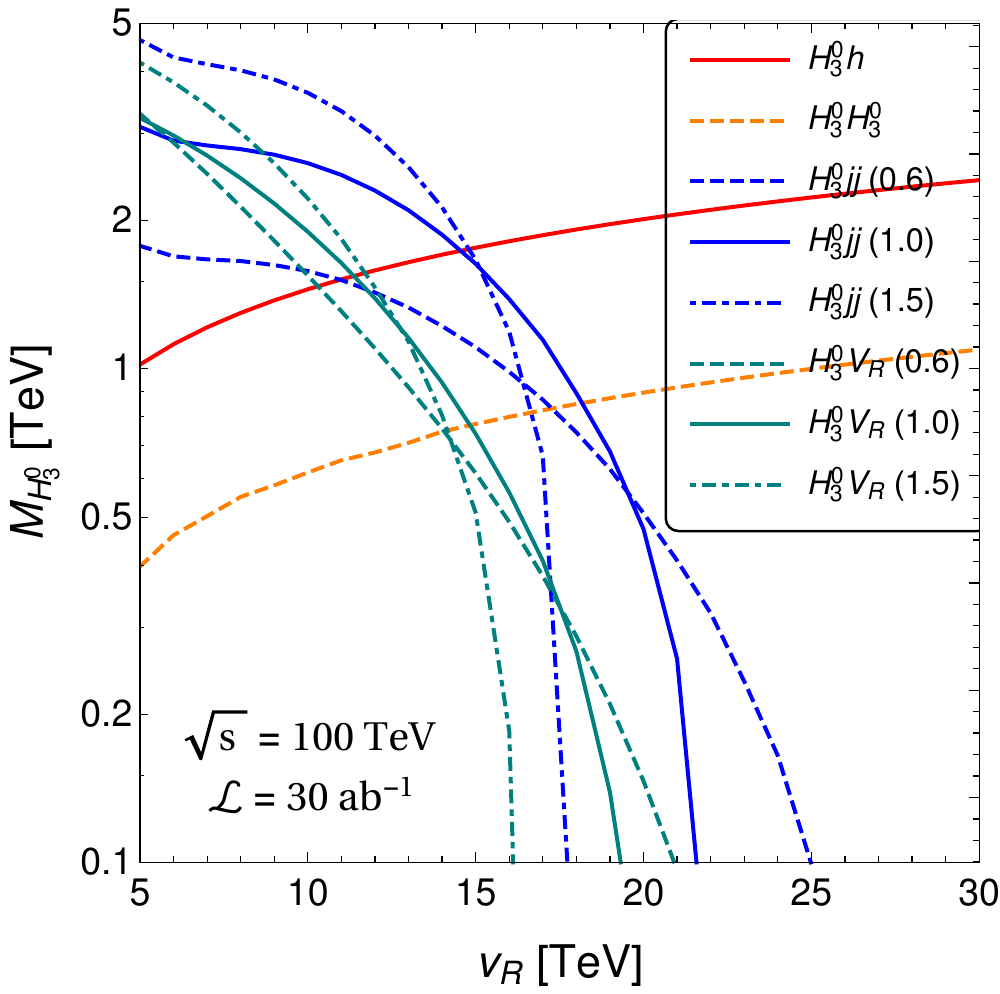}
  \includegraphics[width=0.495\textwidth]{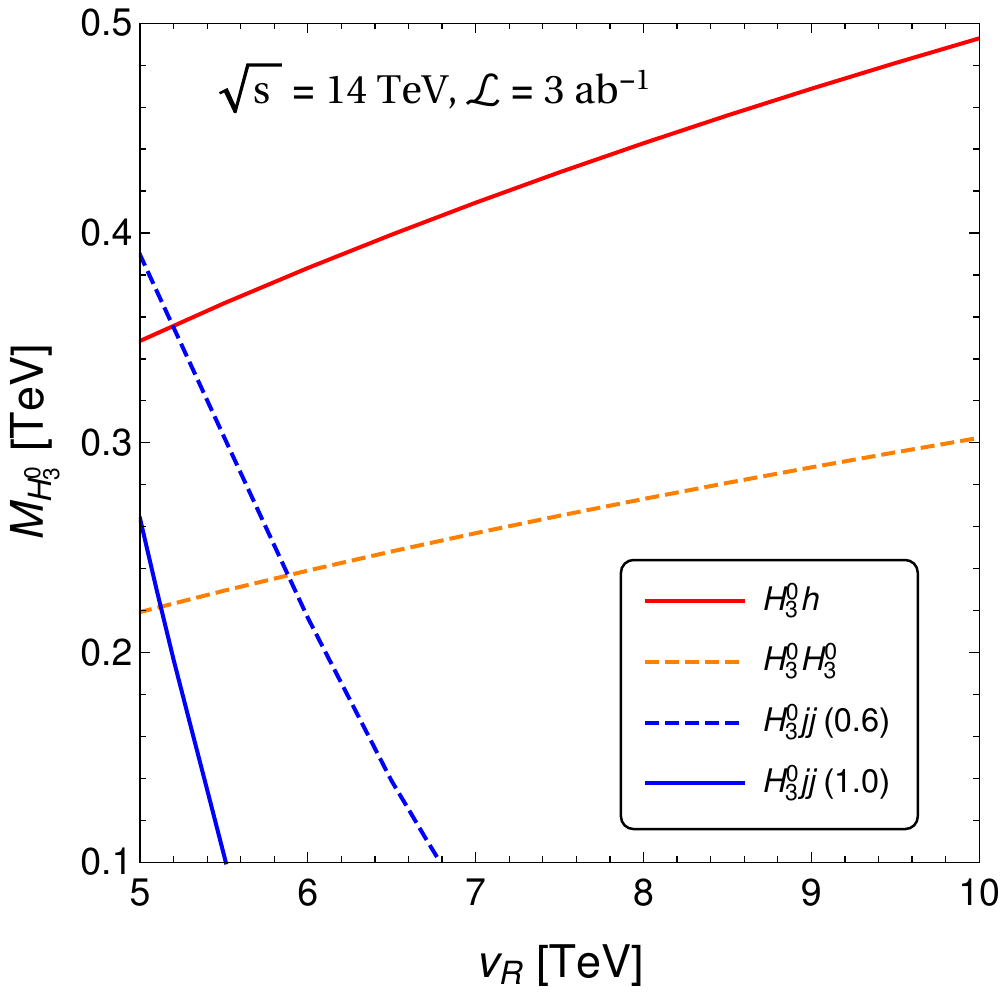}
  \caption{$3\sigma$ sensitivity of the neutral hadrophobic scalar $H_3^0$ in the minimal LR model at $\sqrt s=100$ (14) TeV collider with an integrated luminosity of 30 (3) ab$^{-1}$. The values in parenthesis are for different ratios of $g_R/g_L$.}
  \label{fig:discovery1}
\end{figure}

All the key discovery channels of $H_3^0$ are collected in Table \ref{tab:bkg2}, wherein we also list the cross sections for the dominant SM backgrounds at $\sqrt{s} = 100$ and 14 TeV. As the hadrophobic scalars can be as light as hundreds GeV scale, we do not apply any special cuts on the invariant mass of the final states. Thus, the sensitivity plots shown in Figure~\ref{fig:discovery1} are rather conservative. In the left (right) panel, we show the $3\sigma$ mass reach for the hadrophobic neutral scalar as a function of the RH scale $v_R$ at $\sqrt s=100$ (14) TeV collider with an integrated luminosity of 30 (3) ab$^{-1}$. We find that for a small coupling $\alpha_1 = 0.01$, the hadrophobic scalar $H_3^0$ can be probed via the Higgs-portal up to a few TeV scale. As mentioned above, the $H_3^0H_3^0$ mode remains sub-dominant to the $H_3^0h$ mode, unless we go to very small $M_{H_3^0}$ values (not shown in Figure~\ref{fig:discovery1}). When the ``mixing'' parameter $\alpha_1$ or the RH scale $v_R$ is larger, the sensitivity in both the Higgs-portal channels can be further improved, due to the enhanced signal rate [cf. Table~\ref{tab:H30}], until we hit the perturbativity bound for the triple Higgs couplings. On the other hand, the VBF channel is suppressed by the heavier gauge boson masses at higher $v_R$, and therefore, is dominant only for smaller $v_R$. In this channel, $H_3^0$ can be probed up to a few TeV range, depending on the RH gauge coupling strength. The Higgsstrahlung channel $pp  \to  H_3^0 W_R  \to  hh^{(*)}jj$ turns out to give a smaller sensitivity, as compared to the VBF channel.

At the $\sqrt s=14$ TeV LHC in the high-luminosity (HL) phase, a $3\sigma$ sensitivity can be achieved only in the sub-TeV mass range, mainly due to the much smaller signal event rate [cf. Figure~\ref{fig:prod-lhc}]. We do not show the sensitivities for the VBF case with $g_R/g_L=1.5$ nor the Higgsstrahlung case at the HL-LHC, since the signal rate in this case is too small, mainly due to the larger RH gauge boson mass suppression. In any case, we are not aware of any direct experimental limits on $H_3^0$ and the sensitivity study presented here should provide some motivation for their future collider searches.


\subsubsection{$H_2^{\pm\pm}$}
For the doubly-charged scalars $H_2^{\pm\pm}$, there are two dominant discovery channels, depending on the mass range being probed. (i) For low masses, it is the DY process $pp\to H_2^{++}H_2^{--}\to \ell^+ \ell^+ \ell^- \ell^-$, where some of the leptons could in principle be of different flavor, thus probing lepton flavor violation. This leptonic channel  is rather clean at hadron colliders, and therefore, constitutes a ``smoking gun" signal for the Higgs sector of the LR model. The most important background for this channel is the SM $ZZ$ production~\cite{ZZat100TeV, Melia:2011tj} whose total cross section is 466 fb at NNLO. By suitably reconstructing the invariant masses of same and opposite-sign charged lepton pairs, the $ZZ$ background can be significantly reduced. (ii) For high masses, it is the VBF process $pp\to H_2^{\pm\pm}jj\to \ell^\pm\ell^\pm jj$, which is a high-energy analog of the $0\nu\beta\beta$ process, thus probing lepton number violation at colliders.\footnote{The higher-order VBF process $H_2^{++} H_2^{--} jj$ is also promising at the FCC-hh~\cite{Bambhaniya:2015wna}. It is interesting to note that this could also stem from $pp \to H_3^0 jj \to H_2^{++} H_2^{--} jj$ with on-shell VBF production of $H_3^0$, provided $M_{H_3^0}> 2M_{H_2^{\pm\pm}}$, which could significantly enhance this signal.} The Higgsstrahlung process $pp\to H_2^{\pm\pm}W_R^\mp\to H_2^{\pm\pm}jj$ gives a sub-dominant contribution to this signal. The SM does not have any same-sign dilepton events with jets and without missing $E_T$, at least at LO; however, there are several SM processes which pose as an irreducible background to {\it inclusive} same-sign dilepton searches in the VBF channel, such as leptonic decays of $WZ,ZZ$ and a smaller contribution from $W^\pm W^\pm$~\cite{Jager:2009xx}.  There are also reducible backgrounds from the opposite-sign lepton pairs produced via DY, $t\bar{t}$, $W^\pm W^\mp$ and $Wt$ decays, where the charge for one of the leptons is wrongly reconstructed. The charge mis-identification rate at the LHC is rather small of order 1.5\%~\cite{ATLAS:2014kca}, and is expected to be of similar order for FCC-hh, depending on the detector material. However, since the opposite-sign dilepton background is huge, one has to take into account both reducible and irreducible backgrounds for the VBF process, which results in a total background of order 1 pb for the $\sqrt s=100$ TeV and 71 fb at $\sqrt s=14$ TeV. All the key discovery channels of $H_2^{\pm\pm}$ and the dominant SM backgrounds at $\sqrt{s} = 100$ and 14 TeV are collected in Table \ref{tab:bkg2}.

\begin{figure}[t]
  \centering
  \includegraphics[width=0.49\textwidth]{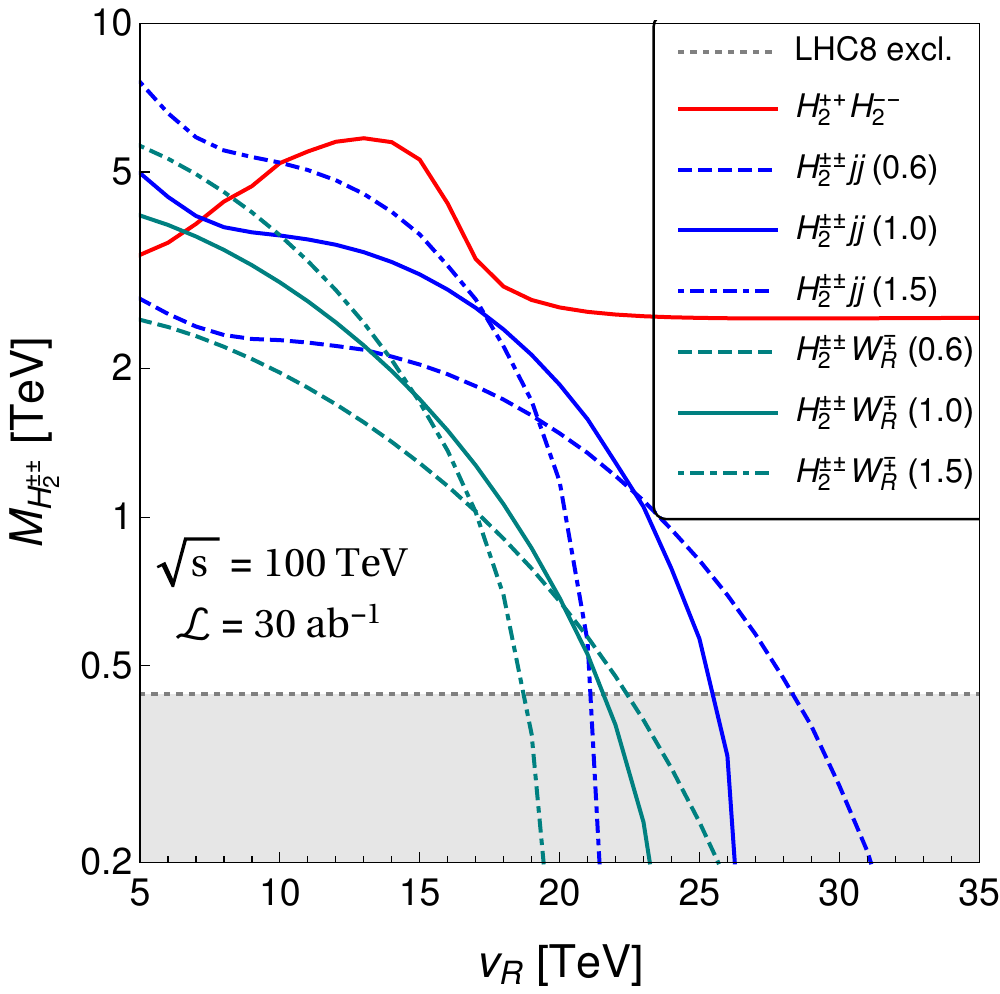}
  \includegraphics[width=0.485\textwidth]{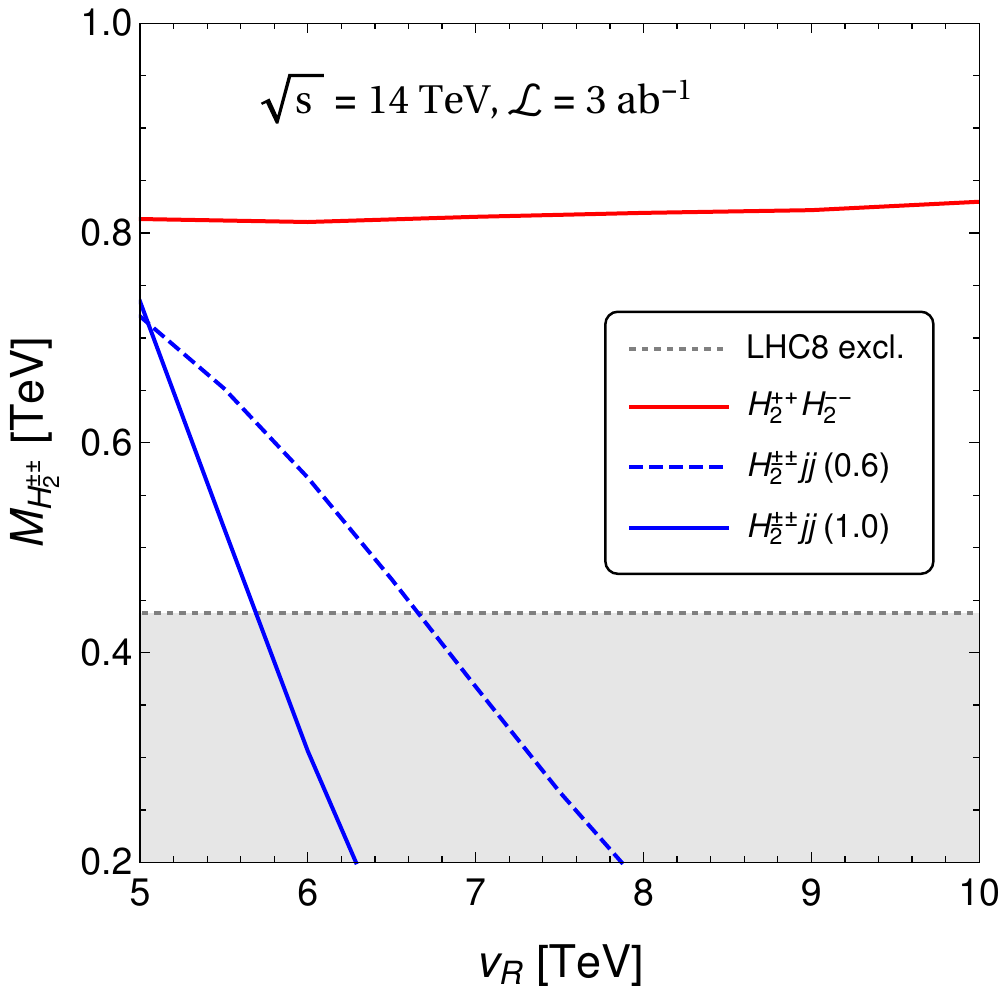}
  \caption{$3\sigma$ sensitivity of the RH doubly-charged scalar $H_2^{\pm\pm}$ in the minimal LR model at $\sqrt s=100$ (14) TeV collider with an integrated luminosity of 30 (3) ab$^{-1}$.  The values in parenthesis are for different ratios of $g_R/g_L$. The gray shaded region is excluded at 95\% C.L. from the $\sqrt s=8$ TeV LHC searches for same-sign dimuon pairs.}
  \label{fig:discovery2}
\end{figure}

The sensitivity reach for $H_2^{\pm\pm}$ in the two dominant channels mentioned above are presented in the left panel  of Figure~\ref{fig:discovery2} for $\sqrt s=$100 TeV collider with an integrated luminosity of 30  ab$^{-1}$. For comparison, we also show the current 95\% C.L. lower limit~\cite{ATLAS:2014kca} on $H_2^{\pm\pm}$ mass from the DY pair-production via $s$-channel $Z$ or photon and the subsequent decay of $H_2^{\pm\pm}$ into same-sign dileptons. The exclusion shown here (gray shaded region) is for the $\mu^\pm \mu^\pm$ channel, which is the most stringent, while the corresponding limits for $e^\pm e^\pm$ and $e^\pm \mu^\pm$ are slightly weaker and not shown here. Also note that we have used the limits on the RH doubly-charged scalars (as applicable to our case), which are weaker than the corresponding limits on LH doubly-charged scalars, due to their different coupling to $Z$.  At the 100 TeV machine, we can probe $H_2^{\pm\pm}$ masses up to a few TeV.  For the DY process, due to the large mass of $Z_R$, the pair production of doubly-charged scalars is dominated by the SM $\gamma$ or $Z$ $s$-channel mediators. Thus, in the absence of the heavy $Z_R$ boson contribution, the sensitivity for $H_2^{\pm\pm}$ is intrinsically independent of the new physics at the RH scale. However, when it comes to the resonance region $M_{Z_R} \simeq 2 M_{H_2^{\pm\pm}}$, the DY signal, and hence, the sensitivity can be improved  significantly. This is the reason why we have a bump around 13 TeV in the right panel of Figure~\ref{fig:discovery1}, which corresponds to $M_{Z_R}\simeq 14$ TeV. Such a resonance-like feature provides a very clear probe of the LR model at future colliders, in combination with the searches for heavy RH gauge bosons which could give us a hint on where to look for this resonance in the new Higgs signal.



For comparison, we also show the $3\sigma$ sensitivities for $H_2^{\pm\pm}$ at the HL-LHC with an integrated luminosity of 3 ab$^{-1}$ in the right panel of Figure~\ref{fig:discovery2}. With conservative treatment of the SM backgrounds in this work, the hadrophobic scalars $H_3^0$ and $H_2^{\pm\pm}$ can only be probed below the TeV scale. Note, however, that the Higgs portals of $H_3^0$ depend on the quartic scalar couplings. A larger value of $\alpha_1$ could enhance the signal rate significantly. On the other hand, for both $H_3^0$ and $H_2^{\pm\pm}$, the VBF channel is rather sensitive to the gauge coupling $g_R$. When $g_R$ is large, e.g. $1.5g_L$, the RH gauge bosons $W_R$ and $Z_R$ become so heavy that it is rather challenging to see the hadrophobic scalars in the VBF mode. The sensitivities in the Higgsstrahlung channel are lower than the VBF ones, and therefore, not shown in the figure. One should keep in mind that all the lines in Figure \ref{fig:discovery2} are based on the rather simple and conservative treatment of the SM backgrounds. Realistic and indicate analysis of the signals and backgrounds could improve largely the sensitivities. In addition, the hadrophobic scalars could also be detected at the LHC with a smaller luminosity, say 300 fb$^{-1}$, which however needs detailed consideration of the backgrounds and is beyond the scope of this paper.

\section{Distinction between the LR and MSSM Higgs sectors} \label{sec:7}

 \begin{table}[t]
  \centering
  \caption[]{A comparison of the dominant collider signals of neutral and charged scalars in the minimal LR model and MSSM. }
  \label{tab:decay2}
  \begin{tabular}{|lll|}
  \hline\hline
  Field & MSSM & LR model \\ \hline
  $H^0_1, A^0_1$  & $b\bar{b},~\tau^+\tau^-$ (high $\tan\beta$) &  $b\bar{b}$ \\
  & $tt$ (low $\tan\beta$) & $W^+_R W^-\to \ell^+\ell^+ 4j$\\ \hline
    $H^+$         &  $t\bar{b}t\bar{b}$,  $t\bar{b}\bar{\tau}\nu$  &    $\bar{t}_Lb_R$ \\
  \hline\hline
  \end{tabular}
\end{table}

The bi-doublet Higgs in the minimal LR model is similar to other popular beyond SM Higgs sectors, such as the MSSM and more generally the 2HDM, which also contain two Higgs doublets. However there is a profound difference between the two models, since in the LR case, the second Higgs doublet, in the limit of $\kappa'=0$, does not contribute to the SM fermion masses and therefore the decay properties are very different, as illustrated in Table~\ref{tab:decay2}. In particular, the $\tau^+\tau^-$ final state is suppressed by either the Dirac Yukawa coupling or the left-right mixing for the neutral bi-doublet scalars $H_1^0/A_1^0$ in the LR model [cf. Table~\ref{tab:fermion}], whereas this is one of the cleanest search channels for the MSSM heavy Higgs sector in the large $\tan\beta$ limit~\cite{Khachatryan:2014wca, ATLAS:2014lxa}. Furthermore, due to the presence of extra gauge fields in our case i.e. $W^\pm_R, Z_R$, new modes which are very different from MSSM appear, e.g. $H^0_1 \to W^+_RW^- $ and $H^+_1\to W^+_RZ$ which have no MSSM analog.  These modes can lead to distinguishing signals in leptonic channels e.g. $\ell^\pm\ell^\pm jj jj$ with $\sim 5\% $ branching ratio. In $3~{\rm ab}^{-1}$ data, this can lead to about 100 signal events and the SM background for these processes is expected to be very small.  Such leptonic final state signals are absent for MSSM Higgs. One can also use the relations between the various partial decay widths as shown in Table~\ref{tab:decay} and Eq.~(\ref{eqn:width}) to distinguish the LR Higgs sector from other 2HDM scenarios. For 2HDM Higgs studies at the 100 TeV collider, see e.g. Ref.~\cite{Baglio:2015wcg, Hajer:2015gka}.

If a positive signal is observed, one can also construct various angular and kinematic observables to distinguish the minimal LR scenario from other models giving similar signals~\cite{Han:2012vk, Chen:2013fna, Dev:2015kca}. For instance, we find from Table~\ref{tab:H1p} that $\bar{t}_Lb_R$ final states are preferred over the $\bar{t}_Rb_L$ final states for $H_1^+$ production, which can be utilized to distinguish it from 2HDM scenarios, including the MSSM.

Another key feature which distinguishes the LR Higgs sector from 2HDM is the presence of the neutral and doubly-charged hadrophobic scalars. A positive signal for any of the doubly-charged scalars discussed above will be a strong evidence for the LR model. Also, the $H_3^0\to hh$ decay mode of the neutral hadrophobic scalar leads to distinct multi-Higgs final states, which are absent in 2HDM scenarios in the so-called {\it alignment limit}, since the $Hhh$ coupling identically vanishes~\cite{Gunion:2002zf, Carena:2013ooa, Dev:2014yca}. As the current LHC Higgs data suggest the couplings of the observed 125 GeV Higgs boson to be close to the SM expectations, thus overwhelmingly favoring the alignment limit for any extended Higgs sector, the multi-Higgs signals listed in Tables~\ref{tab:bkg1} and \ref{tab:bkg2} will provide another unique way to distinguish the LR model from generic 2HDMs.

\section{Summary} \label{sec:8}
We have presented a detailed exploration of the collider signals of the new Higgs bosons of the minimal TeV scale Left-Right model for neutrino masses. We analyze all the dominant production and decay modes of the heavy Higgs bosons in the model at a future 100 TeV collider, such as the FCC-hh/SPPC, as well as at the HL-LHC. FCNC constraints in the minimal model make the 100 TeV collider a unique machine to probe the heavy bi-doublet Higgs bosons of the model. We also discuss how this model can be distinguished from other extended Higgs sectors, such as in the MSSM, whose Higgs sector partially overlaps with that of the minimal LR model, albeit with different couplings. We find that the bi-doublet neutral and singly-charged Higgs scalars can be effectively probed at a 100 TeV collider up to masses of 15 TeV and 7 TeV, respectively, independent of the other model parameters. The sensitivity reach for new hadrophobic neutral and doubly-charged Higgs bosons can go up to a few TeV, depending on the RH scale $v_R$ and the gauge coupling $g_R$.  Some of the considerations here can be further improved once better estimates of the higher-order QCD corrections are taken into account in discussing production cross sections and also more sophisticated simulations are performed to optimize the selection cuts and signal sensitivity. Thus, the results presented here can be taken as an initial guide in the exploration of the heavy Higgs sector of the minimal LR model at future colliders. Our hope is that this will provide a motivation to seriously probe the possibility that neutrino masses could owe their origin to new physics at the TeV scale and supplement any positive results that emerge from the LHC run II.

\section*{Acknowledgments}
Y.Z. would like to thank Qing-Hong Cao for the valuable discussions, and Jean-Marie Fr\`{e}re, Wouter Dekens and Julian Heeck for reading the manuscript and for their enlightening comments.
The work of B.D. is supported by the DFG grant RO 2516/5-1. B.D. would also like to acknowledge partial support from the TUM University Foundation Fellowship and the DFG cluster of excellence ``Origin and Structure of the Universe" during the earlier stages of this work. The work of R.N.M. is supported in part by the US National Science Foundation Grant No. PHY-1315155. Y.Z. would like to thank the IISN and Belgian Science Policy (IAP VII/37) for support.

\appendix
\section{Couplings involving the SM and heavy LR Higgs sector}
\label{app:A}

In this appendix we calculate all the couplings of the SM and heavy Higgs bosons from the bidoublet $\Phi$ and the triplet $\Delta_R$ in the minimal LR model to the SM fermions, vector bosons and among themselves. Our results are collected in Tables \ref{tab:triple} to \ref{tab:gauge2}.

\begin{table}[t!]
  \centering
  \caption[]{Trilinear scalar couplings in the minimal LR model.}
  \label{tab:triple}
  \begin{tabular}{lll}
  \hline\hline
  couplings & $\mathcal{O} (v_R)$ & $\mathcal{O} (\kappa,\, \xi v_R)$ \\ \hline
  $hhh$ & & $ \frac{1}{2\sqrt2} \left( 4\lambda_1 - \frac{\alpha_1^2}{\rho_1} \right)\kappa + \sqrt2 \left( 4\lambda_4 - \frac{\alpha_1\alpha_2}{\rho_1} \right)\xi \kappa $ \\ \hline
  $H^0_1 hh$ && $ \sqrt2 \left[ 3 \lambda _4  +\alpha _1 \alpha _2 \left(\frac{2}{\alpha _3-4 \rho _1}-\frac{1}{\rho _1}\right)\right] \kappa$ \\
  $H^0_3 hh$ & $ \frac{1}{\sqrt{2}} \alpha _1  v_R $ & $ 2\sqrt{2} \alpha _2 \xi v_R $ \\ \hline
  $h H^0_1 H^0_1$ && $ \sqrt2 \left[
   \lambda _1 +4 \lambda _2 +2 \lambda _3 +\frac{8 \alpha _2^2}{\alpha _3-4 \rho _1}-\frac{\alpha _1 \left(\alpha _1+\alpha _3\right)}{4\rho _1} \right] \kappa $ \\
  $h H^0_3 H^0_3$ && $ -\sqrt2 \left[ \alpha _1 - \frac{\alpha _1^2}{2\rho _1} +\frac{8 \alpha _2^2}{\alpha _3-4 \rho _1}  \right] \kappa$ \\
  $h H^0_1 H^0_3$ & $ 2 \sqrt{2} \alpha _2 v_R $ & $ \sqrt{2} \alpha _3 \xi v_R $ \\ \hline
  $h A^0_1 A^0_1$ && $\sqrt2 \left[ \left(\lambda _1-4 \lambda _2+2 \lambda _3\right)-\frac{\alpha _1 \left(\alpha _1+\alpha _3\right)}{4\rho_1}  \right]   \kappa$ \\
  $H_1^0 H^0_1 H^0_1$ && $ \sqrt{2}  \left[ \lambda _4 +\frac{2 \alpha _2 \left(\alpha _1+\alpha _3\right)}{\alpha _3-4 \rho _1} \right] \kappa$ \\
  $H_3^0 H^0_3 H^0_3$ & $ \sqrt{2} \rho_1 v_R $ \\ \hline
  $H_1^0 H^0_3 H^0_3$ && $ -\sqrt2 \alpha_2 \left[ 2- \frac{\alpha _1}{\rho _1}+\frac{4 \alpha _1+\alpha _3}{\alpha _3-4 \rho _1} \right] \kappa$ \\
  $H_3^0 H^0_1 H^0_1$ & $ \frac{1}{\sqrt{2}} \left(\alpha _1+\alpha _3 \right) v_R $ & $ -2\sqrt{2} \alpha _2 \xi v_R $ \\ \hline
  $H_1^0 A^0_1 A^0_1$ && $\sqrt2 \left[ \lambda _4  +\frac{2 \alpha _2 \left(\alpha _1+\alpha _3\right)}{\alpha _3-4 \rho _1} \right] \kappa $ \\
  $H_3^0 A^0_1 A^0_1$ & $\frac{1}{\sqrt{2}} \left(\alpha _1+\alpha _3 \right) v_R$ & $-2 \sqrt{2} \alpha _2 \xi v_R$ \\ \hline
  $h H^+_1 H^-_1$ && $\frac{1}{\sqrt2} \left[ \alpha _3+4 \lambda _1 -\frac{\alpha _1 \left(\alpha _1+\alpha _3\right)}{\rho _1}  \right] \kappa$ \\ \hline
  $H_1^0 H_1^+ H_1^-$ && $2\sqrt2 \left[ \lambda _4 + \frac{2 \alpha _2 \left(\alpha _1+\alpha _3\right)}{\alpha _3-4 \rho _1} \right] \kappa$ \\
  $H_3^0 H_1^+ H_1^-$ & $\sqrt{2} \left( \alpha _1+\alpha _3\right) v_R$ & $-4 \sqrt{2} \alpha _2 \xi v_R $ \\ \hline
  $h H^{++}_2 H^{--}_2$ && $\sqrt2 \left[ \alpha _3-\frac{2 \alpha _1 \rho _2}{\rho _1} \right] \kappa $ \\ \hline
  $H_1^0 H_2^{++} H_2^{--}$ && $2\sqrt2 \alpha_2 \kappa \, \frac{\alpha _3+8 \rho _2}{\alpha _3-4 \rho _1} $ \\
  $H_3^0 H_2^{++} H_2^{--}$ & $2 \sqrt{2} \left(\rho _1+2 \rho _2\right)v_R $ \\
  \hline\hline
  \end{tabular}
\end{table}

Taking the third order derivatives of the potential (\ref{eqn:potential}) with respect to the physical scalar states leads us to the trilinear couplings among the SM Higgs and the heavy beyond SM states. All these couplings are collected in Table~\ref{tab:triple}, which are uniformly written in the form of
\begin{eqnarray}
\mathcal{L} \ \supset \ - \lambda_{s_i s_j s_k} s_i s_j s_k \, ,
\end{eqnarray}
with $s_i$ standing for all the eight physical scalars. All the couplings are expanded in terms of the small parameters $\epsilon, \, \xi,\, \alpha \ll 1$ and truncated to the order of $\epsilon v_R \simeq \kappa = v_{\rm EW}$. One exception is the trilinear coupling for the light SM Higgs, which is calculated up to the order of $\epsilon^2 v_R$. The scalar trilinear couplings are presented in Table~\ref{tab:triple}, with the values at different orders ($v_R$ and $\epsilon v_R = \kappa$) separately shown. From this, one can readily see which couplings are potentially large and which are relatively suppressed. Some couplings vanish at the order of $\kappa$ and are not listed in Table~\ref{tab:triple}. 

\begin{table}[t!]
  \begin{center}
  \caption[]{Quartic scalar couplings in the minimal LR model.}
  \label{tab:quartic}
  \begin{tabular}{lll}
  \hline\hline
  couplings & $\mathcal{O} ({\rm quartic \, couplings})$ & $\mathcal{O} (\epsilon \times {\rm quartic \, couplings})$ \\ \hline
  $hhhh$ & $\frac14 \lambda _1$ & $ \lambda _4 \xi + \left(2 \lambda
   _2+\lambda _3\right) \xi ^2 +\frac{\alpha _1^2  \left(\alpha _1-2 \lambda _1\right) \epsilon^2}{16 \rho _1^2} $ \\ \hline
  $h hh H_1^0$ & $\lambda_4$ & $2 ( 2\lambda_2 + \lambda_3 ) \xi$ \\
  $hhh H_3^0 $ && $-  \left[ \frac{\alpha _1 \left(\alpha _1-2 \lambda _1\right)}{4 \rho _1}+\frac{4 \alpha _2
   \lambda _4}{\alpha _3-4 \rho _1}\right] \epsilon$ \\ \hline
  $hh H_1^0 H_1^0$ & $\frac12 \left(\lambda _1+4 \lambda _2+2 \lambda _3\right)$ \\
  $hh H_3^0 H_3^0 $ & $\frac14 \alpha_1$ & $\alpha_2 \xi$ \\
  $hh A_1^0 A_1^0$ & $\frac12 \left(\lambda _1-4 \lambda _2+2 \lambda _3\right)$ \\
  $hh H_1^0 H_3^0$ & & $\epsilon \left[ \frac{2 \alpha _2 \left(\alpha _1-2 \left(\lambda _1+4 \lambda _2+2 \lambda _3\right)\right)}{\alpha _3-4
   \rho _1}+\frac{\alpha _1 \left(3 \lambda _4-2 \alpha _2\right)}{2\rho _1} \right]$ \\ \hline
  $h H_3^0 H_3^0 H_3^0$ & & $\left[ \frac{\alpha _1 \left(\alpha _1-2 \rho _1\right)}{4 \rho _1}-\frac{4 \alpha _2^2}{\alpha _3-4 \rho _1}\right] \epsilon$ \\
  $h H_1^0 \{  H_1^0 H_1^0 ,\, A_1^0 A_1^0 \}$ & $\lambda_4$ & $-2 ( 2\lambda_2 + \lambda_3 ) \xi$ \\
  $h H_1^0 H_3^0 H_3^0 $ & $\alpha_2$ & $\frac12 \alpha_3 \xi$ \\
  $h H_3^0 A_1^0 A_1^0$ & & $\left[ -\frac{\alpha _1 \left(\alpha _1+\alpha _3-2 \lambda _1+8 \lambda _2-4 \lambda _3\right)}{4 \rho
   _1}-\frac{4 \alpha _2 \lambda _4}{\alpha _3-4 \rho _1} \right] \epsilon$ \\
  $h H_1^0 H_1^0 H_3^0$ & & $ \left[ \frac{4 \alpha _2 \left(2 \alpha _2-3 \lambda _4\right)}{\alpha _3-4 \rho _1}-\frac{\alpha _1 \left(\alpha
   _1+\alpha _3-2 \left(\lambda _1+4 \lambda _2+2 \lambda _3\right)\right)}{4 \rho _1} \right] \epsilon$ \\ \hline
  $H_3^0 H_3^0 H_3^0 H_3^0$ & $\frac14 \rho_1$ \\
  $H_1^0 H_1^0 H_1^0 H_1^0$, $A_1^0 A_1^0 A_1^0 A_1^0$ & $\frac14 \lambda_1$ & $-\lambda_4 \xi$ \\
  $H_1^0 H_1^0 A_1^0 A_1^0$ & $\frac12 \lambda_1$ & $-2\lambda_4 \xi$ \\
  $\{ H_1^0 H_1^0 ,\, A_1^0 A_1^0 \} H_3^0 H_3^0$ & $\frac14 (\alpha_1 + \alpha_3)$ & $-\alpha_2 \xi$ \\ \hline
  $H_1^0 H_1^0 H_1^0 H_3^0$ &  & $\frac{1}{2} \left[ \frac{4 \alpha _2 \left(\alpha _1+\alpha _3-2 \lambda _1\right)}{\alpha _3-4 \rho
   _1}+\frac{\alpha _1 \lambda _4}{\rho _1}\right] \epsilon$ \\
  $H_1^0 H_3^0 H_3^0 H_3^0$ &  & $\frac{1}{2} \alpha _2 \left[ \frac{\alpha _1}{\rho _1}-\frac{2 \left(2 \alpha _1+\alpha _3\right)}{\alpha
   _3-4 \rho _1}-2\right] \epsilon$ \\
  $H_1^0 H_3^0 A_1^0 A_1^0$ &  & $ \left[ \frac{2 \alpha _2 \left(\alpha _1+\alpha _3-2 \lambda _1\right)}{\alpha _3-4 \rho _1}+\frac{\alpha _1
   \lambda _4}{2\rho _1} \right] \epsilon$ \\
  \hline\hline
  $H^+_1 H^+_1 H^-_1 H^-_1$ & $\lambda_1$ & $-4\lambda_4 \xi$ \\ \hline
  $h h H^+_1 H^-_1$ & $\lambda_1$ \\ \hline
  $h H_1^0 H_1^+ H_1^-$ & $2\lambda_4$ & $-4 (2\lambda_2 + \lambda_3) \xi$ \\
  $h H_3^0 H_1^+ H_1^-$ & & $\frac{1}{2} \left[ \alpha _3 -\frac{16 \alpha _2 \lambda _4}{\alpha _3-4 \rho _1}-\frac{\alpha _1
   \left(\alpha _1+\alpha _3-2 \lambda _1\right)}{\rho _1} \right]\epsilon $ \\ \hline
  $\{ H_1^0 H_1^0 ,\, A_1^0 A_1^0 \} H_1^+ H_1^-$ & $\lambda_1$ & $-4 \lambda_4 \xi$ \\
  $H_3^0 H_3^0 H_1^+ H_1^-$ & $\frac{1}{2} ( \alpha_1 + \alpha_3 )$ & $-2 \alpha_2 \xi$ \\
  $H_1^0 H_3^0 H_1^+ H_1^-$ & & $\left[ \frac{4 \alpha _2 \left(\alpha _1+\alpha _3-2 \lambda _1\right)}{\alpha _3-4 \rho _1}+\frac{\alpha _1
   \lambda _4}{\rho _1} \right]\epsilon $ \\
  \hline\hline
  $H_2^{++} H_2^{++} H_2^{--} H_2^{--}$ & $\rho_1$ \\ \hline
  $hh H_{2}^{++} H_{2}^{--} $ & $\frac12 ( \alpha_1 + \alpha_3 )$ & $2 \alpha_2 \xi$ \\ \hline
  $h H_1^0 H_{2}^{++} H_{2}^{--} $ & $2\alpha_2$ & $- \alpha_3 \xi$ \\
  $h H_3^0 H_{2}^{++} H_{2}^{--} $ && $\frac{1}{2} \left[ \frac{\alpha _1 \left(\alpha _1+\alpha _3-2 \left(\rho _1+2 \rho
   _2\right)\right)}{\rho _1}-\frac{16 \alpha _2^2}{\alpha _3-4 \rho _1}\right] \epsilon  $  \\   \hline
  $\{ H_1^0 H_1^0 ,\, A_1^0 A_1^0 \} H_{2}^{++} H_{2}^{--} $ & $\frac12 \alpha_1$ & $-2 \alpha_2 \xi$ \\
  $H_3^0 H_3^0 H_{2}^{++} H_{2}^{--} $ & $\rho_1 + 2 \rho_2$  \\ \hline
  $H_1^0 H_3^0 H_{2}^{++} H_{2}^{--} $ && $\frac{\alpha _2 \left(\alpha _1 \left(\alpha _3-8 \rho _1\right)+8 \rho _1 \left(\rho _1+2 \rho
   _2\right)\right)\epsilon }{\rho _1 \left(\alpha _3-4 \rho _1\right)}$ \\ \hline
  $H_1^{+} H_1^{-} H_2^{++} H_2^{--}$ & $\alpha_1 $ & $-4 \alpha_2 \xi$ \\ \hline
  $h H_1^{-} H_1^{-} H_2^{++}$ && $- \frac{1}{2\sqrt2} \alpha_3 \epsilon$   \\
  \hline\hline
  \end{tabular}
  \end{center}
\end{table}

The quartic couplings can be obtained in a much similar way, by taking the fourth order derivatives of the potential (\ref{eqn:potential}). The couplings are written in the form of
\begin{eqnarray}
\mathcal{L} \ \supset \ - \lambda_{s_i s_j s_k s_l} s_i s_j s_k s_l \, .
\end{eqnarray}
For completeness we list all the non-vanishing quartic couplings up to order $\epsilon$ in Table~\ref{tab:quartic}. Among the large number of quartic couplings, the phenomenologically most interesting ones are the SM-like $\lambda_{hhhh}$ and the new coupling $\lambda_{H_1^0 hhh}$, with the latter relevant for the processes $pp \rightarrow H_1^0 \rightarrow hhh$ at the 100 TeV collier. The coupling $\lambda_{A_1^0 hhh}$ is only non-zero up to the order of $\xi \alpha$ and $\lambda_{H_3^0 hhh}$ at the order of $\epsilon$, and are thus less  interesting for the LR Higgs phenomenology. Note that the SM Higgs self coupling can be measured with 40\% accuracy at 100 TeV collider with 3 ab$^{-1}$ integrated luminosity~\cite{Barr:2014sga}, whereas future lepton colliders could improve the accuracy to about 10-20\%~\cite{Baur:2009uw}; for a review, see e.g. Ref.~\cite{Dittmaier:2012nh}. These precision measurements will provide another way to test the deviations from the SM value as predicted in the minimal LR model.

Couplings of the neutral and charged scalars to the SM up- and down-type quarks are summarized in Table~\ref{tab:fermion}, where $\widehat{Y}_{u,\,d}$ are diagonal Yukawa coupling matrix in the SM and $V_{L,\,R}$ are the left and right-handed quark mixing matrices.  These couplings are proportional to the quark masses or their linear combinations, with flavor mixings potentially involved. The SM Yukawa couplings are reduced by a universal factor of $\mathcal{O} (\epsilon^2)$, due to mixing of the SM Higgs to the heavy scalar $H_3^0$ at the order of $\epsilon$ [cf.~Eq.~(\ref{eqn:higgsmixing})]. As a result of the same scalar mixing, the couplings of scalar $H_3^0$ to the quarks are at the order of $\mathcal{O} (\epsilon)$, and hence we call it {\it hadrophobic}. The bi-doublet scalars can potentially couple to the fermions in such a manner that they lead to dangerous FCNC effects, and thus constrained to be heavier than 8-10 TeV~\cite{Zhang:2007da}.

The couplings to the charged leptons and neutrinos are somewhat different due to the fact that the neutrinos are allowed to obtain Majorana masses. 
In the type-I seesaw case, using the Casas-Ibarra parametrization~\cite{Casas:2001sr} for the Dirac mass matrix
\begin{eqnarray}
m_D \ = \ i \: M_N^{1/2} \: O \: m_\nu^{1/2} \, ,
\end{eqnarray}
where $O$ is an arbitrary (complex) orthogonal matrix, we extract the Yukawa couplings in the neutrino sector: $Y_{\nu N}=m_D/\kappa$.
All the Yukawa couplings in the leptonic sector are collected in Table~\ref{tab:fermion}, where $\widehat{Y}_e$ are the diagonal charged lepton Yukawa coupling matrix in the SM, $U_L$ the effective low energy mixing matrix for the left-handed neutrinos in the basis of diagonal and positive-definite charged lepton mass matrix we adopt, and $U_R$ the mixing matrix among the heavy RH neutrinos.

It is straightforward to obtain the couplings of the SM and heavy scalars to the SM gauge bosons $\gamma$, the $W$ and $Z$, as well as to the heavy gauge bosons $W_R$ and $Z_R$. These are collected in Tables~\ref{tab:gauge} and \ref{tab:gauge2}.

\begin{table}[t!]
  \centering
  \caption[]{Couplings of the scalars in the minimal LR model to fermions. $\widehat{Y}_{u,d,e}$ are the diagonal Yukawa coupling matrices in the SM, $Y_{\nu N} = m_D/\kappa$, and $f$ are the Yukawa coupling in Eq.~(\ref{eqn:Lyukawa}).}
  \label{tab:fermion}
  \begin{tabular}{ll}
  \hline\hline
  couplings & values \\ \hline
  $h  \bar{u} u$ & $\frac{1}{\sqrt2} \widehat{Y}_U \left(1-\frac{\alpha _1^2 \epsilon ^2}{8 \rho _1^2}\right)$ \\
  $H_1^0 \bar{u} u$ & $- \sqrt2 \xi \widehat{Y}_U (1+ i \alpha   ) + \frac{1}{\sqrt2} \left( V_L \widehat{Y}_D V_R^\dagger \right) \left(1+2 \xi ^2-\frac{8 \alpha _2^2 \epsilon ^2}{\left(\alpha _3-4 \rho _1\right){}^2}\right) $ \\
  $H_3^0 \bar{u} u$ & $\frac{1}{\sqrt2} \widehat{Y}_U \left(\frac{\alpha _1 \epsilon }{2 \rho _1}+\frac{4 \alpha _2 \xi  \epsilon }{\alpha _3-4 \rho _1}\right)
   - \frac{1}{\sqrt2} \left( V_L \widehat{Y}_D V_R^\dagger \right) \left(\frac{4 \alpha _2 \epsilon }{\alpha _3-4 \rho
   _1} +\frac{\alpha _1 \xi  \epsilon }{2 \rho _1}\right)
  $ \\
  $A_1^0\bar{u} u$ & $+ \sqrt2 i  \xi  \widehat{Y}_U(1+i\alpha ) -\frac{i}{\sqrt2} \left( V_L \widehat{Y}_D V_R^\dagger \right) \left(1+2 \xi ^2\right) $ \\  \hline
  $h \bar{d}d$ & $\frac{1}{\sqrt2} \widehat{Y}_D \left(1-\frac{\alpha _1^2 \epsilon ^2}{8 \rho _1^2}\right)$ \\
  $H_1^0 \bar{d}d$ &
  $- \sqrt2 \xi  \widehat{Y}_D (1-i\alpha)
   + \frac{1}{\sqrt2} \left( V_L^\dagger \widehat{Y}_U V_R \right) \left(1+2 \xi ^2-\frac{8 \alpha _2^2 \epsilon ^2}{\left(\alpha _3-4 \rho _1\right){}^2}\right)$ \\
  $H_3^0 \bar{d}d$  & $\frac{1}{\sqrt2} \widehat{Y}_D \left(\frac{\alpha _1 \epsilon }{2 \rho   _1} +\frac{4 \alpha _2 \xi  \epsilon }{\alpha _3-4 \rho _1}\right)
  -\frac{1}{\sqrt2} \left( V_L^\dagger \widehat{Y}_U V_R \right) \left(\frac{4 \alpha _2 \epsilon }{\alpha
   _3-4 \rho _1}+\frac{\alpha _1 \xi  \epsilon }{2 \rho _1}\right)$ \\
  $A_1^0 \bar{d}d$ & $-\sqrt2 i \xi \widehat{Y}_D (1-i\alpha )
  +\frac{i}{\sqrt2}  \left( V_L^\dagger \widehat{Y}_U V_R \right)\left(1+2 \xi ^2\right) $ \\ \hline
  $H_1^+ \bar{u}_L d_R$ & $\frac{1}{\sqrt2} \left( \widehat{Y}_U V_R \right) \left(1+2 \xi ^2-\frac{\epsilon ^2}{4}\right) -\sqrt2 \xi (1-i\alpha )  \left( V_L \widehat{Y}_D \right)$ \\
  $H_1^+ \bar{u}_R d_L$ & $- \frac{1}{\sqrt2} \left( V_R \widehat{Y}_D \right) \left(1+2 \xi ^2 - \frac14 \epsilon ^2\right)+ \sqrt2 \xi  \left( \widehat{Y}_U V_L \right)$ \\
  \hline\hline
  $h \bar{e}e$ & $\frac{1}{\sqrt2} \widehat{Y}_e \left(1-\frac{\alpha _1^2 \epsilon ^2}{8 \rho _1^2}\right)$ \\
  $H_1^0 \bar{e}e$ &
  $- \sqrt2 \xi  \widehat{Y}_e (1-i\alpha)
   + \frac{1}{\sqrt2} Y_{\nu N} \left(1+2 \xi ^2-\frac{8 \alpha _2^2 \epsilon ^2}{\left(\alpha _3-4 \rho _1\right){}^2}\right)$ \\
  $H_3^0 \bar{e}e$  & $\frac{1}{\sqrt2} \widehat{Y}_e \left(\frac{\alpha _1 \epsilon }{2 \rho   _1} +\frac{4 \alpha _2 \xi  \epsilon }{\alpha _3-4 \rho _1}\right)
  - \frac{1}{\sqrt2} Y_{\nu N} \left(\frac{4 \alpha _2 \epsilon }{\alpha
   _3-4 \rho _1}+\frac{\alpha _1 \xi  \epsilon }{2 \rho _1}\right)$ \\
  $A_1^0 \bar{e}e$ & $- \sqrt2 i \xi \widehat{Y}_e (1-i\alpha )
  + \frac{i}{\sqrt2} Y_{\nu N} \left(1+2 \xi ^2\right) $ \\ \hline
  $h N N$ & $\frac{1}{\sqrt2} f \left(-\frac{  \alpha _1\epsilon}{2 \rho _1}\right)$ \\
  $H_1^0 N N$ & $\frac{1}{\sqrt2} f \left( -\frac{4 \alpha _2\epsilon}{4 \rho _1-\alpha _3} \right)$ \\
  $H_3^0 N N$ & $\frac{1}{\sqrt2} f \left( 1 -\frac{  \alpha _1^2\epsilon^2}{8 \rho _1^2} -\frac{8 \alpha _2^2\epsilon^2}{(4 \rho _1-\alpha _3)^2} \right)$ \\ \hline
  $H_1^+ \bar{\nu} e_R$ & $ \frac{1}{\sqrt2} \left( U_L^T Y_{\nu N} \right) \left(1+2 \xi ^2-\frac{\epsilon ^2}{4}\right) -\sqrt2 \xi \left( U_L^T \widehat{Y}_e \right) (1-i\alpha )$ \\
  $H_1^+ \bar{N} e_L$ & $- \frac{1}{\sqrt2} \left( U_R^T \widehat{Y}_e \right) \left(1+2 \xi ^2 - \frac14 \epsilon ^2\right)
  +\sqrt2 \xi  \left( U_R^T Y_{\nu N} \right)$ \\
  $H_1^{+} N e_R$ & $- f \left( \frac{1}{\sqrt2} \epsilon \right)$ \\
  $H_2^{++} e_R e_R$ & $- \frac{1}{\sqrt2} f$ \\
  \hline\hline
  \end{tabular}
\end{table}

\begin{table}[t!]
  \centering
  \caption[]{Couplings of the scalars in the minimal LR model to gauge bosons up to the order of $\mathcal{O} (\epsilon^2 ,\, \sin\zeta_W,\, \sin\zeta_Z)$. At the tree level both the couplings of form $H_1^0 Z\gamma$ and $H_3^0 Z\gamma$ vanish, as expected from gauge invariance.}
  \label{tab:gauge}
  \small
  \begin{tabular}{ll}
  \hline\hline
  couplings & values $\times g_{\mu\nu}$ \\ \hline
  $h  W^+ W^{-}$ & $\frac{g_L^2 \kappa}{\sqrt2} \left[ 1-\frac{\alpha _1^2 \epsilon ^2}{8 \rho _1^2} \right]$ \\
  $H_1^0  W^+ W^{-}$ & $\frac{g_L^2 \kappa}{\sqrt2} \left[ -\frac{2 g_R}{g_L} \sin\zeta_W \right]$ \\
  $H_3^0  W^+ W^{-}$ & $\frac{g_L^2 \kappa}{\sqrt2} \left[ \frac{\alpha _1 \epsilon }{2 \rho _1}-\frac{4 \alpha _2 \xi  \epsilon }{\alpha _3-4 \rho _1} \right]$ \\
  $A_1^0  W^+ W^{-}$ & $\frac{g_L^2 \kappa}{\sqrt2} \left[ \mathcal{O} (  \xi^3 \sin\alpha )  \right]$ \\
  \hline
  $h  Z Z^{}$ & $\frac{g_L^2 \kappa}{2\sqrt2\cos^2\theta_w} \left[ 1-\frac{\alpha _1^2 \epsilon ^2}{8 \rho _1^2} -2 \sin\theta_w \cot\phi \sin\zeta_Z \right]$ \\
  $H_1^0 Z Z^{}$ & $\frac{g_L^2 \kappa}{2\sqrt2\cos^2\theta_w} \left[ \frac{64 \alpha _2 \sin ^2\theta_w }{(\alpha _3-4 \rho _1)\cos^22\phi} \sin^2\zeta_Z \right]$ \\
  $H_3^0 Z Z^{}$ & $\frac{g_L^2 \kappa}{2\sqrt2\cos^2\theta_w} \left[ \frac{\alpha _1 \epsilon }{2 \rho _1}-\frac{4 \alpha _2 \xi  \epsilon }{\alpha _3-4 \rho _1} \right]$ \\
  $A_1^0  Z Z^{}$ & $\frac{g_L^2 \kappa}{2\sqrt2\cos^2\theta_w} \left[ \mathcal{O} (  \xi^3 \sin\alpha )  \right]$ \\
  \hline
  $H_1^+   W^{-} Z$ & $\frac{g_L^2 \kappa}{\sqrt2} \left[ - \frac{\sin\theta_w}{\cos^2 \theta_w \sin \phi} \sin\zeta_W \right]$ \\
  $H_2^{++} W^{-} W^{-}$ & $-2 g_R^2 v_R \sin^2\zeta_W$ \\ \hline
  $H_1^+   W^{-} \gamma$ & $\frac{g_L g_R \kappa}{\sqrt2} \left[ -\sin\theta_w \sin\zeta_W \right]$ \\
  \hline\hline
  $h  W_{R }^+ W_R^{-}$ & $\frac{g_R^2 v_R \epsilon}{\sqrt2} \left[1 -\frac{\alpha _1}{\rho _1}-\frac{\alpha _1^2 \epsilon ^2}{8 \rho _1^2} \right]$ \\
  $H_1^0  W_{R }^+ W_R^{-}$ & $\frac{g_R^2 v_R }{\sqrt2} \left[\frac{8 \alpha _2 \epsilon }{\alpha _3-4 \rho _1} + \frac{2g_L \epsilon}{g_R} \sin\zeta_W \right]$ \\
  $H_3^0  W_{R }^+ W_R^{-}$ & $\frac{g_R^2 v_R }{\sqrt2} \left[ 2 -\frac{16 \alpha _2^2 \epsilon ^2}{\left(\alpha
   _3-4 \rho _1\right){}^2}-\frac{\alpha _1  \left(\alpha _1-2 \rho _1\right) \epsilon ^2}{4 \rho _1^2}  \right]$ \\
   $A_1^0  W_{R }^+ W_R^{-}$ & $\frac{g_R^2 v_R }{\sqrt2} \left[ \mathcal{O} (\xi^3 \epsilon \sin\alpha) \right]$ \\
  \hline
  $h Z_{R} Z_R^{\mu}$ & $\frac{\sqrt2 g_R^2 v_R }{\cos^2\phi} \left[ -\frac{\alpha_1\epsilon}{2\rho_1} + \frac14 \epsilon\cos^4\phi \right]$ \\
  $H_1^0  Z_{R } Z_R^{}$ & $\frac{\sqrt2 g_R^2 v_R }{\cos^2\phi} \left[ \frac{4 \alpha _2 \epsilon }{\alpha _3-4 \rho _1} \right]$ \\
  $H_3^0  Z_{R } Z_R^{}$ & $\frac{\sqrt2 g_R^2 v_R }{\cos^2\phi} \left[ 1+ \left( -\frac{\alpha _1^2}{8\rho _1^2}-\frac{8 \alpha _2^2}{\left(\alpha _3-4 \rho _1\right){}^2} + \frac{ \alpha _1 \cos ^4\phi}{8\rho _1} \right) \epsilon^2  \right]$ \\
  $A_1^0  Z_{R } Z_R^{}$ & $\frac{\sqrt2 g_R^2 v_R }{\cos^2\phi} \left[ \mathcal{O} ( \xi^3 \epsilon \sin\alpha ) \right]$ \\
  \hline
  $H_1^+  W_{R }^{-} Z_R^{}$ & $\frac{g_L^2 \kappa}{\sqrt2} \left[ - \frac{\tan\theta_w (1+\sin^2\phi)}{\sin^2\phi\cos\phi} + \frac{\sin\theta_w}{\cos^2\theta_w \sin\phi} \sin\zeta_Z \right]$ \\
  $H_2^{++} W^{-}_{R } W^{-}_R$ & $-2 g_R^2 v_R$ \\ \hline
  $H_1^+  W_{R }^{-} \gamma$ & $\frac{g_L^2 \kappa}{\sqrt2} \left[ \frac{\sin^2\theta_w}{\cos\theta_w \sin\phi} \left( \frac14 \epsilon^2 +2\xi^2 \right) \right]$ \\
  \hline\hline
  $h  W_{}^+ W_R^{-}$ & $\frac{ g_L g_R \kappa }{\sqrt2} \left[
   -2\xi (1+i\alpha) + \left( \frac{g_R}{g_L} (1-\frac{\alpha_1}{\rho_1}) - \frac{g_L}{g_R} \right) \sin\zeta_W \right]$ \\
  $H_1^0  W_{}^+ W_R^{-}$ & $\frac{ g_L g_R \kappa }{\sqrt2} \left[
   -\left( 1 -\frac{8 \alpha _2^2 \epsilon ^2}{\left(\alpha _3-4 \rho _1\right){}^2}-2 \xi ^2 \right) - \frac{g_R}{g_L}\frac{8 \alpha _2}{4 \rho _1-\alpha _3} \sin\zeta_W   \right]$ \\
  $H_3^0  W_{}^+ W_R^{-}$ & $\frac{ g_L g_R v_R}{\sqrt2} \left[  \frac{4 \alpha _2 \epsilon ^2}{\alpha _3-4 \rho _1}  - \frac{2g_R}{g_L} \sin\zeta_W \right]$ \\
  $A_1^0  W_{}^+ W_R^{-}$ & $\frac{ g_L g_R \kappa}{\sqrt2} \left[ -i \left(1-2 \xi ^2\right) \right]$ \\
  \hline
  $h  Z_{} Z_R^{}$ & $\frac{ g_L^2 \kappa }{\sqrt2} \left[
   - \frac{\sin\theta_w \cot\phi}{\cos^2\theta_w} \left( 1 - \frac{\alpha_1 \epsilon^2}{8\rho_1^2} \right)
   + \left(- \frac{1+\sin^2\theta_w}{\cos^2\theta_w} + \frac{\tan^2\theta_w}{\sin^2\phi} \right) \sin\zeta_Z \right]$ \\
  $H_1^0  Z_{} Z_R^{}$ & $\frac{ g_L^2 \kappa }{\sqrt2} \left[
  \frac{16 \alpha _2  }{\alpha
   _3-4 \rho _1} \frac{\tan ^2\theta _w }{\sin^2\phi\cos^2\phi} \sin \zeta _Z \right]$ \\
  $H_3^0  Z_{} Z_R^{}$ & $\frac{ g_L^2 v_R }{\sqrt2} \left[
  -\frac{\alpha _1 }{2 \rho _1} \frac{\sin\theta_w}{\cos^2\theta_w \sin\phi} \epsilon ^2 + \frac{4\tan^2\theta_w}{\sin^2\phi \cos^2\phi} \sin\zeta_Z \right]$ \\
  $A_1^0  Z_{} Z_R^{}$ & $\frac{ g_L^2 \kappa }{\sqrt2} \left[
  \mathcal{O} (\xi^3 \sin\alpha) \right]$ \\
  \hline
  $H_1^+  W_{\mu}^- Z_R^{}$ & $\frac{ g_L g_R \kappa }{\sqrt2} \left[ - \frac{\tan\theta_w ( 1+\sin^2\phi )}{\sin\phi\cos\phi} \sin\zeta_W \right]$ \\
  $H_1^+  W_{R }^- Z^{}$ & $\frac{ g_L g_R \kappa }{\sqrt2} \left[ -\frac{1}{\cos\theta_w} +\frac{ \cos \theta_w}{4} \left(8 \xi ^2+\epsilon ^2\right) - \frac{\tan\theta_w ( 1+\sin^2\phi )}{\sin\phi\cos\phi} \sin\zeta_Z \right]$ \\
   \hline
  $H_2^{++} W^{-}_{} W^{-}_R$ & $-2 g_R^2 v_R \sin\zeta_W$ \\
  \hline\hline
  \end{tabular}
\end{table}

\begin{table}[t!]
  \centering
  \caption[]{Couplings of the scalars in the minimal LR model to the gauge bosons up to the order of $\mathcal{O} (\epsilon^2 ,\, \sin\zeta_W,\, \sin\zeta_Z)$. $k_{1\mu}$ and $k_{2\mu}$ are the momenta for the first and second scalars in each of the couplings pointing inward to the vertex.}
  \label{tab:gauge2}
  \begin{tabular}{ll}
  \hline\hline
  couplings & values $\times (k_{1\mu} - k_{2\mu} )$ \\ \hline
  $H_1^+ h  W_{}^- $ & $- \frac12 g_R \sin\zeta_W$ \\
  \hline
  $H_1^+ H_1^0  W_{}^- $ & $\frac{1}{2} g_L \left[ 1 - \frac14\epsilon ^2 \left( 1 +\frac{32 \alpha _2^2}{\left(\alpha _3-4 \rho
   _1\right){}^2}\right)\right]$ \\


  $H_1^+ H_3^0  W_{}^- $ & $- g_L \left( \frac{2 \alpha _2 \epsilon }{\alpha _3-4 \rho
   _1} +\frac{\alpha _1 \xi  \epsilon }{4 \rho _1} \right)$ \\
  \hline
  $H_1^+ A_1^0  W_{}^- $ & $ - \frac{i}{2} g_L \left(1-\frac14 \epsilon ^2\right)$ \\

  \hline
  $H_2^{++} H_1^-  W_{}^- $ & $- \frac{1}{\sqrt2} g_R \epsilon \sin\zeta_W$ \\

  \hline\hline
  $\begin{array}{l}
  \{
  h h ,\,
  H_1^0 H_1^0 ,\,
  H_3^0 H_3^0 ,\,
  A_1^0 A_1^0
  \} Z 
  \end{array}$ & $0$ \\ \hline
  $\{ H_1^0 h ,\, H_3^0 h ,\, H_3^0 H_1^0 \} Z$ & $\mathcal{O} (\xi \epsilon \sin\alpha)$ \\
  $A_1^0 hZ$ & $\mathcal{O} (\xi \epsilon^2)$ \\
  $A_1^0 H_1^0 Z$ & $ \frac{ig_L}{\cos\theta_w} \left[ -\frac12 + \frac{4 \alpha_2^2 \epsilon^2}{(\alpha_3 - 4\rho_1)^2} + \frac{\sin\theta_w \cos\phi}{2\sin\phi}\sin\zeta_Z \right]$  \\
  $A_1^0 H_3^0 Z$ & $\frac{ i g_L }{\cos\theta_w} \left[ \frac{2\alpha _2\epsilon}{\alpha _3-4 \rho _1} + \frac{\alpha_1 \xi \epsilon}{4\rho_1} \right]$ \\ \hline
  $H_1^{+} H_1^{-} Z$ & $\frac{g_L }{\cos\theta_w} \left[ \frac12 \cos2\theta_w - \frac14 \epsilon^2  + \frac12 \sin\theta_w \cot\phi \sin\zeta_Z \right]$ \\
  \hline
  $H_2^{++} H_2^{--} Z$ & $   \frac{g_L} {\cos\theta_w} \left[ -2 \sin^2\theta_w + \sin\theta_w (\cot\phi-\tan\phi) \sin\zeta_Z \right]$ \\
  \hline
  $H_1^{+} H_1^{-} \gamma$ & $e$ \\
  $H_2^{++} H_2^{--} \gamma$ & $2e$ \\
  \hline\hline
  $H_1^+ h  W_{R}^- $ & $- \frac12 g_R \left[ 1 -2 \xi ^2 - \frac14 \left( 1 -\frac{2\alpha _1 }{\rho _1} +\frac{\alpha _1^2 }{2 \rho _1^2} \right) \epsilon^2 \right]$ \\
  \hline
  $H_1^+ H_1^0  W_{R}^- $ & $g_R \left( \xi -i \alpha  \xi +\frac{2 \alpha _2 \epsilon ^2}{\alpha _3-4 \rho _1}\right) - \frac12 g_L \sin\zeta_W$ \\


  $H_1^+ H_3^0  W_{R}^- $ & $\frac12 g_R \left[   \left(1-\frac{\alpha _1}{2 \rho _1}\right)\epsilon -\frac{4 \alpha _2 \xi \epsilon}{\alpha _3-4 \rho _1} \right]$ \\
  $H_1^+ A_1^0  W_{R}^- $ & $i g_R (\xi -i\alpha\xi) -\frac{i}{2} g_L \sin\zeta_W $ \\

  \hline
  $H_2^{++} H_1^-  W_{R}^- $ & $- \frac{1}{\sqrt2} g_R \epsilon $ \\

  \hline\hline
  $\begin{array}{l}
  \{ h h ,\,
  H_1^0 H_1^0 ,\,
  H_3^0 H_3^0 ,\,
  A_1^0 A_1^0
  \} Z_R 
  \end{array}$ & $0$ \\ \hline
  $\{ H_1^0 h ,\, H_3^0 h ,\, H_3^0 H_1^0 \} Z_R$ & $\mathcal{O} (\xi \epsilon \sin\alpha)$ \\
  $A_1^0 h Z_R$ & $\mathcal{O} (\xi \epsilon^2)$ \\
  $A_1^0 H_1^0 Z_R$ & $ \frac{ig_L}{\cos\theta_w} \left[
  \frac12 \sin\theta_w \cot\phi -\frac{4 \alpha _2^2 \epsilon ^2 \sin \theta_w \cot\phi}{\left(\alpha _3-4 \rho
   _1\right){}^2} +\frac12 \sin\zeta_Z \right]$  \\
  $A_1^0 H_3^0 Z_R$ & $\frac{ i g_L }{\cos\theta_w} \left[ - \frac{2\alpha _2\epsilon \sin\theta_w \cot\phi}{\alpha _3-4 \rho _1} - \frac{\alpha_1 \xi \epsilon \sin\theta_w \cot\phi}{4\rho_1} \right]$ \\ \hline
  $H_1^{+} H_1^{-} Z_R$ & $\frac{g_L }{\cos\theta_w} \left[ \frac12 \sin\theta_w \cot\phi - \frac{\sin\theta_w (1+\sin^2\phi)}{4\sin\phi\cos\phi} \epsilon^2  - \frac12 \cos2\theta_w \sin\zeta_Z \right]$ \\
  \hline
  $H_2^{++} H_2^{--} Z_R$ & $   \frac{g_L} {\cos\theta_w} \left[  \sin\theta_w (\cot\phi-\tan\phi) + 2 \sin^2\theta_w \sin\zeta_Z \right]$ \\
  \hline\hline
  \end{tabular}
\end{table}

\section{Dominant decay widths of the heavy Higgs bosons}
\label{app:B}

In this appendix we list all the potentially dominant decay channels and their widths for the bi-doublet and hadrophobic heavy Higgs states in the minimal LR model. For the neutral bi-doublet scalars, the decay widths of dominant decay channels are, respectively,
\begin{align}
& \Gamma( H_1^0 / A_1^0 \to b\bar{b} )  \ \simeq \ \frac{3 \alpha_3^{1/2} y_t^2 v_R}{16\pi} \beta_1^3 (m_b,\, M_{H_1^0}) \,, \\
& \quad \Gamma( H_1^0 \to H_3^0 h )  \ \simeq \ \frac{\alpha_2^2 \alpha_3^{-1/2} v_R}{2\pi}
\beta_2 (M_h,\, M_{H_3^0},\, M_{H_1^0}) \Theta (M_{H_1^0} - M_h - M_{H_3^0} ) \,, \\
& \Gamma( H_1^0 / A_1^0 \to W^\pm W_R^{\mp} )  \ \simeq \ \frac{\alpha_3^{3/2} v_R}{32\pi}
\beta_2 (M_W,\, M_{W_R},\, M_{H_1^0}) \Theta (M_{H_1^0} - M_W - M_{W_R} ) \nonumber \\
& \qquad \qquad \qquad \times \left[ \beta_2^4 (M_W,\, M_{W_R},\, M_{H_1^0}) f_2^2 (M_W,\, M_{W_R},\, M_{H_1^0}) + \frac{8 M_W^2 M_{W_R}^2}{M_{H_1^0}^4} \right] \,,
\end{align}
where $\Theta(x)$ is the Heavyside function and the prefactors are explicitly written as functions of the RH scale $v_R$, the quartic couplings and Yukawa couplings, up to the LO of the heavy vector and scalar boson masses as given in Eqs.~(\ref{eqn:gbosonmass}) and \ref{eqn:gbosonmassZ}, (\ref{eqn:scalarmass}), (\ref{eqn:scalarmass2}) and (\ref{eqn:scalarmass3}) and the relevant couplings as given in Tables~\ref{tab:triple} to \ref{tab:gauge2}. The velocities of decay products are defined as
\begin{eqnarray}
\beta_1 (m,\, M) & \ \equiv \ & \left[ 1 - \frac{4m^2}{M^2} \right]^{1/2} \,, \\
\label{eqn:beta2}
\beta_2 (m_1,\, m_2,\, M) & \ \equiv \ & \left[ 1 - \frac{2(m_1^2 + m_2^2)}{M^2} + \frac{(m_1^2 - m_2^2)^2}{M^4} \right]^{1/2} \, ,
\end{eqnarray}
and the dependent function
\begin{eqnarray}
\label{eqn:f2}
f_2 (m_1,\, m_2,\, M) \ \equiv \ \frac12 \left[ 1 + \sqrt{\left( 1 + \frac{4m_1^2}{M^2 \beta_2^2 (m_1,\, m_2,\, M)} \right) \left( 1 + \frac{4m_2^2}{M^2 \beta_2^2 (m_1,\, m_2,\, M)} \right) } \right] \,. \nonumber \\
\end{eqnarray}
In the limit of $m_{1,\,2} \ll M$, the function $f_2 \to 1$. The three-body decay width of $H_1^0$ into the SM Higgs given by
\begin{eqnarray}
\Gamma (H_1^0 \to hhh) \ \simeq \ \frac{3 \lambda_4^2 M_{H_1^0}}{256 \pi^3}
\end{eqnarray}
is generally much smaller than the two-body channels for large $v_R\gg \kappa$.

For the singly-charged heavy scalars, the dominant decay widths are given by
\begin{align}
& \Gamma( H_1^\pm \to t\bar{b} (\bar{t}b) ) \ \simeq \ \frac{3 \alpha_3^{1/2} y_t^2 v_R}{32\pi}
\beta_2^3 (m_b,\, m_t,\, M_{H_1^\pm}) f_2 (m_b,\, m_t,\, M_{H_1^\pm}) \,, \\
& \Gamma( H_1^\pm \to Z W_R^{\pm} )  \ \simeq \ \frac{\alpha_3^{3/2} v_R}{64\pi}
\beta_2 (M_Z,\, M_{W_R},\, M_{H_1^\pm}) \Theta (M_{H_1^\pm} - M_Z - M_{W_R} ) \nonumber \\
& \qquad \qquad \qquad \qquad \times \left[ \beta_2^4 (M_Z,\, M_{W_R},\, M_{H_1^\pm}) f_2^2 (M_Z,\, M_{W_R},\, M_{H_1^\pm}) + \frac{8 M_Z^2 M_{W_R}^2}{M_{H_1^\pm}^4} \right] \,, \\
& \Gamma( H_1^\pm \to h W_R^{\pm} )  \ \simeq \ \frac{\alpha_3^{3/2} v_R}{64\pi}
\beta_2 (M_h,\, M_{W_R},\, M_{H_1^\pm}) \Theta (M_{H_1^\pm} - M_h - M_{W_R} ) \nonumber \\
&\qquad \qquad \qquad \qquad \qquad \times \left[ \beta_2^4 (M_h,\, M_{W_R},\, M_{H_1^\pm}) - \frac{16 M_h^2 M_{W_R}^2}{M_{H_1^\pm}^4} \right] \,.
\end{align}

For the hadrophobic scalar $H_3^0$, we list all the potential dominant channels below:
\begin{align}
& \Gamma( H_3^0 \to hh ) \ \simeq \  \frac{\alpha_1^{2} \rho_1^{-1/2} v_R}{32\pi}
\beta_1 (M_h,\, M_{H_3^0}) \,, \\
& \Gamma( H_3^0 \to H_1^0 h )  \ \simeq \ \frac{\alpha_2^{2} \rho_1^{-1/2} v_R}{4\pi}
\beta_2 (M_h,\, M_{H_1^\pm},\, M_{H_3^0}) \Theta (M_{H_3^0} - M_h - M_{H_1^{0}}) \,, \\
& \Gamma( H_3^0 \to H_1^0 H_1^0 / A_1^0 A_1^0 / H_1^+ H_1^- )  \ \simeq \ \frac{ ( \alpha_1 + \alpha_3 )^{2} \rho_1^{-1/2} \delta_H v_R}{32\pi}
\beta_1 (M_{H_1^0},\, M_{H_3^0}) \Theta (M_{H_3^0} - 2 M_{H_1^{0}}) \,, \\
& \Gamma( H_3^0 \to H_2^{++} H_2^{--} ) \ \simeq \ \frac{\rho_1^{-1/2} (\rho_1 + 2\rho_2)^2 v_R}{4\pi} \beta_1 (M_{H_2^{\pm\pm}},\, M_{H_3^0}) \Theta (M_{H_3^0} - 2 M_{H_2^{\pm\pm}}) \,, \\
& \Gamma( H_3^0 \to V_R V_R ) \ \simeq \ \frac{\rho_1^{3/2} \delta_V v_R}{8\pi} \beta_1 (M_{V_R},\, M_{H_3^0}) \left( 1 - \frac{4 M_{V_R}^2}{M_{H_3^{0}}^{2}} +  \frac{12 M_{V_R}^4}{M_{H_3^{0}}^{4}} \right) \Theta (M_{H_3^0} - 2 M_{V_R}) \,,
\label{eqn:H3toNN} \\
& \Gamma( H_3^{0} \to N_i N_i ) \ \simeq \ \frac{3 \rho_1^{1/2} f^2 v_R}{8\pi} \beta_1^3 (M_{N},\, M_{H_3^{0}}) f_2 (M_N,\, M_N,\, M_{H_3^0}) \Theta (M_{H_3^{0}} - 2 M_{N}) \,,
\end{align}
where the factor $\delta_H = 1$ for $H_1^0$ and $A_1^0$, and 2 for the charged scalar $H_1^\pm$, $V_R=W_R, \, Z_R$ with the factor $\delta_V = 2$ for $W_R$ and 1 for $Z_R$. It is obvious that the bi-doublet decay channels of $H_3^0$ are  universally determined by the quartic coupling combination $(\alpha_1 + \alpha_3)$. For the RH neutrino channel, we assume for simplicity the three neutrino states $N_i$ have a degenerate mass $M_N$, and $f$ is the Yukawa coupling in Eq.~(\ref{eqn:Lyukawa}).

For the doubly-charged scalars, we have the following two dominant partial decay widths:
\begin{align}
& \Gamma( H_2^{\pm\pm} \to \ell^{\pm} \ell^{\pm} ) \ \simeq \ \frac{3\rho_2^{1/2} v_R}{8\pi} \\
& \Gamma( H_2^{\pm\pm} \to W_R^{\pm} W_R^{\pm} ) \ \simeq \ \frac{\rho_2^{3/2} v_R}{\pi} \beta_1 (M_{W_R},\, M_{H_2^{\pm\pm}}) \left( 1 - \frac{4 M_{W_R}^2}{M_{H_2^{\pm\pm}}^{2}} +  \frac{12 M_{W_R}^4}{M_{H_2^{\pm\pm}}^{4}} \right) \Theta (M_{H_2^{\pm\pm}} - 2 M_{W_R}) \,. \nonumber \\
\end{align}



\begin{thebibliography}{99}


\bibitem{Aad:2012tfa}
  G.~Aad {\it et al.} [ATLAS Collaboration],
  Phys.\ Lett.\ B {\bf 716}, 1 (2012)
  [arXiv:1207.7214 [hep-ex]].

\bibitem{Chatrchyan:2012xdj}
  S.~Chatrchyan {\it et al.} [CMS Collaboration],
  Phys.\ Lett.\ B {\bf 716}, 30 (2012)
  [arXiv:1207.7235 [hep-ex]].



\bibitem{fcc-hh} \url{https://fcc.web.cern.ch/Pages/default.aspx}

\bibitem{Tang:2015qga}
  J.~Tang {\it et al.},
  arXiv:1507.03224 [physics.acc-ph].

\bibitem{Arkani-Hamed:2015vfh}  N.~Arkani-Hamed, T.~Han, M.~Mangano and L.~T.~Wang,
  arXiv:1511.06495 [hep-ph].


\bibitem{Baglio:2015wcg}
 J.~Baglio, A.~Djouadi and J.~Quevillon,
  arXiv:1511.07853 [hep-ph].


\bibitem{type1a} P. Minkowski, Phys. Lett. B {\bf 67}, 421 (1977).

\bibitem{type1b} R. N. Mohapatra and G. Senjanovi\'{c}, Phys. Rev. Lett. {\bf 44}, 912 (1980).

\bibitem{type1c} T. Yanagida, Conf. Proc. C {\bf 7902131}, 95 (1979).

\bibitem{type1d} M. Gell-Mann, P. Ramond and R. Slansky, Conf. Proc.
{\bf C790927}, 315 (1979)  [arXiv:1306.4669 [hep-th]].

\bibitem {type1e}
S.~L.~Glashow,
  NATO Sci.\ Ser.\ B {\bf 61}, 687 (1980).

\bibitem{Dev:2013oxa}
  C.~H.~Lee, P.~S.~B.~Dev and R.~N.~Mohapatra,
  Phys.\ Rev.\ D {\bf 88}, no. 9, 093010 (2013)
  [arXiv:1309.0774 [hep-ph]].

\bibitem{Drewes:2013gca}
  M.~Drewes,
  Int.\ J.\ Mod.\ Phys.\ E {\bf 22}, 1330019 (2013)
  [arXiv:1303.6912 [hep-ph]].

\bibitem{Deppisch:2015qwa}
  F.~F.~Deppisch, P.~S.~B.~Dev and A.~Pilaftsis,
  New J.\ Phys.\  {\bf 17}, no. 7, 075019 (2015)
  [arXiv:1502.06541 [hep-ph]].



\bibitem{LR1} J. C. Pati and A. Salam, Phys. Rev. D {\bf 10}, 275 (1974).

\bibitem{LR2} R. N. Mohapatra and J. C. Pati, Phys. Rev. D {\bf 11} 2558 (1975).

\bibitem{LR3} G. Senjanovi\'{c} and R. N. Mohapatra, Phys. Rev. D {\bf 12} 1502 (1975).






\bibitem{Gunion:1986im} J.~F.~Gunion, B.~Kayser, R.~N.~Mohapatra, N.~G.~Deshpande, J.~Grifols, A.~Mendez, F.~I.~Olness and P.~B.~Pal,
  PRINT-86-1324 (UC,DAVIS).

\bibitem{Gunion:1989in} J.~F.~Gunion, J.~Grifols, A.~Mendez, B.~Kayser and F.~I.~Olness,
  Phys.\ Rev.\ D {\bf 40}, 1546 (1989).

\bibitem{Deshpande:1990ip}
  N.~G.~Deshpande, J.~F.~Gunion, B.~Kayser and F.~I.~Olness,
  Phys.\ Rev.\ D {\bf 44}, 837 (1991).

\bibitem{Barenboim:2001vu}
  G.~Barenboim, M.~Gorbahn, U.~Nierste and M.~Raidal,
  Phys.\ Rev.\ D {\bf 65}, 095003 (2002)
  [hep-ph/0107121].

\bibitem{Polak:1991vf}
  J.~Polak and M.~Zralek,
  Phys.\ Lett.\ B {\bf 276}, 492 (1992).

\bibitem{Azuelos:2004mwa}
  G.~Azuelos, K.~Benslama and J.~Ferland,
  J.\ Phys.\ G {\bf 32}, no. 2, 73 (2006)
  [hep-ph/0503096].

\bibitem{Jung:2008pz}
  D.~W.~Jung and K.~Y.~Lee,
  Phys.\ Rev.\ D {\bf 78}, 015022 (2008)
  [arXiv:0802.1572 [hep-ph]].

\bibitem{Bambhaniya:2013wza}
  G.~Bambhaniya, J.~Chakrabortty, J.~Gluza, M.~Kordiaczynska and R.~Szafron,
  JHEP {\bf 1405}, 033 (2014)
  [arXiv:1311.4144 [hep-ph]].



\bibitem{Dutta:2014dba}
  B.~Dutta, R.~Eusebi, Y.~Gao, T.~Ghosh and T.~Kamon,
  Phys.\ Rev.\ D {\bf 90}, 055015 (2014)
  [arXiv:1404.0685 [hep-ph]].

\bibitem{Bambhaniya:2014cia}
  G.~Bambhaniya, J.~Chakrabortty, J.~Gluza, T.~Jeli\'{n}ski and M.~Kordiaczynska,
  Phys.\ Rev.\ D {\bf 90}, no. 9, 095003 (2014)
 [arXiv:1408.0774 [hep-ph]].


 \bibitem{Maiezza:2015lza} A.~Maiezza, M.~Nemevsek and F.~Nesti,
  Phys.\ Rev.\ Lett.\  {\bf 115}, 081802 (2015)
  [arXiv:1503.06834 [hep-ph]].

\bibitem{Bambhaniya:2015wna} G.~Bambhaniya, J.~Chakrabortty, J.~Gluza, T.~Jelinski and R.~Szafron,
  Phys.\ Rev.\ D {\bf 92}, no. 1, 015016 (2015)
  [arXiv:1504.03999 [hep-ph]].





\bibitem{Beall:1981ze} G.~Beall, M.~Bander and A.~Soni,
  Phys.\ Rev.\ Lett.\  {\bf 48}, 848 (1982).

\bibitem{Branco:1982wp}
  G.~C.~Branco, J.~M.~Frere and J.~M.~Gerard,
  Nucl.\ Phys.\ B {\bf 221}, 317 (1983).

\bibitem{Ecker:1983uh}
  G.~Ecker, W.~Grimus and H.~Neufeld,
  Phys.\ Lett.\ B {\bf 127}, 365 (1983)
  [Phys.\ Lett.\ B {\bf 132}, 467 (1983)].


\bibitem{Bigi:1983bpa}
  I.~I.~Y.~Bigi and J.~M.~Frere,
  Phys.\ Lett.\ B {\bf 129}, 469 (1983)
  [Phys.\ Lett.\ B {\bf 154}, 457 (1985)].

\bibitem{Babu:1993hx}
  K.~S.~Babu, K.~Fujikawa and A.~Yamada,
  Phys.\ Lett.\ B {\bf 333}, 196 (1994)
  [hep-ph/9312315].

\bibitem{Ball:1999mb}
  P.~Ball, J.~M.~Frere and J.~Matias,
  Nucl.\ Phys.\ B {\bf 572}, 3 (2000)
  [hep-ph/9910211].

\bibitem{Zhang:2007da}
 H. An, X. Ji, R. N. Mohapatra, and Y. Zhang, 
Nucl. Physc. {\bf B 802}, 247 (2008) [arXiv:0712.4218 [hep-ph]].


\bibitem{Maiezza:2010ic}
A.~Maiezza, M.~Nemevsek, F.~Nesti and G.~Senjanovic,
  Phys.\ Rev.\ D {\bf 82}, 055022 (2010)
  [arXiv:1005.5160 [hep-ph]].

\bibitem{Blanke:2011ry}
  M.~Blanke, A.~J.~Buras, K.~Gemmler and T.~Heidsieck,
  JHEP {\bf 1203}, 024 (2012)
  [arXiv:1111.5014 [hep-ph]].

\bibitem{Bertolini:2014sua}
  S.~Bertolini, A.~Maiezza and F.~Nesti,
  Phys.\ Rev.\ D {\bf 89}, no. 9, 095028 (2014)
  [arXiv:1403.7112 [hep-ph]].

\bibitem{Bernard:2015boz}
  V.~Bernard, S.~Descotes-Genon and L.~V.~Silva,
  arXiv:1512.00543 [hep-ph].









\bibitem{Mohapatra:2013cia}
  R.~N.~Mohapatra and Y.~Zhang,
  Phys.\ Rev.\ D {\bf 89}, no. 5, 055001 (2014)
  [arXiv:1401.0018 [hep-ph]].

\bibitem{Keung:1983uu}
  W.~Y.~Keung and G.~Senjanovic,
  Phys.\ Rev.\ Lett.\  {\bf 50}, 1427 (1983).

\bibitem{Ferrari:2000sp}
  A.~Ferrari, J.~Collot, M.~L.~Andrieux, B.~Belhorma, P.~de Saintignon, J.~Y.~Hostachy, P.~Martin and M.~Wielers,
  Phys.\ Rev.\ D {\bf 62}, 013001 (2000).

\bibitem{Schmaltz:2010xr}
  M.~Schmaltz and C.~Spethmann,
  JHEP {\bf 1107}, 046 (2011)
  [arXiv:1011.5918 [hep-ph]].

\bibitem{Nemevsek:2011hz}
  M.~Nemevsek, F.~Nesti, G.~Senjanovic and Y.~Zhang,
  Phys.\ Rev.\ D {\bf 83}, 115014 (2011)
  [arXiv:1103.1627 [hep-ph]].

\bibitem{Chen:2011hc}
  C.~Y.~Chen and P.~S.~B.~Dev,
  Phys.\ Rev.\ D {\bf 85}, 093018 (2012)
  [arXiv:1112.6419 [hep-ph]].


\bibitem{Chakrabortty:2012pp}
  J.~Chakrabortty, J.~Gluza, R.~Sevillano and R.~Szafron,
  JHEP {\bf 1207}, 038 (2012)
  [arXiv:1204.0736 [hep-ph]].

\bibitem{Das:2012ii}
  S.~P.~Das, F.~F.~Deppisch, O.~Kittel and J.~W.~F.~Valle,
  Phys.\ Rev.\ D {\bf 86}, 055006 (2012)
  [arXiv:1206.0256 [hep-ph]].

\bibitem{AguilarSaavedra:2012gf}
  J.~A.~Aguilar-Saavedra and F.~R.~Joaquim,
  Phys.\ Rev.\ D {\bf 86}, 073005 (2012)
  [arXiv:1207.4193 [hep-ph]].

\bibitem{Han:2012vk}
  T.~Han, I.~Lewis, R.~Ruiz and Z.~g.~Si,
  Phys.\ Rev.\ D {\bf 87}, no. 3, 035011 (2013)
  [Phys.\ Rev.\ D {\bf 87}, no. 3, 039906 (2013)]
  [arXiv:1211.6447 [hep-ph]].

\bibitem{Chen:2013fna}
  C.~Y.~Chen, P.~S.~B.~Dev and R.~N.~Mohapatra,
  Phys.\ Rev.\ D {\bf 88}, 033014 (2013)
  [arXiv:1306.2342 [hep-ph]].

\bibitem{Rizzo:2014xma}
  T.~G.~Rizzo,
  Phys.\ Rev.\ D {\bf 89}, no. 9, 095022 (2014)
  [arXiv:1403.5465 [hep-ph]].



\bibitem{Gluza:2015goa}
  J.~Gluza and T.~Jelinski,
  Phys.\ Lett.\ B {\bf 748}, 125 (2015)
  [arXiv:1504.05568 [hep-ph]].

\bibitem{Ng:2015hba}
  J.~N.~Ng, A.~de la Puente and B.~W.~P.~Pan,
  JHEP {\bf 1512}, 172 (2015)
  [arXiv:1505.01934 [hep-ph]].

\bibitem{Dev:2015kca}
  P.~S.~B.~Dev, D.~Kim and R.~N.~Mohapatra,
  JHEP {\bf 1601}, 118 (2016)
  [arXiv:1510.04328 [hep-ph]].

\bibitem{Khachatryan:2014dka}
  V.~Khachatryan {\it et al.} [CMS Collaboration],
  Eur.\ Phys.\ J.\ C {\bf 74}, no. 11, 3149 (2014)
  [arXiv:1407.3683 [hep-ex]].

\bibitem{Aad:2015xaa}
  G.~Aad {\it et al.} [ATLAS Collaboration],
  JHEP {\bf 1507}, 162 (2015)
  [arXiv:1506.06020 [hep-ex]].

\bibitem{Aad:2014vgg}
  G.~Aad {\it et al.} [ATLAS Collaboration],
  JHEP {\bf 1411}, 056 (2014)
  [arXiv:1409.6064 [hep-ex]].

\bibitem{Aad:2015kna}
  G.~Aad {\it et al.} [ATLAS Collaboration],
  Eur.\ Phys.\ J.\ C {\bf 76}, no. 1, 45 (2016)
  [arXiv:1507.05930 [hep-ex]].

\bibitem{Aad:2015agg}
  G.~Aad {\it et al.} [ATLAS Collaboration],
  JHEP {\bf 1601}, 032 (2016)
  [arXiv:1509.00389 [hep-ex]].

\bibitem{neutral1}
  The ATLAS Collaboration,
  ATLAS-CONF-2015-061.

\bibitem{Khachatryan:2014wca}
  V.~Khachatryan {\it et al.} [CMS Collaboration],
  JHEP {\bf 1410}, 160 (2014)
  [arXiv:1408.3316 [hep-ex]].


\bibitem{Khachatryan:2015tha}
  V.~Khachatryan {\it et al.} [CMS Collaboration],
  arXiv:1510.01181 [hep-ex].

\bibitem{Aad:2014kga}
  G.~Aad {\it et al.} [ATLAS Collaboration],
  JHEP {\bf 1503}, 088 (2015)
  [arXiv:1412.6663 [hep-ex]].

\bibitem{Aad:2015nfa}
  G.~Aad {\it et al.} [ATLAS Collaboration],
  Phys.\ Rev.\ Lett.\  {\bf 114}, no. 23, 231801 (2015)
  [arXiv:1503.04233 [hep-ex]].



\bibitem{Aad:2015typ}
  G.~Aad {\it et al.} [ATLAS Collaboration],
  arXiv:1512.03704 [hep-ex].


\bibitem{CMS:2014cdp}
  The CMS Collaboration,
  CMS-PAS-HIG-14-020.

\bibitem{Khachatryan:2015qxa}
  V.~Khachatryan {\it et al.} [CMS Collaboration],
  JHEP {\bf 1511}, 018 (2015)
  [arXiv:1508.07774 [hep-ex]].

\bibitem{ATLAS:2014kca}
  G.~Aad {\it et al.} [ATLAS Collaboration],
  JHEP {\bf 1503}, 041 (2015)
  [arXiv:1412.0237 [hep-ex]].

\bibitem{CMS:2016cpz}
  The CMS Collaboration,
  CMS-PAS-HIG-14-039.

\bibitem{Riazuddin:1981hz}
  Riazuddin, R.~E.~Marshak and R.~N.~Mohapatra,
  Phys.\ Rev.\ D {\bf 24}, 1310 (1981).

\bibitem{Pal:1983bf}
  P.~B.~Pal,
  Nucl.\ Phys.\ B {\bf 227}, 237 (1983).

\bibitem{Mohapatra:1992uu}
  R.~N.~Mohapatra,
  Phys.\ Rev.\ D {\bf 46}, 2990 (1992).

\bibitem{Cirigliano:2004mv}
V.~Cirigliano, A.~Kurylov, M.~J.~Ramsey-Musolf and P.~Vogel,
  Phys.\ Rev.\ D {\bf 70}, 075007 (2004)
  [hep-ph/0404233].

\bibitem{Cirigliano:2004tc}
  V.~Cirigliano, A.~Kurylov, M.~J.~Ramsey-Musolf and P.~Vogel,
  Phys.\ Rev.\ Lett.\  {\bf 93}, 231802 (2004)
  [hep-ph/0406199].

\bibitem{Bajc:2009ft}
  B.~Bajc, M.~Nemevsek and G.~Senjanovic,
  Phys.\ Lett.\ B {\bf 684}, 231 (2010)
  [arXiv:0911.1323 [hep-ph]].

\bibitem{Tello:2010am}
V.~Tello, M.~Nemevsek, F.~Nesti, G.~Senjanovic and F.~Vissani,
  Phys.\ Rev.\ Lett.\  {\bf 106}, 151801 (2011)
  [arXiv:1011.3522 [hep-ph]].

\bibitem{Barry:2013xxa}
  J.~Barry and W.~Rodejohann,
  JHEP {\bf 1309}, 153 (2013)
  [arXiv:1303.6324 [hep-ph]].


\bibitem{Vasquez:2015una}
  J.~C.~Vasquez,
  JHEP {\bf 1509}, 131 (2015)
  [arXiv:1504.05220 [hep-ph]].

\bibitem{Awasthi:2015ota}
  R.~L.~Awasthi, P.~S.~B.~Dev and M.~Mitra,
  Phys.\ Rev.\ D {\bf 93}, no. 1, 011701 (2016)
  [arXiv:1509.05387 [hep-ph]].

\bibitem{Bambhaniya:2015ipg}
  G.~Bambhaniya, P.~S.~B.~Dev, S.~Goswami and M.~Mitra,
  arXiv:1512.00440 [hep-ph].

\bibitem{Ecker:1983dj}
  G.~Ecker, W.~Grimus and H.~Neufeld,
  Nucl.\ Phys.\ B {\bf 229}, 421 (1983).

\bibitem{Frere:1991jt}
  J.~M.~Frere, J.~Galand, A.~Le Yaouanc, L.~Oliver, O.~Pene and J.~C.~Raynal,
  Phys.\ Rev.\ D {\bf 45}, 259 (1992).

\bibitem{Maiezza:2014ala}
  A.~Maiezza and M.~Nemevsek,
  Phys.\ Rev.\ D {\bf 90}, no. 9, 095002 (2014)
  [arXiv:1407.3678 [hep-ph]].

\bibitem{Nieves:1986uk}
  J.~F.~Nieves, D.~Chang and P.~B.~Pal,
  Phys.\ Rev.\ D {\bf 33}, 3324 (1986).



\bibitem{Nemevsek:2012iq}
  M.~Nemevsek, G.~Senjanovic and V.~Tello,
  Phys.\ Rev.\ Lett.\  {\bf 110}, no. 15, 151802 (2013)
  [arXiv:1211.2837 [hep-ph]].

\bibitem{Mohapatra:1980yp} R.~N.~Mohapatra and G.~Senjanovic,
  Phys.\ Rev.\ D {\bf 23}, 165 (1981).

\bibitem{Mohapatra:1981pm}
  R.~N.~Mohapatra and J.~D.~Vergados,
  Phys.\ Rev.\ Lett.\  {\bf 47}, 1713 (1981).

\bibitem{Picciotto:1982qe}
  C.~E.~Picciotto and M.~S.~Zahir,
  Phys.\ Rev.\ D {\bf 26}, 2320 (1982).


\bibitem{Hirsch:1996qw}
  M.~Hirsch, H.~V.~Klapdor-Kleingrothaus and O.~Panella,
  Phys.\ Lett.\ B {\bf 374}, 7 (1996)
  [hep-ph/9602306].

\bibitem{Arnold:2010tu}
  R.~Arnold {\it et al.} [SuperNEMO Collaboration],
  Eur.\ Phys.\ J.\ C {\bf 70}, 927 (2010)
  [arXiv:1005.1241 [hep-ex]].


\bibitem{Chakrabortty:2012mh} J.~Chakrabortty, H.~Z.~Devi, S.~Goswami and S.~Patra,
  JHEP {\bf 1208}, 008 (2012)
  [arXiv:1204.2527 [hep-ph]].




\bibitem{Dev:2013vxa}
  P.~S.~B.~Dev, S.~Goswami, M.~Mitra and W.~Rodejohann,
  Phys.\ Rev.\ D {\bf 88}, 091301 (2013)
  [arXiv:1305.0056 [hep-ph]].



\bibitem{Huang:2013kma}
  W.~C.~Huang and J.~Lopez-Pavon,
  Eur.\ Phys.\ J.\ C {\bf 74}, 2853 (2014)
  [arXiv:1310.0265 [hep-ph]].

\bibitem{Dev:2014xea}
  P.~S.~B.~Dev, S.~Goswami and M.~Mitra,
  Phys.\ Rev.\ D {\bf 91}, no. 11, 113004 (2015)
  [arXiv:1405.1399 [hep-ph]].

\bibitem{Mahajan:2014nca}
  N.~Mahajan,
  Phys.\ Rev.\ D {\bf 90}, no. 3, 035015 (2014)
  [arXiv:1406.2606 [hep-ph]].

\bibitem{Ge:2015yqa}
  S.~F.~Ge, M.~Lindner and S.~Patra,
  JHEP {\bf 1510}, 077 (2015)
  [arXiv:1508.07286 [hep-ph]].

\bibitem{Borah:2015ufa}
  D.~Borah and A.~Dasgupta,
  JHEP {\bf 1511}, 208 (2015)
  [arXiv:1509.01800 [hep-ph]].

\bibitem{Marshak:1979fm}
  R.~E.~Marshak and R.~N.~Mohapatra,
  Phys.\ Lett.\ B {\bf 91}, 222 (1980).

\bibitem{Davidson:1978pm}
  A.~Davidson,
  Phys.\ Rev.\ D {\bf 20}, 776 (1979).

\bibitem{CMP}  D.~Chang, R.~N.~Mohapatra and M.~K.~Parida,
  Phys.\ Rev.\ Lett.\  {\bf 52}, 1072 (1984).

\bibitem{type2a} M. Magg and C. Wetterich, Phys. Lett. {\bf B 94}, 61 (1980).

\bibitem{type2b}  J.~Schechter and J.~W.~F.~Valle,
  Phys.\ Rev.\ D {\bf 22}, 2227 (1980).

\bibitem{type2c} T. P. Cheng and L.-F. Li, Phys. Rev. {\bf D 22}, 2860 (1980).

\bibitem{type2d} G. Lazarides, Q. Shafi and C. Wetterich, Nucl. Phys. {\bf B 181}, 287 (1981).






\bibitem{Georgi:1977wk}
  H.~Georgi and S.~Weinberg,
  Phys.\ Rev.\ D {\bf 17}, 275 (1978).

\bibitem{Brehmer:2015cia}
  J.~Brehmer, J.~Hewett, J.~Kopp, T.~Rizzo and J.~Tattersall,
  JHEP {\bf 1510}, 182 (2015)
  [arXiv:1507.00013 [hep-ph]].

\bibitem{Deppisch:2015cua}
  F.~F.~Deppisch, L.~Graf, S.~Kulkarni, S.~Patra, W.~Rodejohann, N.~Sahu and U.~Sarkar,
  Phys.\ Rev.\ D {\bf 93}, no. 1, 013011 (2016)
  [arXiv:1508.05940 [hep-ph]].

\bibitem{Patra:2015bga}
  S.~Patra, F.~S.~Queiroz and W.~Rodejohann,
  Phys.\ Lett.\ B {\bf 752}, 186 (2016)
  [arXiv:1506.03456 [hep-ph]].

\bibitem{Aydemir:2015nfa}
  U.~Aydemir, D.~Minic, C.~Sun and T.~Takeuchi,
  Int.\ J.\ Mod.\ Phys.\ A {\bf 31}, no. 01, 1550223 (2016)
  [arXiv:1509.01606 [hep-ph]].



\bibitem{Aad:2015owa}
  G.~Aad {\it et al.} [ATLAS Collaboration],
  JHEP {\bf 1512}, 055 (2015)
  [arXiv:1506.00962 [hep-ex]].

\bibitem{Deppisch:2014zta}
  F.~F.~Deppisch, T.~E.~Gonzalo, S.~Patra, N.~Sahu and U.~Sarkar,
  Phys.\ Rev.\ D {\bf 91}, no. 1, 015018 (2015)
  [arXiv:1410.6427 [hep-ph]].

\bibitem{Dobrescu:2015qna}
  B.~A.~Dobrescu and Z.~Liu,
  Phys.\ Rev.\ Lett.\  {\bf 115}, no. 21, 211802 (2015)
  [arXiv:1506.06736 [hep-ph]].

\bibitem{Gao:2015irw}
  Y.~Gao, T.~Ghosh, K.~Sinha and J.~H.~Yu,
  Phys.\ Rev.\ D {\bf 92}, no. 5, 055030 (2015)
  [arXiv:1506.07511 [hep-ph]].

\bibitem{Dev:2015pga}
  P.~S.~B.~Dev and R.~N.~Mohapatra,
  Phys.\ Rev.\ Lett.\  {\bf 115}, no. 18, 181803 (2015)
  [arXiv:1508.02277 [hep-ph]].

\bibitem{Coloma:2015una}
  P.~Coloma, B.~A.~Dobrescu and J.~Lopez-Pavon,
  Phys.\ Rev.\ D {\bf 92}, no. 11, 115023 (2015)
  [arXiv:1508.04129 [hep-ph]].



\bibitem{Dobrescu:2015jvn}
  B.~A.~Dobrescu and P.~J.~Fox,
  arXiv:1511.02148 [hep-ph].

\bibitem{Sajjad:2015urz}
  A.~Sajjad,
  arXiv:1511.02244 [hep-ph].

\bibitem{Das:2015ysz}
  K.~Das, T.~Li, S.~Nandi and S.~K.~Rai,
  Phys.\ Rev.\ D {\bf 93}, no. 1, 016006 (2016)
  [arXiv:1512.00190 [hep-ph]].

\bibitem{Aydemir:2015oob}
  U.~Aydemir,
  arXiv:1512.00568 [hep-ph].

\bibitem{Das:2016akd}
  A.~Das, N.~Nagata and N.~Okada,
  arXiv:1601.05079 [hep-ph].

\bibitem{Shu:2016exh}
  J.~Shu and J.~Yepes,
  arXiv:1601.06891 [hep-ph].



\bibitem{Frere:2008ct}
  J.~M.~Frere, T.~Hambye and G.~Vertongen,
  JHEP {\bf 0901}, 051 (2009)
  [arXiv:0806.0841 [hep-ph]].

\bibitem{Dev:2014iva}
  P.~S.~B.~Dev, C.~H.~Lee and R.~N.~Mohapatra,
  Phys.\ Rev.\ D {\bf 90}, no. 9, 095012 (2014)
  [arXiv:1408.2820 [hep-ph]].

\bibitem{Dev:2015vra}
  P.~S.~B.~Dev, C.~H.~Lee and R.~N.~Mohapatra,
  J.\ Phys.\ Conf.\ Ser.\  {\bf 631}, no. 1, 012007 (2015)
  [arXiv:1503.04970 [hep-ph]].

\bibitem{Dhuria:2015cfa}
  M.~Dhuria, C.~Hati, R.~Rangarajan and U.~Sarkar,
  Phys.\ Rev.\ D {\bf 92}, no. 3, 031701 (2015)
  [arXiv:1503.07198 [hep-ph]].

\bibitem{ATLAS:2015}
  The ATLAS collaboration,
  ATLAS-CONF-2015-081.

\bibitem{CMS:2015dxe}
  The CMS Collaboration,
  CMS-PAS-EXO-15-004.

\bibitem{Dasgupta:2015pbr}
  A.~Dasgupta, M.~Mitra and D.~Borah,
  arXiv:1512.09202 [hep-ph].

\bibitem{Deppisch:2016scs}
  F.~F.~Deppisch, C.~Hati, S.~Patra, P.~Pritimita and U.~Sarkar,
  arXiv:1601.00952 [hep-ph].

\bibitem{Berlin:2016hqw}
  A.~Berlin,
  arXiv:1601.01381 [hep-ph].

\bibitem{Babu:1988mw}
  K.~S.~Babu and R.~N.~Mohapatra,
  Phys.\ Rev.\ Lett.\  {\bf 62}, 1079 (1989).

\bibitem{Babu:1989rb}
  K.~S.~Babu and R.~N.~Mohapatra,
  Phys.\ Rev.\ D {\bf 41}, 1286 (1990).

\bibitem{Dev:2015vjd}
  P.~S.~B.~Dev, R.~N.~Mohapatra and Y.~Zhang,
  arXiv:1512.08507 [hep-ph].

\bibitem{Mohapatra:2014qva}
  R.~N.~Mohapatra and Y.~Zhang,
  JHEP {\bf 1406}, 072 (2014)
  [arXiv:1401.6701 [hep-ph]].




\bibitem{Atre:2009rg} A.~Atre, T.~Han, S.~Pascoli and B.~Zhang,
  JHEP {\bf 0905}, 030 (2009)
  [arXiv:0901.3589 [hep-ph]].

\bibitem{Dev:2012zg} P.~S.~B.~Dev, R.~Franceschini and R.~N.~Mohapatra,
  Phys.\ Rev.\ D {\bf 86}, 093010 (2012) [arXiv:1207.2756 [hep-ph]].

\bibitem{Cely:2012bz}
  C.~G.~Cely, A.~Ibarra, E.~Molinaro and S.~T.~Petcov,
  Phys.\ Lett.\ B {\bf 718}, 957 (2013)
  [arXiv:1208.3654 [hep-ph]].

\bibitem{Dev:2013wba}
  P.~S.~B.~Dev, A.~Pilaftsis and U.~k.~Yang,
  Phys.\ Rev.\ Lett.\  {\bf 112}, no. 8, 081801 (2014)
  [arXiv:1308.2209 [hep-ph]].

\bibitem{Das:2014jxa}
  A.~Das, P.~S.~B.~Dev and N.~Okada,
  Phys.\ Lett.\ B {\bf 735}, 364 (2014)
  [arXiv:1405.0177 [hep-ph]].

\bibitem{Alva:2014gxa}
  D.~Alva, T.~Han and R.~Ruiz,
  JHEP {\bf 1502}, 072 (2015)
  [arXiv:1411.7305 [hep-ph]].

\bibitem{Antusch:2015mia}
  S.~Antusch and O.~Fischer,
  JHEP {\bf 1505}, 053 (2015)
  [arXiv:1502.05915 [hep-ph]].

\bibitem{Banerjee:2015gca}
  S.~Banerjee, P.~S.~B.~Dev, A.~Ibarra, T.~Mandal and M.~Mitra,
  Phys.\ Rev.\ D {\bf 92}, 075002 (2015)
  [arXiv:1503.05491 [hep-ph]].

\bibitem{Izaguirre:2015pga}
  E.~Izaguirre and B.~Shuve,
  Phys.\ Rev.\ D {\bf 91}, no. 9, 093010 (2015)
  [arXiv:1504.02470 [hep-ph]].

\bibitem{Gago:2015vma} A.~M.~Gago, P.~Hernandez, J.~Jones-Perez, M.~Losada and A.~Moreno Briceno,
  Eur.\ Phys.\ J.\ C {\bf 75}, no. 10, 470 (2015)
  [arXiv:1505.05880 [hep-ph]].

\bibitem{Asaka:2015oia}
  T.~Asaka and T.~Tsuyuki,
  Phys.\ Rev.\ D {\bf 92}, no. 9, 094012 (2015)
  [arXiv:1508.04937 [hep-ph]].

\bibitem{Das:2015toa}
  A.~Das and N.~Okada,
  Phys.\ Rev.\ D {\bf 93}, 033003 (2016)
  [arXiv:1510.04790 [hep-ph]].

\bibitem{deGouvea:2015euy}
  A.~de Gouvea and A.~Kobach,
  Phys.\ Rev.\ D {\bf 93}, 033005 (2016)
  [arXiv:1511.00683 [hep-ph]].

\bibitem{Antusch:2015gjw}
  S.~Antusch, E.~Cazzato and O.~Fischer,
  arXiv:1512.06035 [hep-ph].

\bibitem{Basso:2015aee}
  L.~Basso,
  arXiv:1512.06381 [hep-ph].

\bibitem{Dev:2016vif}
  P.~S.~B.~Dev and A.~Ibarra,
  arXiv:1601.01658 [hep-ph].


\bibitem{Belyaev:2012qa}
  A.~Belyaev, N.~D.~Christensen and A.~Pukhov,
  Comput.\ Phys.\ Commun.\  {\bf 184}, 1729 (2013)
  [arXiv:1207.6082 [hep-ph]].


\bibitem{Buckley:2014ana}
  A.~Buckley, J.~Ferrando, S.~Lloyd, K.~Nordstr\"{o}m, B.~Page, M.~R\"{u}fenacht, M.~Sch\"{o}nherr and G.~Watt,
  Eur.\ Phys.\ J.\ C {\bf 75}, 132 (2015)
  [arXiv:1412.7420 [hep-ph]].



\bibitem{Dulat:2015mca}
  S.~Dulat {\it et al.},
  arXiv:1506.07443 [hep-ph].

\bibitem{Djouadi:2005gi}
  A.~Djouadi,
  Phys.\ Rept.\  {\bf 457}, 1 (2008)
  [hep-ph/0503172].

\bibitem{Ball:2014uwa}
  R.~D.~Ball {\it et al.} [NNPDF Collaboration],
  JHEP {\bf 1504}, 040 (2015)
  [arXiv:1410.8849 [hep-ph]].

\bibitem{lhchiggs} \url{https://twiki.cern.ch/twiki/bin/view/LHCPhysics/HiggsEuropeanStrategy}

\bibitem{Harlander:2012pb}
  R.~V.~Harlander, S.~Liebler and H.~Mantler,
  Comput.\ Phys.\ Commun.\  {\bf 184}, 1605 (2013)
  [arXiv:1212.3249 [hep-ph]].

\bibitem{Degrande:2015vpa}
  C.~Degrande, M.~Ubiali, M.~Wiesemann and M.~Zaro,
  JHEP {\bf 1510}, 145 (2015)
  [arXiv:1507.02549 [hep-ph]].

\bibitem{Baglio:2014uba}
  J.~Baglio {\it et al.},
  arXiv:1404.3940 [hep-ph].

\bibitem{Basecq:1988cv}
  J.~Basecq and D.~Wyler,
  Phys.\ Rev.\ D {\bf 39}, 870 (1989).

\bibitem{Papaefstathiou:2015paa}
  A.~Papaefstathiou and K.~Sakurai,
  JHEP {\bf 1602}, 006 (2016)
  [arXiv:1508.06524 [hep-ph]].

\bibitem{Chen:2015gva}
  C.~Y.~Chen, Q.~S.~Yan, X.~Zhao, Y.~M.~Zhong and Z.~Zhao,
  Phys.\ Rev.\ D {\bf 93}, no. 1, 013007 (2016)
  [arXiv:1510.04013 [hep-ph]].

\bibitem{Torrielli:2014rqa}
  P.~Torrielli,
  arXiv:1407.1623 [hep-ph].


\bibitem{Zat100TeV}
\url{https://indico.cern.ch/event/437912/contribution/0/attachments/1164669/1678539/FCC-SM.pdf}

\bibitem{ZZat100TeV}
\url{https://indico.cern.ch/event/437912/session/1/contribution/6/attachments/1166328/1681749/talk.pdf}

\bibitem{Yong-Bai:2015xna}
  S.~Yong-Bai, Z.~Ren-You, M.~Wen-Gan, L.~Xiao-Zhou, Z.~Yu and G.~Lei,
  JHEP {\bf 1510}, 186 (2015)
  [arXiv:1507.03693 [hep-ph]].

\bibitem{Binoth:2008kt}
  T.~Binoth, G.~Ossola, C.~G.~Papadopoulos and R.~Pittau,
  JHEP {\bf 0806}, 082 (2008)
  [arXiv:0804.0350 [hep-ph]].

\bibitem{Lazopoulos:2007ix}
  A.~Lazopoulos, K.~Melnikov and F.~Petriello,
  Phys.\ Rev.\ D {\bf 76}, 014001 (2007)
  [hep-ph/0703273].


\bibitem{Melia:2011tj}
  T.~Melia, P.~Nason, R.~Rontsch and G.~Zanderighi,
  JHEP {\bf 1111}, 078 (2011)
  doi:10.1007/JHEP11(2011)078
  [arXiv:1107.5051 [hep-ph]].

\bibitem{Jager:2009xx}
  B.~Jager, C.~Oleari and D.~Zeppenfeld,
  Phys.\ Rev.\ D {\bf 80}, 034022 (2009)
  [arXiv:0907.0580 [hep-ph]].

\bibitem{delAguila:2008cj}
  F.~del Aguila and J.~A.~Aguilar-Saavedra,
  Nucl.\ Phys.\ B {\bf 813}, 22 (2009)
  [arXiv:0808.2468 [hep-ph]].

\bibitem{Khachatryan:2014wca}
  V.~Khachatryan {\it et al.} [CMS Collaboration],
  JHEP {\bf 1410}, 160 (2014)
  [arXiv:1408.3316 [hep-ex]].

\bibitem{ATLAS:2014lxa}
  The ATLAS collaboration [ATLAS Collaboration],
  ATLAS-CONF-2014-049.

\bibitem{Hajer:2015gka}
  J.~Hajer, Y.~Y.~Li, T.~Liu and J.~F.~H.~Shiu,
  JHEP {\bf 1511}, 124 (2015)
  [arXiv:1504.07617 [hep-ph]].

\bibitem{Gunion:2002zf}
  J.~F.~Gunion and H.~E.~Haber,
  Phys.\ Rev.\ D {\bf 67}, 075019 (2003)
  [hep-ph/0207010].

\bibitem{Carena:2013ooa}
  M.~Carena, I.~Low, N.~R.~Shah and C.~E.~M.~Wagner,
  JHEP {\bf 1404}, 015 (2014)
  [arXiv:1310.2248 [hep-ph]].

\bibitem{Dev:2014yca}
  P.~S.~B.~Dev and A.~Pilaftsis,
  JHEP {\bf 1412}, 024 (2014)
  [arXiv:1408.3405 [hep-ph]].



\bibitem{Barr:2014sga}
  A.~J.~Barr, M.~J.~Dolan, C.~Englert, D.~E.~Ferreira de Lima and M.~Spannowsky,
  JHEP {\bf 1502}, 016 (2015)
  [arXiv:1412.7154 [hep-ph]].

\bibitem{Baur:2009uw}
  U.~Baur,
  Phys.\ Rev.\ D {\bf 80}, 013012 (2009)
  [arXiv:0906.0028 [hep-ph]].

\bibitem{Dittmaier:2012nh}
  S.~Dittmaier and M.~Schumacher,
  Prog.\ Part.\ Nucl.\ Phys.\  {\bf 70}, 1 (2013)
  [arXiv:1211.4828 [hep-ph]].

\bibitem{Casas:2001sr}
  J.~A.~Casas and A.~Ibarra,
  Nucl.\ Phys.\ B {\bf 618}, 171 (2001)
  [hep-ph/0103065].

\end{thebibliography}
\end{document}